\documentclass[12pt]{article}
\usepackage[margin=1in]{geometry} 
\usepackage{dsfont}
\usepackage{amsmath,amsthm,amssymb,hyperref}
\usepackage{multirow}
\numberwithin{equation}{section}
\usepackage{mdframed,tcolorbox}
\usepackage{float}
\usepackage{graphicx,psfrag,epsf, subcaption, enumitem}
\usepackage{mathrsfs}
\usepackage[sort,round,authoryear]{natbib}
\usepackage{rotating} 

\usepackage{setspace}
\usepackage{algorithm}
\usepackage{algorithmic}

\newtheorem{remark}{Remark}
\newtheorem{prop}{Proposition}

\newtheorem{lemma}{Lemma}
\newtheorem{theorem}{Theorem}
\newtheorem{example}{Example}
\newtheorem{assumption}{Assumption} 

\usepackage{titlesec}
\usepackage{booktabs,graphicx}
\titleformat{\section}
{\normalfont\large\bfseries}
{\thesection}{1em}{}

\titleformat{\subsection}
{\normalfont\normalsize\bfseries}
{\thesubsection}{1em}{}

\title{Spectral Regularization for Regression Specification Testing}
\author{%
  \begin{tabular}{@{}c@{}}
    Shiwei Sang\textsuperscript{1}
    \quad
    Shao-Bo Lin\textsuperscript{2,}\thanks{%
      Corresponding authors: Shao-Bo Lin (\texttt{sblin1983@gmail.com})
      and Xuehu Zhu (\texttt{zhuxuehu@xjtu.edu.cn}).}
    \quad
    Xuehu Zhu\textsuperscript{1,}\footnotemark[1]\\[0.6em]
    {\small \textsuperscript{1}School of Mathematics and Statistics}\\
    {\small \textsuperscript{2}Center for Intelligent Decision-Making and Machine Learning,}\\
    {\small School of Management}\\
    {\small Xi'an Jiaotong University, Xi'an, China}
  \end{tabular}%
}
\date{}
\begin{document}
	
	\maketitle
	
	\begin{abstract}
		While conventional regression specification tests rely on
		paired  observations, obtaining reliable
		responses is often more costly and time-consuming than
		collecting covariates.
		In this paper, we develop a kernel-embedding approach to
		parametric regression specification testing using both
		covariate--response pairs and auxiliary covariate data available from various sources.
		The proposed statistic combines two building blocks,
		fitted residuals from labeled observations and a
		Tikhonov-regularized integral operator estimated from auxiliary
		covariates, to enhance sensitivity to specific model departures.
		We establish a standard normal null limit for the studentized
		statistic, yielding an asymptotically pivotal test.
		Our power analysis characterizes its detection boundary under
		suitable regularity conditions and establishes a uniform
		detection advantage over the oracle maximum moment
		restriction (MMR) benchmark.
		We further derive a minimax lower bound to assess the
		optimality of the proposed test.
		Simulation studies and a real-data analysis illustrate
		the finite-sample performance and practical use of our method.
		
	\end{abstract}
	
	\section{Introduction}
	\label{sec: introduction}
	As one of the most powerful tools in empirical research, regression analysis enables practitioners to understand and quantify the relationship between the response and covariates. In many applied fields--particularly economics and finance--parametric regression models are widely adopted for their interpretability and policy relevance~\citep{Brooks_2019}. Despite their practical appeal, the reliability of the resulting analyses depends critically on
	whether the parametric models are correctly specified. 
	This concern has motivated a large literature on regression specification testing; see~\citet{gonzalez2013updated} for a comprehensive review.
	
	Formally, let $Y\in\mathbb{R}$ denote the response variable and $X\in\mathcal{X}\subset\mathbb{R}^p$ the covariate vector. Write $m(\cdot)=\mathbb{E}(Y\mid X=\cdot)$ for the unknown regression function, and consider a prespecified parametric family
	$\mathcal{G}(\Theta):=\{g(\cdot,\theta):\theta\in\Theta\subset\mathbb{R}^d\}$.
	The conditional mean specification problem is to test
	\begin{equation}
		\label{equ: testing problem}
		\begin{aligned}
			H_0: 
			\mathbb{P}\{m(X)=g(X,\theta_0)\}=1,
			\quad\exists \, \theta_0\in\Theta,
			\quad \mathrm{versus} \quad 
			H_1:
			\mathbb{P}\{m(X)\neq g(X,\theta)\}>0,
			\quad\forall \,\theta\in\Theta.
		\end{aligned}
	\end{equation} 
	Existing approaches to testing~\eqref{equ: testing problem}
	are commonly classified into local-smoothing and
	global-smoothing tests.
	Local-smoothing tests use nonparametric smoothing to construct
	measures of departure from the parametric specification;
	representative examples include
	\citet{hardle1993comparing,zheng1996consistent,fan1996consistent,
		fan2001generalized,dette1999consistent,zhang2004power}.
	Global-smoothing tests typically assess residual-based
	unconditional moment restrictions indexed by rich classes
	of weighting functions, using suitable functionals of the corresponding
	empirical processes~\citep{bierens1982consistent,
		bierens1990consistent,su1991lack,stute1997nonparametric,
		stute1998model}.

	Despite their different constructions, these classical tests
	were developed for supervised settings and rely on fully
	observed covariate--response pairs.
	In many modern applications, however, reliable responses are costly
	or time-consuming to obtain~\citep{zrnic2024cross}.
	For example, ascertaining   clinical outcomes from
	electronic health records often requires manual chart review,
	so validated outcomes are available for relatively few
	patients~\citep{hou2023surrogate}.
	Similarly, in lending, borrower information is available
	when loans are issued, whereas repayment outcomes require
	subsequent follow-up~\citep{kim2019ensemble}.
	Covariate data, by contrast, are often more readily   available from
	a variety of sources, such as records with missing  responses
	in the original dataset,  historical or external databases,
	and synthetic samples produced by generative models or simulators. 
	The use of unlabeled covariates has been studied in
	semi-supervised inference~\citep{zhang2019semi,
		zrnic2024cross,kim2025semi},
	covariate-shift adaptation~\citep{sugiyama2008direct},
	and unsupervised domain adaptation
	\citep{bendavid2010theory,ganin2016domain}. Although these developments highlight the statistical value of
	auxiliary data,   their use in regression
	specification testing has received  less attention.

	This work studies kernel-embedding approaches
	to the parametric specification problem
	in~\eqref{equ: testing problem}
	within the reproducing kernel Hilbert space (RKHS) framework,
	using both covariate--response pairs and auxiliary covariates.
	Our starting point is recent advances in minimax kernel
	distribution testing~\citep{balasubramanian2021optimality,
		hagrass-gof-test,hagrass-two-sample-test,hagrass2026ksd}.
	These studies have shown that tests based on maximum mean
	discrepancy (MMD) with a fixed kernel can be suboptimal under
	$L^2$-type separations.
	This limitation arises because the squared MMD assigns
	smaller weights to directions associated with smaller kernel
	eigenvalues, an effect that can be mitigated by spectral
	regularization through a transformation of the underlying
	kernel operators.
	In practice, implementing this regularization typically
	involves estimating the operator using auxiliary samples.
	Since a similar spectral weighting appears in the maximum
	moment restriction (MMR) formulation of kernel conditional
	moment tests~\citep{muandet2020kernel}, the same strategy
	may also improve the detection of regression misspecification. 
	
	To evaluate this possibility, we adopt a minimax
	framework~\citep{ingster1987minimax,ingster2012nonparametric,
		guerre2002optimal} and quantify misspecification by the mean squared distance  between the true regression function and its best parametric
	approximation. Specifically, define the conditional
	residual function
	\begin{equation}
		\label{equ: conditional residual function}
		\mathbf f(\cdot)
		:=m(\cdot)-g(\cdot,\theta^\star),
		\qquad
		\theta^\star
		:=\arg\min_{\theta\in\Theta}
		\mathbb E\{\epsilon^2(Z,\theta)\},
	\end{equation}
	where $\epsilon(Z,\theta):=Y-g(X,\theta)$ and
	$Z:=(X^\top,Y)^\top$. We consider the composite alternatives
	\begin{equation}
		\label{equ: H1Deltan}
		H_1(\mathcal F,\Delta_n)
		:=
		\left\{
		\mathbf f\in\mathcal F\subset L^2(P_X):
		\|\mathbf f\|_{L^2(P_X)}^2\geq\Delta_n
		\right\},
	\end{equation}
	where $P_X$ is the marginal distribution of the labeled covariates.
	At a fixed significance level, the \emph{detection boundary}
	of a test describes the smallest order of the separation
	$\Delta_n$ needed to attain a prescribed nontrivial power uniformly
	over this alternative class. The best achievable order among all valid
	tests is the \emph{minimax-optimal detection boundary}.
	
	Although auxiliary samples have been shown   useful in kernel
	distribution testing~\citep{hagrass-gof-test,
		hagrass-two-sample-test,hagrass2026ksd,
		mukherjee2025randomfeatures}, extending these benefits to
	regression problems is not straightforward.
	Unlike goodness-of-fit testing against a fixed null
	distribution, our null hypothesis in~\eqref{equ: testing problem}
	encompasses an entire parametric family of regression functions.
	Fitting this family introduces an additional estimation error
	that interacts with regularization and must be controlled for both null calibration and power uniformly
	over alternatives.
	A further difficulty arises from differences in the data
	available for regularization.
	Existing approaches use known population spectral
	quantities~\citep{balasubramanian2021optimality},
	estimate the relevant operator from null reference
	samples~\citep{hagrass-gof-test},
	or use independent subsamples of the observed
	data~\citep{hagrass-two-sample-test,hagrass2026ksd,
		mukherjee2025randomfeatures}.
	In our setting, however, auxiliary covariates come from various sources, and their distribution is generally unknown
	and may differ from that of the labeled covariates.
	The regularized statistic must therefore remain sensitive
	to model misspecification despite this distributional difference.
	These challenges motivate us to investigate whether
	spectral regularization informed by auxiliary covariates
	can improve testing performance and, in particular, yield minimax-optimal   tests in the regression setting. 
	\subsection{Contributions and Overview}
	Our main contributions are summarized as follows:
	
	\medskip
	\noindent
	$\bullet$ \textbf{Methodological novelty.} At the population level, we construct a discrepancy from
	the residual mean embedding using Tikhonov regularization
	of a kernel integral operator.
	To obtain the empirical counterpart, we estimate this operator
	using auxiliary covariates and combine it with fitted residuals
	from the labeled observations, yielding the
	\emph{spectral regularized moment restriction} (SRMR) statistic.
    
	
	\medskip
	\noindent
	$\bullet$ \textbf{Theoretical guarantees.}  
	We establish a pivotal standard normal limit for the
	studentized statistic under the null, allowing calibration
	using normal quantiles
	(Theorem~\ref{theorem: asymptotic null distribution}).
	For alternatives whose conditional residuals  lie in a source ball defined through the integral
	operator, we establish a strict improvement in uniform
	detection over oracle MMR.
	Specifically, we identify  separation conditions in which our
	test is uniformly consistent while oracle MMR lacks
	nontrivial uniform power
	(Theorem~\ref{theorem: uniform power}).
	We further derive a minimax lower bound to assess the
	optimality of the achieved detection rate
	(Theorem~\ref{theorem: minimax optimality}).
	
	\medskip
	\noindent
	$\bullet$ \textbf{Practical relevance.}   We examine the finite-sample implications of these results
	through simulation studies, investigating the effects of
	spectral regularization across different settings and
	comparing the proposed test with classical specification tests.
	We further analyze the Los Angeles homelessness data to
	illustrate how our method  can be used in assessing
	regression model specifications.
	
	\medskip
	
	The rest of the paper is organized as follows.
	In Section~\ref{sec: statistic construction}, we review the
	RKHS background and MMR formulation, and present the
	construction of the proposed statistic.
	In Section~\ref{sec: test construction and theoretical properties},
	we present the testing procedure and establish its validity
	and power properties.
	In Section~\ref{sec: numerical studies},
	we examine the finite-sample performance of the proposed test
	and illustrate its application to real data.  In Section~\ref{sec: conclusion}, we summarize our findings and discuss
directions for future research. In
Appendices~\ref{sec: error analysis}--\ref{sec: addtional lemmas and proofs},
we provide the error analysis and proofs of the main results.
        \subsection{Notation}
	For a finite nonnegative Borel measure $P$ on $\mathcal X$
	and $1\leq q< \infty$, let $L^q(P)$ be the space of
	measurable  functions with finite norm
	$\|h\|_{L^q(P)}:=
	\{\int_{\mathcal X}|h(x)|^q\,dP(x)\}^{1/q}$. For $q = 2$,
	the inner product on $L^2(P)$ is
	$\langle h_1,h_2\rangle_{L^2(P)}
	:=\int_{\mathcal X}h_1(x)h_2(x)\,dP(x)$.
	The space $L^\infty(P)$ is equipped with the essential
	supremum norm
	$\|h\|_{L^\infty(P)}
	:=\operatorname*{ess\,sup}_{x\sim P}|h(x)|$. 	 	For a probability measure $P$, $P^{\otimes n}$ denotes
	its $n$-fold product measure. We write $\mathbb{E}(\cdot)$, $\mathrm{Var}(\cdot)$, and $\mathrm{Cov}(\cdot)$ for expectation, variance, and covariance, respectively.
	We write $\mathbb E(\cdot\mid X)$ and
	$\mathrm{Var}(\cdot\mid X)$ for conditional expectation
	and variance given $X$, respectively.
	A subscript $P$, as in $\mathbb E_P$ and $\mathbb P_P$,
	specifies the underlying probability measure. We write $\xrightarrow{p}$ and $\leadsto$
	for convergence in probability and weak convergence,
	respectively.
	For a positive deterministic sequence $a_n$,
	$V_n=o_p(a_n)$ means $V_n/a_n\xrightarrow{ p}0$,
	whereas $V_n=O_p(a_n)$ means that $V_n/a_n$ is bounded
	in probability.
	 
	 For a vector $\mathbf a$ and a matrix $A$, write
		$\mathbf a^\top$ and $A^\top$ for their transposes,
		$a_j$ for the $j$-th coordinate of $\mathbf a$,
		and $A_{ij}$ for the $(i,j)$-th entry of $A$.
		The Euclidean norm of $\mathbf a$ is denoted by $\|\mathbf a\|_2$. 	Let $\mathcal H,\mathcal H_1,\mathcal H_2$ denote real Hilbert spaces.
		For a bounded linear operator $A:\mathcal H_1\to\mathcal H_2$,
		the unsubscripted norm $\|A\|$ denotes the operator norm.
		The same convention applies to matrices and bounded bilinear maps.
		We write $A^\ast$ for the adjoint of $A$ and
		$\operatorname{Tr}(A)$ for its trace, whenever defined.
		For Hilbert--Schmidt operators $A$ and $B$, define
		$\langle A,B\rangle_{HS}:=\operatorname{Tr}(A^\ast B)$
		and $\|A\|_{HS}:=\{\operatorname{Tr}(A^\ast A)\}^{1/2}$.
		For $u,v\in\mathcal H$, the rank-one operator
		$u\otimes_{\mathcal H}v$ is defined by
		$	(u\otimes_{\mathcal H}v)h
		:=u\langle v,h\rangle_{\mathcal H}$ for $ h\in\mathcal H,$
		where $\langle\cdot,\cdot\rangle_{\mathcal H}$ denotes
		the inner product on $\mathcal H$.
		The subscript is omitted when the space is clear.
		For self-adjoint operators $A$ and $B$ on $\mathcal H$,
		$A\preceq B$ means
		$\langle(B-A)h,h\rangle_{\mathcal H}\geq0$
		for every $h\in\mathcal H$.
		For a symmetric matrix $A$, $\lambda_{\min}(A)$ denotes
		its smallest eigenvalue. For matrices $A$ and $B$ of the same dimension, $A\circ B$ denotes their Hadamard product. For an event $E$, $\mathds{1}\{E\}$ denotes its indicator. 
		The Kronecker delta is $\delta_{ij}:=\mathds{1}\{i=j\}$.
		For $a\in\mathbb R$, let $a_+:=\max\{a,0\}$.
		For a finite set $S$, $|S|$ denotes its cardinality.
		For positive quantities $a$ and $b$, write $a\lesssim b$
		if $a\leq Cb$ for a constant $C>0$, and write
		$a\asymp b$ if both $a\lesssim b$ and $b\lesssim a$ hold.
		For $f\in\mathcal H$ and a subspace $\mathcal T\subseteq\mathcal H$,
		write $f\perp\mathcal T$ if
		$\langle f,h\rangle_{\mathcal H}=0$ for every $h\in\mathcal T$. Let $\operatorname{span}\{f_1,\ldots,f_n\}$ denote
		the linear space spanned by $f_1,\ldots,f_n$.

	\section{Spectral Regularized Kernel Statistics with Auxiliary Covariates}
	\label{sec: statistic construction}
	
	\subsection{Kernel Embeddings and Maximum Moment Restriction}
	Let $K:\mathcal X\times\mathcal X\to\mathbb R$ be a Mercer
	(continuous, symmetric, and positive semidefinite) kernel satisfying
	$\sup_{x\in\mathcal X}K(x,x)\leq\kappa^2$ for some $\kappa>0$.
	Denote by $(\mathcal H_K,\langle\cdot,\cdot\rangle_K)$ the
	reproducing kernel Hilbert space (RKHS) induced by $K$, namely,
	the completion of the linear span of
	$\{K_x:x\in\mathcal X\}$ under the inner product induced by $K$,
	where $K_x:=K(x,\cdot)$. The reproducing property gives
	$\langle K_x,h\rangle_K=h(x)$ for every $h\in\mathcal H_K$.
	
	The kernel $K$ is said to be \emph{integrally strictly positive
		definite} (ISPD) with respect to $P_X$ if
	\begin{equation}
		\label{equ: ISPD}
		\iint_{\mathcal X\times\mathcal X}
		f(x)K(x,x')f(x')\,dP_X(x)\,dP_X(x')>0,
		\qquad
		\forall f\in L^2(P_X)\setminus\{0\}.
	\end{equation}
	This property yields an equivalent RKHS characterization of
	the hypotheses in~\eqref{equ: testing problem}, as formalized
	in the following lemma~\citep{muandet2020kernel}.
	
	\begin{lemma}
		\label{lemma: RKHS_equivalence}
		Let $(\mathcal H_K,\langle\cdot,\cdot\rangle_K)$ be the RKHS
		induced by a bounded Mercer kernel $K$ satisfying
		\eqref{equ: ISPD}. For every fixed $\theta\in\Theta$ such that
		$\mathbb E\{\epsilon^2(Z,\theta)\}<\infty$, define
		$	\mu(\theta)
		:=\mathbb E\{\epsilon(Z,\theta)K_X\}\in\mathcal H_K$.
		Then
		\[
		\mathbb E\{\epsilon(Z,\theta)\mid X\}=0
		\quad P_X\text{-a.s.}
		\iff
		\sup_{\substack{h\in\mathcal H_K\\ \|h\|_K\leq1}}
		\big|\mathbb E\{\epsilon(Z,\theta)h(X)\}\big|=0
		\iff
		\mu(\theta)=0,
		\]
		where $\|\cdot\|_K$ denotes the RKHS norm induced by
		$\langle\cdot,\cdot\rangle_K$.
	\end{lemma}
	
	The element $\mu(\theta)\in\mathcal H_K$ is called the
	\emph{conditional moment embedding} (CMME). Its RKHS norm,
	$\mathrm{MMR}(\theta):=\|\mu(\theta)\|_K$, is called the
	\emph{maximum moment restriction} (MMR)~\citep{muandet2020kernel}.
	By the reproducing property, the squared MMR admits the representation
	\begin{equation}
		\label{equ: MMR}
		\mathrm{MMR}^2(\theta)
		:=\|\mu(\theta)\|_K^2
		=
		\mathbb E\left\{
		\epsilon(Z_1,\theta)K(X_1,X_2)
		\epsilon(Z_2,\theta)
		\right\},
	\end{equation}
	where $Z_1$ and $Z_2$ are independent copies of $Z$. Given independent and identically distributed (i.i.d.) observations
	$\{Z_i\}_{i=1}^n=\{(X_i^\top,Y_i)^\top\}_{i=1}^n$,
	an unbiased estimator of $\mathrm{MMR}^2(\theta)$ 
	is the second-order $U$-statistic
	\begin{equation}
		\label{equ: MMR U-statistic}
		\widehat{\mathrm{MMR}}^2(\theta)
		:=
		\frac{1}{n(n-1)}
		\sum_{i\neq j}
		\epsilon(Z_i,\theta)K(X_i,X_j)\epsilon(Z_j,\theta).
	\end{equation}
	Under $H_0$, treating the null parameter $\theta_0$ as known,  
	\citet{muandet2020kernel} established that
	\[
	n\widehat{\mathrm{MMR}}^2(\theta_0)
	\leadsto
	\sum_{k\geq1}\tau_k(W_k^2-1)
	=:W,
	\qquad
	W_k\stackrel{\mathrm{i.i.d.}}{\sim}N(0,1),
	\]
	where the  weights  $\{\tau_k\}_{k\geq1}$  depend on the kernel, the parametric model and the data distribution. Accordingly, for a given significance level $\alpha\in(0,1)$,
	an asymptotically $\alpha$-level test is defined by
	\begin{equation}
		\label{equ: MMR test}
		T_{\mathrm{MMR}}^{\mathrm{or}}
		:=
		\mathds{1}\left\{
		n\widehat{\mathrm{MMR}}^2(\theta_0)>q_{1-\alpha,W}
		\right\},
	\end{equation}
	where $q_{1-\alpha,W}$ denotes the $(1-\alpha)$-quantile of $W$.
	The superscript ``$\mathrm{or}$'' denotes an oracle test, as in practical implementation, the parameter is unknown and  requires additional estimation.
	\subsection{Test Statistics with Auxiliary Data}
	Our primary interest is to understand how auxiliary covariate data
	can be incorporated into kernel-embedding statistics and
	whether they can improve the parametric specification test.   Although the MMR-based test  provides a natural benchmark for testing \eqref{equ: testing problem}, with a fixed kernel, however,
	the statistic in~\eqref{equ: MMR U-statistic} is constructed
	entirely from covariate--response pairs and does not directly use
	auxiliary covariate data. To motivate our construction, we compare the spectral
	representations of the squared MMR and the squared
	$L^2(P_X)$-norm of the conditional residual  $\mathbf{f} $ in \eqref{equ: conditional residual function}.
	
	We first introduce some operator notation. For a probability
	measure $P$ on $\mathcal X$, define the canonical inclusion
	$I_{K,P}:\mathcal H_K\to L^2(P)$ by $h\mapsto[h]_P$,
	where $[h]_P$ denotes the equivalence class of $h$ in $L^2(P)$.
	Its adjoint $I_{K,P}^\ast:L^2(P)\to\mathcal H_K$    is given by $I_{K,P}^\ast h
	=
	\int_{\mathcal X}K_xh(x)\,dP(x)$.
	Define
	\begin{equation}
		\label{equ: integral operators}
		\mathcal L_{K,P}:=I_{K,P}I_{K,P}^\ast,
		\qquad
		L_{K,P}:=I_{K,P}^\ast I_{K,P}.
	\end{equation}
	Then $\mathcal{L}_{K,P}$ and $L_{K,P}$ are referred to as \emph{integral operators} on $L^2(P)$ and $\mathcal{H}_K$, respectively. For notational convenience, we write
	$I_{K,a}:=I_{K,P_a}$,
	$I_{K,a}^{\ast}:=I_{K,P_a}^{\ast}$,
	$\mathcal L_{K,a}:=\mathcal L_{K,P_a}$, and
	$L_{K,a}:=L_{K,P_a}$ for $a\in\{X,0\}$,
	where $P_0$ will be introduced in
	Assumption~\ref{assumption: auxiliary covariate distribution}.	Using this notation, the squared MMR at the pseudo-true parameter $\theta^\star$ admits the expansion
	\begin{equation}
		\label{equ: MMR expansion}
		\begin{aligned}
			\mathrm{MMR}^2(\theta^\star)
			&=\left\|I_{K,X}^\ast\mathbf f\right\|_K^2 =\left\langle\mathcal L_{K,X}\mathbf f,
			\mathbf f\right\rangle_{L^2(P_X)} =\sum_{k\geq1}\sigma_{X,k}
			\left\langle\mathbf f,\phi_{X,k}\right\rangle_{L^2(P_X)}^2,
		\end{aligned}
	\end{equation}
	where 
	$\{(\sigma_{X, k},\phi_{X, k})\}_{k\geq1}$ are normalized
	eigenpairs of $\mathcal L_{K,X}$, with
	$\{\phi_{X, k}\}_{k\geq1}$ forming an orthonormal basis of
	$L^2(P_X)$. Although  the squared MMR is positive
	for any $\mathbf{f} \neq 0$, it weights the Fourier coefficients $ \langle\mathbf f,\phi_{X,k} \rangle_{L^2(P_X)}$ by the eigenvalue sequence  
	$\{\sigma_{X,k}\}_{k\geq 1}$, which tend to zero as $k$ increases. Compared with 
	\begin{equation}
		\label{equ: L2-norm}
		\|\mathbf f\|_{L^2(P_X)}^2
		= 
		\sum_{k\geq1}
		\langle\mathbf f,\phi_{X,k}\rangle_{L^2(P_X)}^2,
	\end{equation}  
	departures concentrated in small-eigenvalue directions receive
	less weight, potentially limiting the sensitivity of MMR-based
	tests under $L^2(P_X)$ separation.
	
	
	These observations motivate modifying the spectral weighting
	of  MMR through  regularization methods~\citep{engl1996regularization,guo2017learning}, so that the resulting
	discrepancy better reflects the $L^2(P_X)$ energy of regression
	departures. Inspired by recent developments in distributional tests~\citep{balasubramanian2021optimality,	hagrass-gof-test,hagrass-two-sample-test,hagrass2026ksd}, we develop this construction with the the aid of auxiliary covariates, while allowing their distributions  $P_0$ to differ from $P_X$, subject to the following assumption.

	%
	
	\begin{assumption}[Auxiliary covariates]
		\label{assumption: auxiliary covariate distribution}
		The auxiliary covariate sample
		$D=\{X_{0\ell}\}_{\ell=1}^N$ consists of i.i.d. observations
		from $P_0$ and is independent of the labeled sample
		$\{Z_i\}_{i=1}^n$. Moreover, $P_0$ is absolutely continuous
		with respect to $P_X$, and
		\[
		\underline c
		\leq w(x):=\frac{dP_0}{dP_X}(x)
		\leq\overline c,
		\qquad P_X\text{-a.s.},
		\]
		where $0<\underline c\leq1\leq\overline c<\infty$.
	\end{assumption}
	
	\begin{remark}
		The density-ratio bounds in Assumption~\ref{assumption: auxiliary covariate distribution} allow covariate shift while ensuring
		that the two distributions have the same support. In particular, $P_0$
		and $P_X$ are mutually absolutely continuous, and their
		$L^2$-norms are equivalent. The special case $P_0=P_X$, with $w=1$ almost surely,
		recovers the standard semi-supervised setting, where labeled
		and unlabeled covariates share the same marginal
		distribution~\citep{zhang2019semi,kim2025semi}.  
	\end{remark}

	Based on Assumption~\ref{assumption: auxiliary covariate distribution},	we implement the spectral adjustment using Tikhonov
	regularization~\citep{engl1996regularization}. Given a fixed $\lambda >0$, define the
	\emph{spectral regularized moment restriction} (SRMR) by
	\begin{equation}
		\label{equ: SRMR}
		\begin{aligned}
			\eta_\lambda(\theta)
			&:=
			\left\|(L_{K,0}+\lambda I)^{-1/2}
			\mu(\theta)\right\|_K^2 =
			\left\langle
			\mu(\theta),
			(L_{K,0}+\lambda I)^{-1}\mu(\theta)
			\right\rangle_K.
		\end{aligned}
	\end{equation}
	Since the operator
	$(L_{K,0}+\lambda I)^{-1}$ is strictly positive for each fixed $\lambda>0$, 
	$\eta_\lambda(\theta)=0$ if and only if $\mu(\theta)=0$.
	The SRMR therefore preserves the RKHS characterization of
	the hypotheses in~\eqref{equ: testing problem} given by
	Lemma~\ref{lemma: RKHS_equivalence}.  	To further explain our construction,  we
	consider its value at $\theta^\star$. A change of measure gives
	\[
	\mu(\theta^\star)
	=
	I_{K,X}^\ast\mathbf f
	=
	\int_{\mathcal X}
	K_x\frac{\mathbf f(x)}{w(x)}\,dP_0(x)
	=
	I_{K,0}^\ast\left(\frac{\mathbf f}{w}\right).
	\]
	Assumption~\ref{assumption: auxiliary covariate distribution}
	ensures that $\mathbf f/w\in L^2(P_0)$ whenever
	$\mathbf f\in L^2(P_X)$. Hence, the spectral decomposition
	of $\mathcal L_{K,0}$ yields
	\begin{equation}
		\label{equ: SRMR expansion}
		\eta_\lambda(\theta^\star)
		=
		\sum_{k\geq1}
		\frac{\sigma_{0,k}}{\sigma_{0,k}+\lambda}
		\left\langle
		\frac{\mathbf f}{w},\,\phi_{0,k}
		\right\rangle_{L^2(P_0)}^2,
	\end{equation}
	where $\{(\sigma_{0,k},\phi_{0,k})\}_{k\geq1}$ are the
	positive eigenpairs of $\mathcal L_{K,0}$, with
	$\{\phi_{0,k}\}_{k\geq1}$ forming an orthonormal basis of
	$L^2(P_0)$. 
	If $K$ is an ISPD kernel,  monotone convergence theorem, together with the density-ratio bounds in Assumption~\ref{assumption: auxiliary covariate distribution} implies
	\begin{equation}
		\label{equ: limit discrepancy}
		\overline c^{-1}\|\mathbf f\|_{L^2(P_X)}^2
		\leq
		\lim_{\lambda\downarrow0}\eta_\lambda(\theta^\star) = \left\|\frac{\mathbf f}{w}\right\|_{L^2(P_0)}^2
		=
		\int_{\mathcal X}\frac{\mathbf f^2(x)}{w(x)}\,dP_X(x)
		\leq
		\underline c^{-1}\|\mathbf f\|_{L^2(P_X)}^2.
	\end{equation}
	Thus, the limiting discrepancy is equivalent to the squared
	$L^2(P_X)$-norm in~\eqref{equ: L2-norm} and equals it when
	$P_0=P_X$, and the attenuation induced by the
	kernel eigenvalues in~\eqref{equ: MMR expansion}
	disappears.
	
	Based on auxiliary samples  $D=\{X_{0\ell}\}_{\ell = 1}^N$, the empirical integral operator $L_{K,D} : \mathcal{H}_K \to \mathcal{H}_K$ for estimating $L_{K,0}$ is defined by
	\begin{equation}
		\label{equ: empirical integral operator}
		L_{K,D}\,h
		:=
		\frac{1}{N}
		\sum_{\ell=1}^N K_{x_{0\ell}}h(x_{0\ell}),
		\qquad h\in\mathcal{H}_K.
	\end{equation}
	Combining $L_{K,D}$ with the labeled sample $\{Z_i\}_{i = 1}^n$ gives the
	fully empirical SRMR statistic
	\begin{equation}
		\label{equ: totally empirical statistic}
		\widehat{\eta}_\lambda(\widehat{\theta})
		:=
		\frac{1}{n(n-1)}
		\sum_{i\neq j}
		\left\langle
		(L_{K,D} + \lambda I)^{-1/2}K_{X_i}, (L_{K,D} + \lambda I)^{-1/2}K_{X_j}
		\right\rangle_K 	\epsilon(Z_i,\widehat{\theta}\,)
		\epsilon(Z_j,\widehat{\theta}\,),
	\end{equation}
	where $\widehat\theta$ is the least square estimator of the pseudo-true parameter $\theta^\star$.
	Although the density ratio $w$ enters the population discrepancy in \eqref{equ: SRMR expansion}, computing this statistic does not require
	estimating it.  The following proposition gives a finite-dimensional
	representation of the statistic.
		\begin{prop}
			\label{prop: computation}
			Define the kernel matrices $	K_{nn}:=\{K(X_i,X_j)\}_{n\times n}$, $	K_{NN}:=\{K(X_{0i},X_{0j})\}_{N \times N}$, $	K_{nN}:=\{K(X_i,X_{0j})\}_{n \times N}$,  and $	R_n:=
			\operatorname{diag}\{
			\epsilon(Z_1,\widehat\theta),\ldots,
			\epsilon(Z_n,\widehat\theta)
			\}$.
			For $\lambda>0$, let $\widehat\eta_\lambda(\widehat\theta)$ be defined
			in~\eqref{equ: totally empirical statistic}. Then
			\[
			\widehat\eta_\lambda(\widehat\theta)
			=
			\frac{
				\mathbf1_n^\top M_n\mathbf1_n
				-\operatorname{Tr}(M_n)
			}{n(n-1)}, \qquad 	M_n
			:=
			\lambda^{-1}
			R_n\left\{
			K_{nn}
			-K_{nN}(K_{NN}+N\lambda I)^{-1}K_{nN}^\top
			\right\}R_n.
			\]
			where  
			$\mathbf1_n$ denotes the   $n$-dimensional all-one vector.
		\end{prop}
		\section{Test Calibration and Detection Properties}
		\label{sec: test construction and theoretical properties}
		We begin by stating the assumptions on the parametric model and the kernel function that will be maintained throughout this paper.
		\begin{assumption}[Parametric regularity and identification]
		\label{assumption: regularities on parametric families}
		The set $\Theta\subset\mathbb R^d$ is convex. For every $x$,
		$g(x,\cdot)$ is twice continuously differentiable on an open set
		containing $\Theta$.
		\begin{itemize}
			\item[(i)]
			Define
			$J_g(x,\theta)=\nabla_\theta g(x,\theta)\in\mathbb R^d$
			and
			$H_g(x,\theta)=\nabla_\theta^2 g(x,\theta)
			\in\mathbb R^{d\times d}$.
			There exist measurable  envelopes $M_J$ and $M_H$
			such that $	\sup_{\theta\in\Theta}\|J_g(x,\theta)\|_2\leq M_J(x), 
			\sup_{\theta\in\Theta}\|H_g(x,\theta)\|\leq M_H(x),$
			and
			$\|M_J\|_{L^2(P_X)}+\|M_H\|_{L^1(P_X)}
			\leq C_g<\infty$.
			
			\item[(ii)]
			At a fixed null parameter $\theta_0\in\Theta$,
			$g(\cdot,\theta_0)\in L^2(P_X)$ and $	Q_0
			:=
			\mathbb E\{
			J_g(X,\theta_0)J_g(X,\theta_0)^\top
			\}$ is nonsingular.
			Moreover, for every $\varepsilon>0$, $ \inf_{\theta\in\Theta: \|\theta-\theta_0\|_2\geq\varepsilon}
			\|g(\cdot,\theta)-g(\cdot,\theta_0)\|_{L^2(P_X)}
			>0.$
		\end{itemize}
	\end{assumption}
		Assumption~\ref{assumption: regularities on parametric families}~\textit{(i)} provides the smoothness and domination conditions needed
		to control the effect of parameter estimation. 
		Related regularity conditions appear in regression specification
		testing; see
		\citet{zheng1996consistent,guerre2002optimal,
			horowitz2006testing,hitomi2023optimal}.
		We do not require the gradient and Hessian to be
		uniformly bounded, as imposed by
		\citet{guerre2002optimal,horowitz2006testing}, and  thus allow unbounded
		covariates and derivatives.
		
		Assumption~\ref{assumption: regularities on parametric families}~\textit{(ii)} imposes local nondegeneracy and  global
		$L^2(P_X)$ separation
		at a  fixed null  parameter $\theta_0$, which is used in the lower-bound
		constructions.   Related identification conditions conditions appear in minimax specification
		testing~\citep{guerre2002optimal,hitomi2023optimal}.
		For example, Assumption M2(ii) of
		\citet{guerre2002optimal} requires $\mathbb E\{
		J_g(X,\theta)J_g(X,\theta)^\top
		\} $
		to be nonsingular for every $\theta\in\Theta$. At an interior null parameter,
		Their Assumptions M1(i)--(iii)
		also imply  the separation condition.

		\begin{assumption}[Conditional noise moments]
		\label{assumption: conditional variance} For $e:=Y - m(X)$, 
		there are constants \(0<c_e\leq C_e<\infty\)  such that
		\begin{equation}
			\label{equ: conditional noise moments}
			c_e\leq\mathbb E(e^2\mid X=x)\leq C_e,
			\qquad
			\mathbb E(e^4\mid X=x)\leq C_{e}^2,
			\qquad P_X\text{-a.s.}.
		\end{equation}
		
	\end{assumption}
		
		Assumption~\ref{assumption: conditional variance} allows
		heteroskedasticity while imposing uniform bounds on the conditional
		variance and fourth moment of the noise. The lower bound prevents variance degeneracy under the null, while the upper bounds control the moments needed for normal approximation, variance estimation, and power analysis. Such assumptions have been widely adopted in regression specification testing~\citep{guerre2005data,jun2009semiparametric,hardle1993comparing,hardle1993optimal}. For example,  \citet{guerre2005data} imposes
		a uniformly bounded conditional moment of order  greater
		than four, together with continuity and a positive lower bound
		for the conditional variance, and therefore implies Assumption~\ref{assumption: conditional variance}.
		
		\begin{assumption}[Parameter estimation under the null]
			\label{assumption: identibility and parametric estimation under H0}
			The estimator $\widehat\theta$ takes values in $\Theta$.
			Under $H_0$,
			$\sqrt n\,\|\widehat\theta-\theta_0\|_2=O_p(1)$.
		\end{assumption}
		
		Assumption~\ref{assumption: identibility and parametric estimation under H0}
		requires a $\sqrt{n}$-consistent estimator of $\theta_0$ under
		the null hypothesis. This condition holds for ordinary least
		squares in linear models and for nonlinear least squares under
		suitable conditions. We provide two concrete examples  in Appendix~\ref{sec: addtional lemmas and proofs}.

			 \begin{assumption}[Kernel regularity]
			\label{assumption: eigenvalues and eigenfunctions}
			The kernel $K$ is continuous, symmetric,
			and ISPD with respect to $P_X$, with
			$\sup_{x\in\mathcal X}K(x,x)\leq\kappa^2$
			for some $\kappa>0$.
			The eigenpairs
			$\{(\sigma_{X,k},\phi_{X,k})\}_{k\geq1}$ of
			$\mathcal L_{K,X}$ satisfy $	\sigma_{X,k}\asymp k^{-1/s}$	for some $0<s<1$, and $
			\sup_{k\geq1}
			\|\phi_{X,k}\|_{L^\infty(P_X)}<\infty.$
			
		\end{assumption}

		Assumption~\ref{assumption: eigenvalues and eigenfunctions}
		requires infinitely many positive eigenvalues with polynomial
		decay and uniformly bounded eigenfunctions  relative
		to $P_X$. The polynomial decay characterizes the capability of the RKHS, with larger $s$ corresponding to slower
		decay; such conditions are widely used in kernel
		learning~\citep{caponnetto2007optimal}.
		The uniform boundedness of eigenfunctions is also assumed by
		\citet{balasubramanian2021optimality} in kernel goodness-of-fit
		testing. Both requirements hold, for example, for periodic
		Sobolev kernels and the Brownian bridge kernel under the uniform
		distribution on $[0,1]$; see \citet{balasubramanian2021optimality} and \citet{cavoretto2015introduction} for details.
		

		\subsection{Null Distribution and Test Calibration} 
		
		We are now in a position to study the null distribution of
		$\widehat{\eta}_{\lambda}(\widehat{\theta})$
		and derive a feasible testing rule.
		The following theorem establishes this limit for the fully
		empirical statistic and provides a consistent variance
		estimator for studentization.
		\begin{theorem}
			\label{theorem: asymptotic null distribution}
			Suppose that Assumptions~\ref{assumption: auxiliary covariate distribution}, \ref{assumption: regularities on parametric families}, \ref{assumption: conditional variance}, \ref{assumption: identibility and parametric estimation under H0}, and \ref{assumption: eigenvalues and eigenfunctions} hold. Under $H_0$, if $\lambda \to 0$, $n,N\to\infty$ such that $n\lambda^{s/2}\to\infty$ and $N\lambda^s / \log(\lambda^{-1}) \to \infty$, there holds
			$$
			n\left[2 \mathbb E\left\{H_{\lambda,0}^2(Z_1,Z_2)\right\}\right]^{-1/2} \widehat{\eta}_{\lambda}(\widehat\theta)\leadsto N(0,1),
			$$
			where
			$$
			H_{\lambda,0}(Z_1,Z_2) = \left\langle (L_{K,0 }+ \lambda I)^{-1/2}K_{X_1},(  L_{K,0}  + \lambda I)^{-1/2}K_{X_2}\right\rangle_K\epsilon(Z_1,\theta_0)\epsilon(Z_2,\theta_0).
			$$
			Furthermore, define
			\begin{equation}
				\label{equ: variance estimator}
				\widehat{v}_{n,N}^2 := \frac{1}{n(n-1)}\left\{ \mathbf{1}_n^{\top}\left(M_n\circ M_n\right)\mathbf{1}_n - \mathrm{Tr}\left(M_n\circ M_n\right)\right\},
			\end{equation}
			where $M_n$ is defined in Proposition~\ref{prop: computation}. Then 
			$$
			\frac{\widehat{v}_{n,N}^2}{ \mathbb E\left\{H_{\lambda,0}^2(Z_1,Z_2) \right \}} = 1+ o_{p}(1).
			$$
		\end{theorem}
		Theorem~\ref{theorem: asymptotic null distribution}
		and Slutsky's lemma yield the asymptotically pivotal test
		\begin{equation}
			\label{equ: T_{SRMR}}
			T_{\mathrm{SRMR}}
			:=
			\mathds{1}\left\{
			\widehat{\eta}_{\lambda}(\widehat{\theta})
			>
			\frac{\sqrt{2}\,z_{1-\alpha}\widehat{v}_{n,N}}{n}
			\right\},
		\end{equation}
		which has asymptotic level $\alpha$ under each fixed null model, where $z_{1-\alpha}$
		denotes the $(1-\alpha)$-th quantile of the standard normal
		distribution. 
		\begin{remark}
			
			Theorem~\ref{theorem: asymptotic null distribution} gives a Gaussian null limit similar to that for moderated MMD in \citet[Theorem~2]{balasubramanian2021optimality}, but establishing this approximation requires more effort.
			Their theorem concerns distribution goodness-of-fit testing
			against a fixed known distribution, and the testing rule
			uses population spectral quantities determined by the kernel
			and that distribution.
			In our setting, the  operator $L_{K,0}$ is instead
			estimated from the auxiliary covariates
			$D=\{X_{0\ell}\}_{\ell=1}^N$, drawn from an unknown
			distribution $P_0$ that may differ from $P_X$.
			In addition, we must control the error introduced by
			$\widehat{\theta}$, which interacts with regularization
			and is absent from the related  
			analyses~\citep{hagrass-gof-test,hagrass2026ksd,
				hagrass-two-sample-test,balasubramanian2021optimality}.
			Thus, the theorem shows that both source of errors do not
			affect the limiting null distribution and justifies normal
			calibration using an estimated variance.
		\end{remark}
		\subsection{Detection Boundaries and Minimax Analysis}
		
		We now turn to study the power properties of the proposed test. As in~\eqref{equ: H1Deltan}, we measure separation
		by the squared $L^2(P_X)$-norm of the conditional residual
		function. To specify the regularity of the alternatives, for
		fixed $r>0$ and $B>0$, consider the bounded source ball
		\begin{equation}
			\label{equ: source ball}
			\mathcal F_{r,B}
			:=
			\left\{
			\mathbf f=\mathcal L_{K,X}^{\,r}u:
			u\in L^2(P_X),\
			\|u\|_{L^2(P_X)}\leq1,\
			\|\mathbf f\|_{L^\infty(P_X)}\leq B
			\right\}.
		\end{equation} 
		The smoothness of the  source ball  is  controlled by the index $r>0$, with larger values imposing
		stronger regularity. For $r\geq1/2$,
		every member admits a representative in $\mathcal H_K$.
		For $0<r<1/2$, the source condition describes an interpolation
		space between $L^2(P_X)$ and the RKHS, allowing functions
		that need not belong to $\mathcal H_K$.
		The  upper bound   $\|u\|_{L^2(P_X)}$ specifies the radius of the
		regularity class over which testing power can be assessed uniformly. Different choices of the radius  change only the constants in the resulting  detection rates.  Hence, we  specify this radius as one in \eqref{equ: source ball} without loss of generality.
		Related source conditions have also been used in minimax kernel
		testing; see
		\citet{hagrass-two-sample-test,hagrass-gof-test,hagrass2026ksd}.
		
		\begin{remark}
			In addition to source regularity, the class in
			\eqref{equ: source ball} imposes a common bound on the
			$L^\infty(P_X)$ norm of the conditional residual functions.
			When $r>s/2$, the uniform boundedness of the eigenfunctions in Assumption~\ref{assumption: eigenvalues and eigenfunctions} gives
			\[
			\|\mathcal L_{K,X}^{\,r}u\|_{L^\infty(P_X)}
			\leq
			\left(\sup_{k\geq1}\|\phi_{X,k}\|_{L^\infty(P_X)}\right)
			\left(\sum_{k\geq1}\sigma_{X,k}^{2r}\right)^{1/2}
			\|u\|_{L^2(P_X)}.
			\]
			The sum is finite because
			$\sigma_{X,k}^{2r}\asymp k^{-2r/s}$.
			Since $\|u\|_{L^2(P_X)}\leq1$, the source condition already
			provides a common supremum-norm bound in this regime. For $r\leq s/2$, this argument no longer yields
			a finite common bound, so the restriction is imposed in \eqref{equ: source ball} explicitly.
			This common amplitude bound provides the uniform residual-moment
			control required to establish our uniform power results
			over the alternative class.
		\end{remark}

		To establish power uniformly over the alternative class for our tests, we  require uniform control of the effect of parameter  estimation.
		We   impose the following assumption.

		\begin{assumption}[Uniform parameter estimation under alternatives]
			\label{assumption: uniform parametric estimation under H1}
			The pseudo-true parameter $\theta^\star$ exists and is unique
			for every data-generating distribution  
			with $\mathbf f\in\mathcal F_{r,B}$.
			Moreover,
			$\sqrt n\,\|\widehat\theta-\theta^\star\|_2=O_p(1)$
			uniformly over this class, that is,
			\begin{equation}
				\label{equ: uniformly consistent estimator}
				\lim_{\nu \to \infty}\limsup_{n\to\infty}
				\sup_{\mathbf f\in\mathcal F_{r,B}}
				\mathbb P\left(
				\sqrt n\,\|\widehat\theta-\theta^\star\|_2>\nu
				\right)
				=0.
			\end{equation}

		\end{assumption}

		Unlike the pointwise $\sqrt{n}$-consistency in
		Assumption~\ref{assumption: identibility and parametric estimation under H0},
		Assumption~\ref{assumption: uniform parametric estimation under H1}
		controls the parameter-estimation error simultaneously over the
		alternative class. This uniformity is needed for the   power
		analysis under the minimax framework. Related conditions
		have been imposed in specification testing; see, for example,
		\citet{guerre2002optimal,horowitz2006testing,hitomi2023optimal}. We also provide two concrete examples   in Appendix~\ref{sec: addtional lemmas and proofs} to illustrate that this
		uniform rate holds under alternatives.
		
		
		\begin{theorem}
			\label{theorem: uniform power}
			Let $0<\alpha<1$, $r>0$, and $B>0$ be fixed.
			Suppose that
			Assumptions~\ref{assumption: auxiliary covariate distribution},
			\ref{assumption: regularities on parametric families},
			\ref{assumption: conditional variance},
			\ref{assumption: identibility and parametric estimation under H0},
			\ref{assumption: eigenvalues and eigenfunctions}, and
			\ref{assumption: uniform parametric estimation under H1}
			hold. Consider testing $H_0$ against
			$H_1(\mathcal F_{r,B},\Delta_n)$. If $\Delta_nn^{2r/(2r+1)} \to 0$, then
			\begin{equation}
				\label{equ: mmr sharp_uniform power}
				\liminf_{n\to\infty}	\sup_{H_1(\mathcal F_{r,B},\Delta_n)}
				\mathbb
				P (T_{\mathrm{MMR}}^{\mathrm{or}}=0) \geq 1-\alpha.
			\end{equation}
			For
			$T_{\mathrm{SRMR}}$ with $	\lambda=n^{-2/(4r+s)}$ and $ {N}{n^{-2s/(4r+s)}(\log  n)^{-1}}\to\infty,$ if $\Delta_n n^{4r/(4r+s)}\to\infty$,  
			then
			\begin{equation}
				\label{equ: uniform Type II error}
				\limsup_{n\to\infty}
				\sup_{H_1(\mathcal F_{r,B},\Delta_n)}
				\mathbb P\left(T_{\mathrm{SRMR}}=0\right)=0.
			\end{equation}
		\end{theorem}
		
		Theorem~\ref{theorem: uniform power} shows that $T_{\mathrm{MMR}}^{\mathrm{or}}$
		cannot attain power uniformly above the significance level
		$\alpha$ when $\Delta_n=o(n^{-2r/(2r+1)})$. In contrast, $T_{\mathrm{SRMR}}$ achieves uniform consistency whenever
		$\Delta_n n^{4r/(4r+s)}\to\infty$ under the stated choices
		of $\lambda$ and $N$. Hence, $T_{\mathrm{SRMR}}$ can detect weaker departures than $T_{\mathrm{MMR}}^{\mathrm{or}}$, since $n^{-4r/(4r+s)} = o(n^{-2r/(2r+1)})$ for $r>0$ and $0<s<1$.
		
		The detection boundary rate $n^{-4r/(4r+s)}$ improves as $r$ increases or $s$ decreases, since stronger
		source regularity and faster eigenvalue decay both make departures
		from the null easier to detect. In kernel regression and inverse problems, Tikhonov regularization  
		is known to exhibit \emph{saturation}, meaning that additional smoothness
		beyond a certain threshold no longer improves the convergence
		rate~\citep{engl1996regularization,gerfo2008spectral,guo2017learning}.
		Theorem~\ref{theorem: uniform power} shows that the present testing
		procedure avoids this limitation, because this rate
		continues to improve with $r$, without saturating at a finite
		level. Intuitively, hypothesis testing only requires the test statistic
		to remain sufficiently large when the signal is strong.
		In our setting, this property is illustrated at the population
		level by \eqref{equ: limit discrepancy}.
		This requirement is weaker than that in estimation problems,
		which typically require accurate approximations.
		Related observations have also been established for spectral regularized  distribution
		tests~\citep{hagrass-gof-test,hagrass-two-sample-test,hagrass2026ksd}.

		\begin{remark}
			\label{remark: parameter selection}
			The choice of $\lambda$ in
			Theorem~\ref{theorem: uniform power} depends on $r$ and $s$,
			which are generally unknown in practice.
			A common approach to address this difficulty is to consider a grid of candidate values
			and combine the resulting tests.
			For example, one can take the maximum of the standardized
			statistics, and calibrate its joint null distribution~\citep{horowitz2001adaptive} or aggregate the individual
			tests using Bonferroni correction~\citep{hagrass-gof-test,
				hagrass-two-sample-test}.
			In the simulation studies, we assess the empirical performance
			of SRMR-based tests aggregated over prespecified grids of
			regularization parameters and kernel bandwidths using Bonferroni
			correction strategy. Since this paper focuses on
			how auxiliary covariate data contribute to the construction
			and power of specification tests, we do not theoretically study the parameter selection strategies.
			
			
			Besides the choice of $\lambda$,   Theorem~\ref{theorem: uniform power} requires a sufficiently
			large auxiliary sample to estimate the integral operator accurately. 
			When $r>s/4$, we have
			$n^{2s/(4r+s)}\log n=o(n)$, so $N\asymp n$ already suffices.
			When $r=s/4$, the condition becomes $N/(n\log n)\to\infty$.
			For $0<r<s/4$, it requires
			$N/(n^{2s/(4r+s)}\log n)\to\infty$, with
			$1<2s/(4r+s)<2$.
			Thus, the sufficient condition permits an auxiliary sample of the
			same order as the labeled sample for smoother alternatives, while
			requiring faster growth for rougher alternatives. 	As discussed earlier, auxiliary covariates may come from records
			with missing responses, historical databases, or independent
			datasets describing related populations~\citep{zhang2019semi,zhang2023double}. 	Simulation and generative modeling may offer an additional avenue.
			For example, a separate pilot subset of the available covariates
			could be used to fit a generative model, from which further
			covariate records are then sampled~\citep{xu2019modeling}.
			Such approaches suggest possible ways to supplement existing
			covariate collections.
		\end{remark}
		
		Having established the detection boundary of $T_{\mathrm{SRMR}}$,  we next derive a lower bound on the worst-case Type II error to assess whether this rate can be further improved. 

		\begin{theorem}
			\label{theorem: minimax optimality}
			Let $0<\alpha<1$, $r>0$, and $B>0$ be fixed.
			Consider the Gaussian regression experiment
			$Y_i=m(X_i)+e_i$, where
			$e_i\mid X_i\sim N(0,1)$.
			Suppose that
			Assumptions~\ref{assumption: regularities on parametric families}
			and~\ref{assumption: eigenvalues and eigenfunctions} hold. Let $T_n$ be any  test satisfying
			$	\limsup_{n\to\infty}\mathbb P(T_n=1)\leq\alpha$
			under each fixed null model.
			If $\Delta_n>0$ satisfies
			$\Delta_n=o(n^{-4r/(4r+s)})$ for $r\geq s/4$,
			or $\Delta_n=o(n^{-1/2})$ for $0<r<s/4$,
			then
			\[
			\liminf_{n\to\infty}
			\sup_{H_1(\mathcal F_{r,B},\Delta_n)}
			\mathbb P(T_n=0)
			\geq1-\alpha.
			\]
		\end{theorem}
		
		Theorem~\ref{theorem: minimax optimality} shows that no
		asymptotically $\alpha$-level test can achieve nontrivial
		uniform power under the stated separation conditions.  For $r\geq s/4$, the separation rate achieved by
		$T_{\mathrm{SRMR}}$ in Theorem~\ref{theorem: uniform power}
		matches this  minimax lower bound. For rough alternatives with $0<r<s/4$, however, the lower bound $n^{-1/2}$
		is smaller than the upper rate $n^{-4r/(4r+s)}$, so the 
		optimality of $T_{\mathrm{SRMR}}$ cannot be guaranteed, and such question is left open.
		\begin{remark}
			The Gaussian noise condition allows explicit likelihood-ratio
			calculations and is commonly adopted in deriving minimax
			lower bounds for regression testing
			\citep{guerre2002optimal,ingster2009minimax,
				ingster2010detection,hitomi2023optimal}.
			This condition is specific to the lower-bound argument and
			is not required for the validity or power results of the
			proposed tests.
		\end{remark}
		
		
		\section{Numerical Experiments}
		\label{sec: numerical studies}
This section examines the finite-sample performance of the
proposed test through simulation studies and an analysis of
Los Angeles homelessness data. 
		\subsection{Simulation Studies}
		\label{sec: simulation studies}
		
 We consider three designs aligned with our theoretical
 setting to examine the gains from spectral regularization
 relative to the MMR benchmark and compare the proposed test
 with existing specification tests.
 To assess performance beyond these designs, we consider more
 general regression models,
 where we also evaluate the aggregation strategy discussed in
 Remark~\ref{remark: parameter selection}.  Unless otherwise stated, the nominal significance level is
 $\alpha=0.05$, and tests requiring bootstrap calibration use
 $B=500$ bootstrap replications.
 Empirical sizes and powers are computed from $2{,}000$
 Monte Carlo replications.

		We test the linear parametric family
		\[
		\mathcal G(\Theta)
		=
		\left\{
		g(x,\theta)
		=
		\theta_1+\sum_{\ell=1}^{p}\theta_{\ell+1}x_\ell:
		\theta\in\mathbb R^{p+1}
		\right\},
		\]
		and generate observations from
		\begin{equation}
			\label{equ: simulation model}
			Y
			=
			1+\sum_{\ell=1}^{p}X_\ell+\mathbf f(X)+e,
		\end{equation}
		where $e\sim N(0,1)$ is independent of $X$, and $\mathbf f$
		is a perturbation function.
		When $\mathbf f=0$, the model satisfies the null hypothesis
		with $\theta=\mathbf1_{p+1}$.
		To construct model departures with the regularity considered in our
		theoretical analysis, we begin with the following three kernels and specify   corresponding perturbation
		functions.  
		\begin{example}[Brownian bridge kernel]
			\label{example: simulation bridge}
			Let $\mathcal X=[0,1]^p$ and
			$P_X=\operatorname{Unif}([0,1]^p)$. When $p=1$,
			the marginal kernel is
			\[
			K_1(t,t')
			=
			\pi^2\{\min(t,t')-tt'\}
			=
			2\sum_{j\geq1}j^{-2}\sin(\pi jt)\sin(\pi jt').
			\]
			Its normalized eigenfunctions are $\sqrt2\sin(\pi jt)$,
			with eigenvalues $j^{-2}$, $j\geq1$.
			We use the departure 
			\[
			\mathbf f(x)
			=
			a\sqrt{\frac25}
			\{\sin(2\pi qx_1)-2\sin(4\pi qx_1)\}
			\prod_{\ell=2}^{p}\sqrt2\sin(\pi x_\ell),
			\]
			where the empty product equals one when $p=1$. 
		\end{example}
		
		\begin{example}[Chebyshev kernel with an arcsine design]
			\label{example: simulation chebyshev}
			Let $\mathcal X=[-1,1]^p$, with independent coordinates
			having the arcsine density $p_{X_\ell}(t)
			= \pi^{-1}(1-t^2)^{-1/2}
			\mathbf1\{-1<t<1\} $. 
			Write $T_j(t)=\cos\{j\arccos(t)\}$, and set
			$u=\arccos(t)/\pi$ and $v=\arccos(t')/\pi$.  When $p=1$,
			the marginal kernel is
			\[
			\begin{aligned}
				K_1(t,t')
				&=
				1+2\sum_{j\geq1}j^{-2}T_j(t)T_j(t') =
				1+\pi^2
				\left\{
				\frac{u^2+v^2}{2}-\max(u,v)+\frac13
				\right\}.
			\end{aligned}
			\]
			Since
			$\arccos(X_\ell)/\pi\sim\operatorname{Unif}([0,1])$,
			its normalized eigenfunctions are $1$ and $\sqrt2T_j(t)$,
			with eigenvalues $1$ and $j^{-2}$, respectively.
			We use 	$\mathbf f(x)=a\sqrt2T_q(x_1)$, with $q\geq2$.
			
		\end{example}
		
		\begin{example}[Normal design with a transformed cosine kernel]
			\label{example: simulation normal cdf}
			Let $\mathcal X=\mathbb R^p$ and $P_X=N(0,I_p)$.
			Write $u=\Phi(t)$ and $v=\Phi(t')$, where $\Phi$ is the
			standard normal distribution function. When $p=1$, define
			\[
			\begin{aligned}
				K_1(t,t')
				&=
				1+2\sum_{j\geq1}
				j^{-2}\cos(\pi ju)\cos(\pi jv) =
				1+\pi^2
				\left\{
				\frac{u^2+v^2}{2}-\max(u,v)+\frac13
				\right\}.
			\end{aligned}
			\]
			Since $\Phi(X_\ell)\sim\operatorname{Unif}([0,1])$,
			its normalized eigenfunctions are $1$ and
			$\sqrt2\cos\{\pi j\Phi(t)\}$, with eigenvalues $1$ and
			$j^{-2}$, respectively.
			We use 
			$\mathbf f(x)=a\sqrt2\cos\{2\pi q\Phi(x_1)\}$. 
		\end{example}
		
\begin{figure}[!htb]
			\centering
			\begin{subfigure}[t]{0.32\textwidth}
				\centering
				\includegraphics[width=\linewidth]{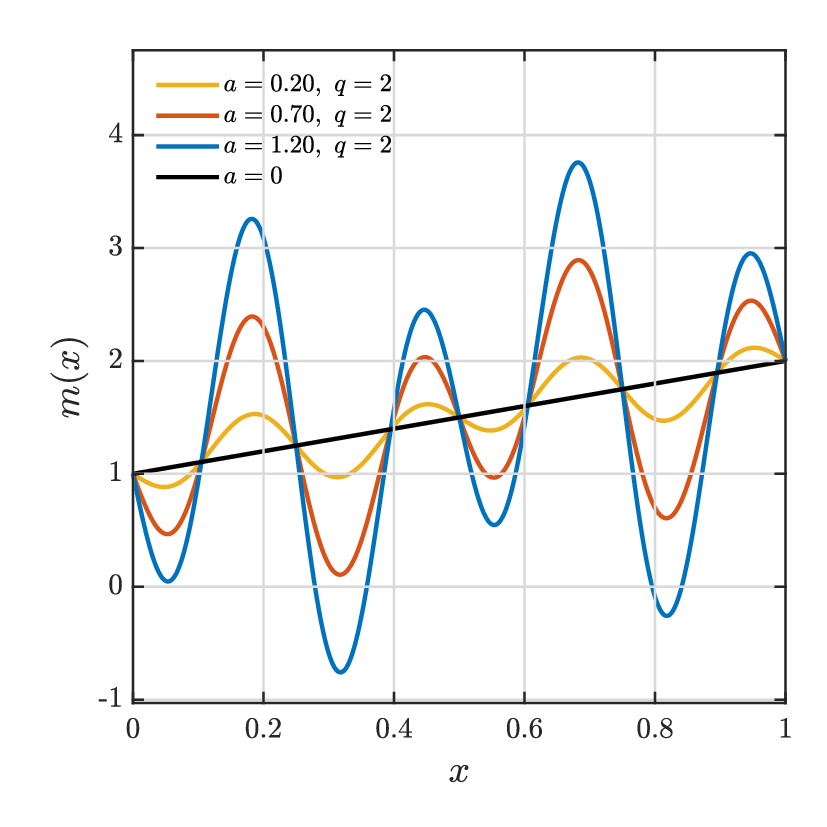}
				\caption{Example~\ref{example: simulation bridge}}
				\label{fig: departures bridge}
			\end{subfigure}
			\hfill
			\begin{subfigure}[t]{0.32\textwidth}
				\centering
				\includegraphics[width=\linewidth]{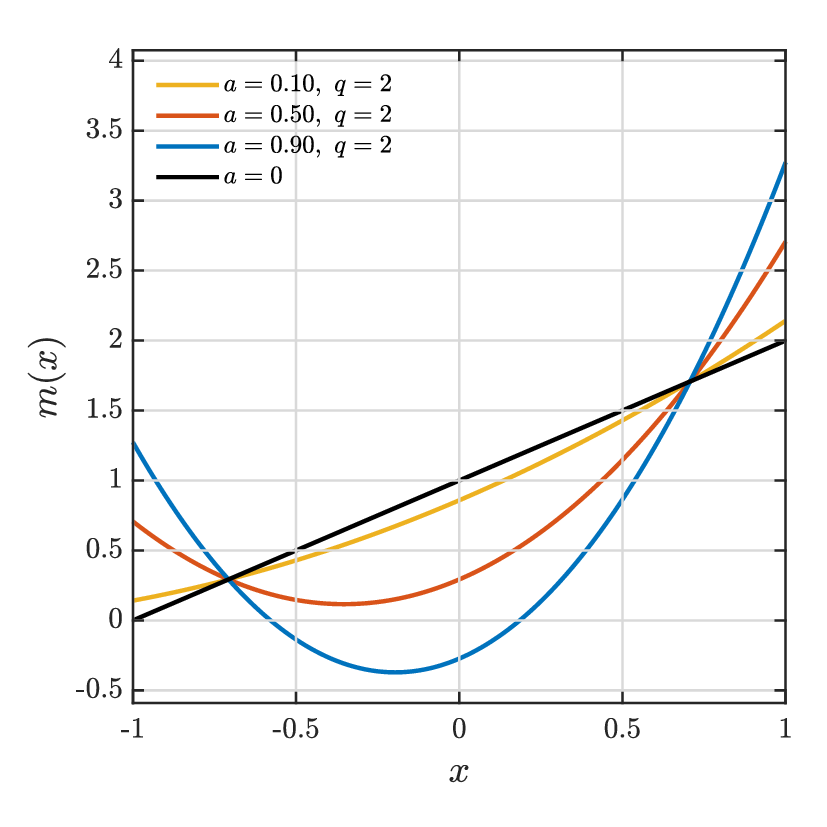}
				\caption{Example~\ref{example: simulation chebyshev}}
				\label{fig: departures chebyshev}
			\end{subfigure}
			\hfill
			\begin{subfigure}[t]{0.32\textwidth}
				\centering
				\includegraphics[width=\linewidth]{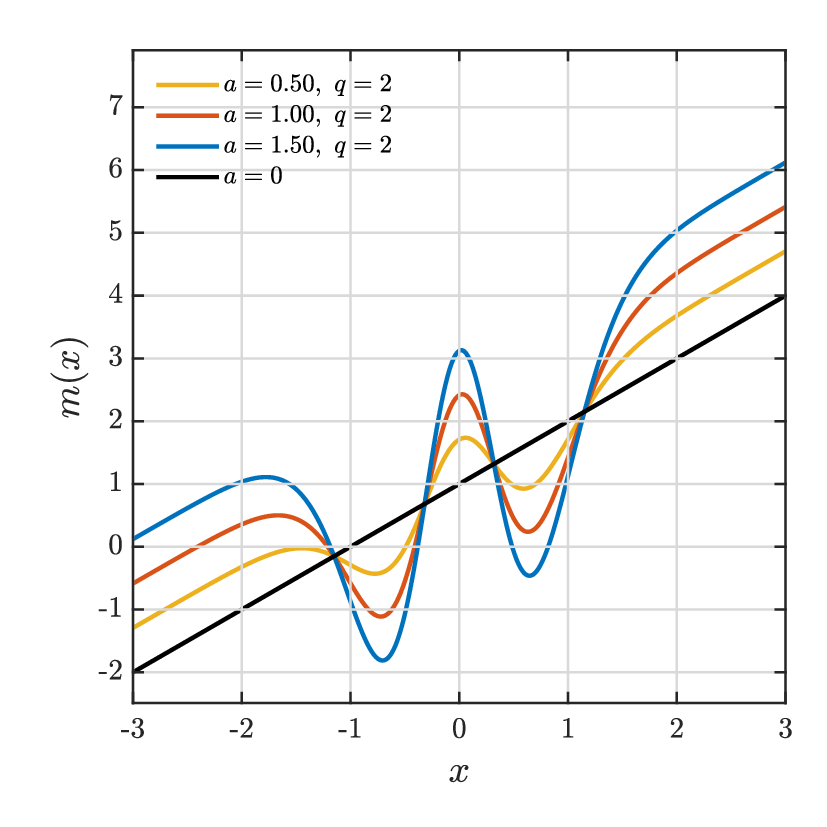}
				\caption{Example~\ref{example: simulation normal cdf}}
				\label{fig: departures normal cdf}
			\end{subfigure}
			\caption{One-dimensional regression functions corresponding to
				Examples~\ref{example: simulation bridge}--%
				\ref{example: simulation normal cdf}.
				Black lines represent the null model $m(x)=1+x$, and
				colored curves represent $m(x)=1+x+\mathbf f(x)$
				for different amplitudes $a$.}
			\label{fig: simulation departures}
		\end{figure}

		The one-dimensional kernels in
		Examples~\ref{example: simulation bridge}--\ref{example: simulation normal cdf}
		are bounded, continuous, and ISPD relative to their respective
		covariate distributions. Their normalized eigenfunctions are
		uniformly bounded, and their eigenvalues decay as $j^{-2}$,
		satisfying Assumption~\ref{assumption: eigenvalues and eigenfunctions}
		with $s=1/2$. For the multivariate experiments, we use the
		tensor-product kernel $K(x,x')=\prod_{\ell=1}^{p}K_1(x_\ell,x'_\ell)$. The perturbations in these examples satisfy
		$\mathbb E\mathbf f(X)=0$ and
		$\mathbb E\{X_\ell\mathbf f(X)\}=0$ for $1\leq\ell\leq p$.
		Consequently, the unique population least-squares projection is
		$g(x,\theta^\star)=1+\sum_{\ell=1}^{p}x_\ell$, so $\mathbf f$
		coincides with the conditional residual function defined
		in~\eqref{equ: conditional residual function}.
		Each perturbation is a finite linear combination of the
		corresponding kernel eigenfunctions, or their tensor products
		when $p>1$. It therefore belongs to the associated RKHS and has
		finite $L^2(P_X)$, supremum, and source norms for every fixed
		$r>0$, as summarized in Table~\ref{tab: simulation source norms}. 
		The parameter $a$ in Table~\ref{tab: simulation source norms}
		controls the amplitude of $\mathbf{f}$, with $a=0$ corresponding to the null
		hypothesis and larger values of $|a|$ producing stronger
		alternative signals. By suitably rescaling this amplitude, each perturbation can be
		placed in the bounded source ball in~\eqref{equ: source ball}. Figure~\ref{fig: simulation departures}
		illustrates the corresponding one-dimensional regression
		functions in these examples.

		\begin{table}[!htb]
			\centering
			\caption{Conditional residual functions and their norms.
				Here, $a\in\mathbb R$, $r>0$, and $q\in\mathbb N^+$.}
			\label{tab: simulation source norms}
			\begingroup
			\renewcommand{\arraystretch}{1.65}
			\setlength{\tabcolsep}{6pt}
			\resizebox{\textwidth}{!}{%
				\begin{tabular}{lcccc}
					\toprule
					& $\mathbf f(x)$
					& $\|\mathbf f\|_{L^2(P_X)}^2$
					& $\|\mathcal L_{K,X}^{-r}\mathbf f\|_{L^2(P_X)}^2$
					& $\|\mathbf f\|_{L^\infty(P_X)}$
					\\
					\midrule
					Example~\ref{example: simulation bridge}
					& $\displaystyle
					a\sqrt{\frac25}
					\{\sin(2\pi qx_1)-2\sin(4\pi qx_1)\}
					\prod_{\ell=2}^{p}\sqrt2\sin(\pi x_\ell)$
					& $a^2$
					& $\displaystyle
					\frac{a^2}{5}\{(2q)^{4r}+4(4q)^{4r}\}$
					& $\displaystyle
					\leq3|a|\sqrt{\frac25}\,2^{(p-1)/2}$
					\\
					Example~\ref{example: simulation chebyshev}
					& $a\sqrt2T_q(x_1)$
					& $a^2$
					& $a^2q^{4r}$
					& $\sqrt2|a|$
					\\
					Example~\ref{example: simulation normal cdf}
					& $a\sqrt2\cos\{2\pi q\Phi(x_1)\}$
					& $a^2$
					& $a^2(2q)^{4r}$
					& $\sqrt2|a|$
					\\
					\bottomrule
				\end{tabular}%
			}
			\endgroup
		\end{table}

		To illustrate the benefits of spectral regularization, we compare our proposed methods with the MMR benchmark. To implement the oracle procedure in~\eqref{equ: MMR test},
		we replace $\theta_0$ by its least-squares estimator
		$\widehat\theta$ and calibrate
		$n\widehat{\mathrm{MMR}}^2(\widehat\theta)$ by the wild bootstrap method~\citep{wildbootstrap, hardle1993comparing}. Specifically, 
		let $\widehat\epsilon_i=Y_i-g(X_i,\widehat\theta)$.
		In each bootstrap replication, we keep the covariates fixed and
		generate $Y_i^*=g(X_i,\widehat\theta)+\widehat\epsilon_i\xi_i^*$,
		where the multipliers are i.i.d., independent of the original
		data, and follow Mammen's distribution
		\[
		\mathbb P\left(\xi_i^*=\frac{1-\sqrt5}{2}\right)
		=\frac{\sqrt5+1}{2\sqrt5},
		\qquad
		\mathbb P\left(\xi_i^*=\frac{1+\sqrt5}{2}\right)
		=\frac{\sqrt5-1}{2\sqrt5}.
		\]
		We refit the parametric model by least squares to the bootstrap
		sample $\{(X_i,Y_i^*)\}_{i=1}^n$ to obtain
		$\widehat\theta^*$,
		and then calculate
		$n\widehat{\mathrm{MMR}}^{2,*}(\widehat\theta^*)
		=(n-1)^{-1}\sum_{i\ne j}
		\widehat\epsilon_i^*K(X_i,X_j)\widehat\epsilon_j^*$,
		where $\widehat\epsilon_i^*=Y_i^*-g(X_i,\widehat\theta^*)$.
		After $B$ independent replications, we reject $H_0$ if
		$n\widehat{\mathrm{MMR}}^2(\widehat\theta)$ exceeds the empirical
		$(1-\alpha)$-quantile of the bootstrap statistics. We denote the resulting test by $T_{\mathrm{MMR}}$. Since the proposed statistics depend on the regularization parameter $\lambda$,
		the auxiliary sample size $N$, and the discrepancy between
		the auxiliary and labeled covariate distributions, we examine their effects separately. 
		
		\medskip
		\noindent
		\textbf{Effect of the regularization parameter.}
		We first examine how the regularization level affects testing
		performance. We set $P_0=P_X$, fix $n=N=500$, and take $\lambda\in\{10^{-6},10^{-5},10^{-4},10^{-3}\}.$
		The simulation results are reported in Table~\ref{tab:effect_lambda}.
		We can see that the empirical Type I error rates are well controlled, while spectral regularization yields substantial power improvement
		over the MMR benchmark. For example, in
		Example~\ref{example: simulation bridge}, at $a=0.4$ and
		$\lambda=10^{-4}$, $T_{\mathrm{SRMR}}$ achieves power $0.984$,
		compared with $0.050$ for $T_{\mathrm{MMR}}$.
		Similarly, in Example~\ref{example: simulation chebyshev},
		at $a=0.2$ and $\lambda=10^{-3}$, the corresponding powers are
		$0.613$ and $0.076$. Since these examples involve different alternative structures,
		the spectral reweighting induced by different values of regularization parameters
		affects their detectability in different ways, leading to varying
		degrees of power improvement.

		\begin{table}[!htb]
			\centering
			\caption{Effect of $\lambda$ on empirical size and power,
				with $n=N=500$ and $P_0=P_X$.}
			\label{tab:effect_lambda}
			\begingroup
			\small
			\renewcommand{\arraystretch}{1.2}
			\setlength{\tabcolsep}{3.5pt}
			\begin{tabular}{@{}ccccccc@{}}
				\toprule
				\multirow[c]{2}{*}{}
				& \multirow[c]{2}{*}{$a$}
				& \multicolumn{4}{c}{$T_{\mathrm{SRMR}}$}
				& \multirow[c]{2}{*}{$T_{\mathrm{MMR}}$}
				\\
				\cmidrule(lr){3-6}
				& & $\lambda=10^{-6}$ & $\lambda=10^{-5}$
				& $\lambda=10^{-4}$ & $\lambda=10^{-3}$ &
				\\
				\midrule
				\multirow[c]{5}{*}{Example~\ref{example: simulation bridge}}
				& 0.0 & 0.057 & 0.055 & 0.047 & 0.038 & 0.051 \\
				& 0.2 & 0.171 & 0.204 & 0.316 & 0.237 & 0.045 \\
				& 0.4 & 0.650 & 0.806 & 0.984 & 0.965 & 0.050 \\
				& 0.6 & 0.974 & 0.998 & 1.000 & 1.000 & 0.059 \\
				& 0.8 & 1.000 & 1.000 & 1.000 & 1.000 & 0.075 \\
				\midrule
				\multirow[c]{5}{*}{Example~\ref{example: simulation chebyshev}}
				& 0.0 & 0.056 & 0.057 & 0.047 & 0.040 & 0.047 \\
				& 0.2 & 0.172 & 0.207 & 0.384 & 0.613 & 0.076 \\
				& 0.4 & 0.664 & 0.836 & 0.992 & 1.000 & 0.284 \\
				& 0.6 & 0.982 & 0.998 & 1.000 & 1.000 & 0.911 \\
				& 0.8 & 1.000 & 1.000 & 1.000 & 1.000 & 1.000 \\
				\midrule
				\multirow[c]{5}{*}{Example~\ref{example: simulation normal cdf}}
				& 0.0 & 0.058 & 0.057 & 0.045 & 0.040 & 0.044 \\
				& 0.2 & 0.165 & 0.214 & 0.367 & 0.466 & 0.057 \\
				& 0.4 & 0.675 & 0.841 & 0.989 & 0.999 & 0.073 \\
				& 0.6 & 0.980 & 0.999 & 1.000 & 1.000 & 0.104 \\
				& 0.8 & 1.000 & 1.000 & 1.000 & 1.000 & 0.140 \\
				\bottomrule
			\end{tabular}
			\endgroup
		\end{table}
		
		\medskip
		\noindent
		\textbf{Effect of the auxiliary sample size.} We examine how the availability of auxiliary covariates affects testing
		performance.

		We first consider the gain from collecting
		additional covariate observations when the labeled sample  is fixed.
	   We set $n=500$ and $\lambda=10^{-4}$, retain $P_0=P_X$, and vary
		$N\in\{50,100,200,500,800\}$.
		The results in Table~\ref{tab:effect_N} indicate reasonable
		Type I error control across the auxiliary sample sizes considered.
		The effect of $N$ on power depends on the underlying alternative
		structure. The improvement is particularly pronounced in
		Example~\ref{example: simulation bridge}: at $a=0.4$, increasing
		$N$ from $50$ to $800$ raises power from $0.583$ to $0.984$. 
		Example~\ref{example: simulation normal cdf} also benefits from
		additional auxiliary observations, with power increasing from
		$0.226$ to $0.376$ at $a=0.2$, and from $0.871$ to $0.993$ at
		$a=0.4$. In contrast, Example~\ref{example: simulation chebyshev}
		is less sensitive to $N$. At $a=0.2$, the empirical powers remain between
		$0.358$ and $0.387$. 
		Across these settings, $T_{\mathrm{SRMR}}$ achieves
		power at least as high as $T_{\mathrm{MMR}}$,
		with clear power improvement before both tests reach empirical power one.
		
		\begin{table}[!htb]
			\centering
			\caption{Effect of $N$ on empirical size and power,
				with $n=500$, $\lambda=10^{-4}$, and $P_0=P_X$.}
			\label{tab:effect_N}
			\begingroup
			\small
			\renewcommand{\arraystretch}{1.2}
			\setlength{\tabcolsep}{5pt}
			\begin{tabular}{@{}cccccccc@{}}
				\toprule
				\multirow[c]{2}{*}{}
				& \multirow[c]{2}{*}{$a$}
				& \multicolumn{5}{c}{$T_{\mathrm{SRMR}}$}
				& \multirow[c]{2}{*}{$T_{\mathrm{MMR}}$}
				\\
				\cmidrule(lr){3-7}
				& & $N=50$ & $N=100$ & $N=200$ & $N=500$ & $N=800$ &
				\\
				\midrule
				\multirow[c]{5}{*}{Example~\ref{example: simulation bridge}}
				& 0.0 & 0.040 & 0.044 & 0.057 & 0.047 & 0.046 & 0.051 \\
				& 0.2 & 0.122 & 0.198 & 0.256 & 0.316 & 0.326 & 0.045 \\
				& 0.4 & 0.583 & 0.827 & 0.935 & 0.984 & 0.984 & 0.050 \\
				& 0.6 & 0.942 & 0.996 & 1.000 & 1.000 & 1.000 & 0.059 \\
				& 0.8 & 0.991 & 1.000 & 1.000 & 1.000 & 1.000 & 0.075 \\
				\midrule
				\multirow[c]{5}{*}{Example~\ref{example: simulation chebyshev}}
				& 0.0 & 0.037 & 0.042 & 0.055 & 0.047 & 0.046 & 0.047 \\
				& 0.2 & 0.385 & 0.358 & 0.366 & 0.384 & 0.387 & 0.076 \\
				& 0.4 & 0.966 & 0.973 & 0.987 & 0.992 & 0.992 & 0.284 \\
				& 0.6 & 1.000 & 1.000 & 1.000 & 1.000 & 1.000 & 0.911 \\
				& 0.8 & 1.000 & 1.000 & 1.000 & 1.000 & 1.000 & 1.000 \\
				\midrule
				\multirow[c]{5}{*}{Example~\ref{example: simulation normal cdf}}
				& 0.0 & 0.036 & 0.043 & 0.053 & 0.045 & 0.044 & 0.044 \\
				& 0.2 & 0.226 & 0.294 & 0.331 & 0.367 & 0.376 & 0.057 \\
				& 0.4 & 0.871 & 0.943 & 0.981 & 0.989 & 0.993 & 0.073 \\
				& 0.6 & 0.991 & 0.999 & 1.000 & 1.000 & 1.000 & 0.104 \\
				& 0.8 & 0.999 & 1.000 & 1.000 & 1.000 & 1.000 & 0.140 \\
				\bottomrule
			\end{tabular}
			\endgroup
		\end{table}

		Then, we  fix the total sample size at $n+N=500$ for	$T_{\mathrm{SRMR}}$ keeping
		$\lambda=10^{-4}$,  where $n$ labeled observations to   compute residuals, and the remaining $N$ covariate observations
	 estimate the integral operator. $T_{\mathrm{MMR}}$
		uses all 500 covariate--response pairs.  
	 Table~\ref{tab:fixed-total-n} shows that, with either 250 or 400 labels,
		$T_{\mathrm{SRMR}}$ matches or exceeds the empirical power of
		$T_{\mathrm{MMR}}$ for every alternative considered. With only 100
		labels, the comparison depends on the alternative. At $a=0.6$, for
		instance, the empirical power of $T_{\mathrm{SRMR}}$ falls below that
		of $T_{\mathrm{MMR}}$ in Example~\ref{example: simulation chebyshev}, but
		remains more powerful in both Example~\ref{example: simulation bridge} and Example~\ref{example: simulation normal cdf}. Hence, even with data splitting, the proposed method can achieve empirical power comparable to or greater than that of the MMR benchmark based on the full labeled sample.
		\begin{table}[!htb]
			\centering
			\caption{Effect of $(n,N)$ on empirical size and power,
			 $\lambda=10^{-4}$, and $P_0=P_X$.}
			\label{tab:fixed-total-n}
			\begingroup
			\small
			\renewcommand{\arraystretch}{1.2}
			\setlength{\tabcolsep}{5pt}
			\begin{tabular}{@{}cccccc@{}}
				\toprule
				& \multirow[c]{2}{*}{$a$}
				& \multicolumn{3}{c}{$T_{\mathrm{SRMR}}$}
				& \multirow[c]{2}{*}{$T_{\mathrm{MMR}}$} \\
				\cmidrule(lr){3-5}
				& & $(100,400)$ & $(250,250)$ & $(400,100)$ & \\
				\midrule
				\multirow[c]{5}{*}{Example~\ref{example: simulation bridge}}
				& 0.0 & 0.038 & 0.051 & 0.049 & 0.051 \\
				& 0.2 & 0.074 & 0.142 & 0.151 & 0.045 \\
				& 0.4 & 0.199 & 0.575 & 0.692 & 0.050 \\
				& 0.6 & 0.497 & 0.956 & 0.980 & 0.059 \\
				& 0.8 & 0.768 & 1.000 & 0.999 & 0.075 \\
				\midrule
				\multirow[c]{5}{*}{Example~\ref{example: simulation chebyshev}}
				& 0.0 & 0.039 & 0.050 & 0.048 & 0.047 \\
				& 0.2 & 0.078 & 0.171 & 0.292 & 0.076 \\
				& 0.4 & 0.260 & 0.710 & 0.915 & 0.284 \\
				& 0.6 & 0.614 & 0.989 & 1.000 & 0.911 \\
				& 0.8 & 0.872 & 1.000 & 1.000 & 1.000 \\
				\midrule
				\multirow[c]{5}{*}{Example~\ref{example: simulation normal cdf}}
				& 0.0 & 0.038 & 0.051 & 0.045 & 0.044 \\
				& 0.2 & 0.070 & 0.164 & 0.228 & 0.057 \\
				& 0.4 & 0.231 & 0.670 & 0.858 & 0.073 \\
				& 0.6 & 0.573 & 0.979 & 0.996 & 0.104 \\
				& 0.8 & 0.866 & 1.000 & 1.000 & 0.140 \\
				\bottomrule
			\end{tabular}
			\endgroup
		\end{table}

		\medskip
		\noindent
		\textbf{Effect of covariate shift.}
		We next examine whether auxiliary observations remain useful when
		their distribution differs from that of the labeled covariates.
		We fix $n=N=500$ and $\lambda=10^{-4}$ and compare $P_0=P_X$
		with the shifted distribution satisfying
		\begin{equation}
			\label{equ: covariate shift}
			\frac{dP_0}{dP_X}(x)
			=
			\frac12+F_{X_1}(x_1),
		\end{equation}
		where $F_{X_1}$ is the marginal distribution function of the first
		coordinate.
		Since the marginal designs are continuous,
		$\mathbb E\{F_{X_1}(X_1)\}=1/2$, so the ratio integrates to one
		and lies between $1/2$ and $3/2$. 
		We denote the test using shifted auxiliary observations by
		$\widetilde T_{\mathrm{SRMR}}$.
		As reported in Table~\ref{tab:effect_shift}, the empirical sizes
		of $T_{\mathrm{SRMR}}$ and $\widetilde T_{\mathrm{SRMR}}$ are
		close, and their powers are also comparable,
		with $T_{\mathrm{SRMR}}$ performing slightly better.
		Both tests achieve higher power than $T_{\mathrm{MMR}}$
		across all reported alternatives, suggesting that auxiliary
		covariates remain useful under the covariate shift.

		\begin{table}[!htb]
			\centering
			\caption{Effect of covariate shift on empirical size and power,
				with $n=N=500$ and $\lambda=10^{-4}$.}
			\label{tab:effect_shift}
			\begingroup
			\small
			\renewcommand{\arraystretch}{1.2}
			\setlength{\tabcolsep}{5pt}
			\begin{tabular}{@{}ccccccc@{}}
				\toprule
				\multirow[c]{2}{*}{}
				& \multirow[c]{2}{*}{Methods}
				& \multicolumn{5}{c}{$a$}
				\\
				\cmidrule(lr){3-7}
				& & 0.0 & 0.2 & 0.4 & 0.6 & 0.8
				\\
				\midrule
				\multirow[c]{3}{*}{Example~\ref{example: simulation bridge}}
				& $T_{\mathrm{SRMR}}$
				& 0.047 & 0.316 & 0.984 & 1.000 & 1.000 \\
				& $\widetilde T_{\mathrm{SRMR}}$
				& 0.045 & 0.290 & 0.969 & 1.000 & 1.000 \\
				& $T_{\mathrm{MMR}}$
				& 0.051 & 0.045 & 0.050 & 0.059 & 0.075 \\
				\midrule
				\multirow[c]{3}{*}{Example~\ref{example: simulation chebyshev}}
				& $T_{\mathrm{SRMR}}$
				& 0.047 & 0.384 & 0.992 & 1.000 & 1.000 \\
				& $\widetilde T_{\mathrm{SRMR}}$
				& 0.047 & 0.372 & 0.989 & 1.000 & 1.000 \\
				& $T_{\mathrm{MMR}}$
				& 0.047 & 0.076 & 0.284 & 0.911 & 1.000 \\
				\midrule
				\multirow[c]{3}{*}{Example~\ref{example: simulation normal cdf}}
				& $T_{\mathrm{SRMR}}$
				& 0.045 & 0.367 & 0.989 & 1.000 & 1.000 \\
				& $\widetilde T_{\mathrm{SRMR}}$
				& 0.043 & 0.331 & 0.982 & 1.000 & 1.000 \\
				& $T_{\mathrm{MMR}}$
				& 0.044 & 0.057 & 0.073 & 0.104 & 0.140 \\
				\bottomrule
			\end{tabular}
			\endgroup
		\end{table}
		
		\medskip
		\noindent
		\textbf{Comparison with existing methods.}
		To further evaluate the performance of the proposed test, we include additional two classical tests in the literature for  comparison. The first is the local-smoothing test of \citet{zheng1996consistent}, denoted by $T_{\mathrm{ZH}}$. For simplicity, we   use the product biweight kernel $ K_h(x,x')
		=
		\prod_{\ell=1}^{p}
		k\!\left\{  (x_\ell-x'_\ell)/h\right\},$ with $ k(t)
		=
		15(1-t^2)^2\mathbf1\{|t|\leq1\}/16$ and bandwidth $h=n^{-1/(2p)}$ for their statistic construction. The second is the global-smoothing test of \citet{stute1997nonparametric},
		denoted by $T_{\mathrm{STU}}$.  We note that both SRMR-based tests and the test of~\citet{zheng1996consistent} are pivotal, and can be calibrated by their limiting distributions. Their finite-sample behavior, however, may depend on the accuracy of  the normal approximation, in addition to the choice of kernel bandwidth
		or regularization parameter. For these local-smoothing  tests,
		\citet{li1998simple} show that wild-bootstrap calibration can
		approximate the null distribution more accurately than asymptotic
		normal calibration.  Hence, 	we report both normal and wild-bootstrap calibration for the local-smoothing tests $T_{\mathrm{SRMR}}$, $\widetilde{T}_{\mathrm{SRMR}}$ and $T_{\mathrm{ZH}}$, while the global-smoothing tests
		$T_{\mathrm{MMR}}$ and $T_{\mathrm{STU}}$ use wild-bootstrap critical values.

		The results are reported in
		Tables~\ref{tab:comparison_p1}--\ref{tab:comparison_p4}.
		In the Calibration column, \textnormal{N} and \textnormal{B}
		denote normal approximation and wild bootstrap, respectively. 
		Under the null hypothesis, wild-bootstrap calibration alleviates
		the underrejection associated with the normal approximation,
		particularly for $T_{\mathrm{ZH}}$.
		For example, in Example~\ref{example: simulation bridge} with
		$p=4$, its empirical size increases from $0.005$ to $0.045$.
		In Example~\ref{example: simulation chebyshev} with $p=2$,
		the corresponding improvement is from $0.025$ to $0.050$.
		The proposed tests also benefit from bootstrap calibration in some settings. For instance, in
		Example~\ref{example: simulation bridge} with $p=2$,
		the empirical size of $\widetilde T_{\mathrm{SRMR}}$
		changes from $0.037$ under normal calibration to $0.048$
		under the wild bootstrap. 
		
		The main advantage of the proposed methods is reflected in their
		power performance. For the univariate case in
		Table~\ref{tab:comparison_p1}, the empirical powers of
		$T_{\mathrm{SRMR}}$, $\widetilde T_{\mathrm{SRMR}}$, and
		$T_{\mathrm{ZH}}$ depend on the alternative structure and signal
		strength, with no test uniformly dominating the others.
		In Example~\ref{example: simulation chebyshev}, the
		bootstrap-calibrated $T_{\mathrm{ZH}}$ is more sensitive to weak
		departures, attaining empirical power $0.580$ at $a=0.2$, compared with
		$0.389$ and $0.425$ for  $T_{\mathrm{SRMR}}$ under normal and bootstrap calibrations, respectively.
		In Example~\ref{example: simulation bridge}, our methods yield
		a clear advantage. At $a=0.4$, the  powers of $T_{\mathrm{SRMR}}$ and $\widetilde{T}_{\mathrm{SRMR}}$ are $0.974$ and
		$0.966$ under normal calibration, and $0.977$ and $0.968$
		under bootstrap calibration, compared with $0.397$ for the
		bootstrap-calibrated $T_{\mathrm{ZH}}$.
		The proposed SRMR-based tests have similar power in these univariate
		examples, and both, together with $T_{\mathrm{ZH}}$, are more
		powerful than $T_{\mathrm{MMR}}$ and $T_{\mathrm{STU}}$. The power advantage of the proposed methods is more pronounced
		in the multivariate settings.
		Tables~\ref{tab:comparison_p2}--\ref{tab:comparison_p4}
		show that the SRMR-based tests outperform the three competing methods
		under both calibrations. For example, in
		Example~\ref{example: simulation normal cdf} with $p=2$ and
		$a=0.4$, $T_{\mathrm{SRMR}}$ and
		$\widetilde T_{\mathrm{SRMR}}$ achieve powers of $0.815$ and
		$0.748$ under normal calibration, and $0.835$ and $0.773$
		under bootstrap calibration. These compare with $0.218$,
		$0.052$, and $0.062$ for the bootstrap-calibrated
		$T_{\mathrm{ZH}}$, $T_{\mathrm{MMR}}$, and
		$T_{\mathrm{STU}}$, respectively.
		For $p=4$ and $a=0.8$ in
		Example~\ref{example: simulation chebyshev}, the two SRMR-based
		tests attain powers of $0.936$ and $0.852$ under normal
		calibration, and $0.951$ and $0.876$ under bootstrap calibration,
		whereas the three competing tests have powers no greater than
		$0.103$. Although  $T_{\mathrm{SRMR}}$ generally gives
		higher power, the test $\widetilde{T}_{\mathrm{SRMR}}$  retains power
		advantages in these comparisons.
		
		
		\begin{table}[!htb]
			\centering
			\caption{Empirical sizes and powers for $p=1$,
				with $n=N=500$, $q=10$, and $\lambda=10^{-4}$.}
			\label{tab:comparison_p1}
			\begingroup
			\small
			\renewcommand{\arraystretch}{1.1}
			\setlength{\tabcolsep}{4pt}
			\begin{tabular}{@{}cccccccc@{}}
				\toprule
				\multirow[c]{2}{*}{Example}
				& \multirow[c]{2}{*}{Methods}
				& \multirow[c]{2}{*}{Calibration}
				& \multicolumn{5}{c}{$a$}
				\\
				\cmidrule(lr){4-8}
				& & & 0.0 & 0.2 & 0.4 & 0.6 & 0.8
				\\
				\midrule
				\multirow[c]{8}{*}{Example~\ref{example: simulation bridge}}
				& \multirow[c]{2}{*}{$T_{\mathrm{SRMR}}$}
				& N & 0.041 & 0.317 & 0.974 & 1.000 & 1.000 \\
				& &  {B}
				&  {0.048}
				&  {0.350}
				&  {0.977}
				&  {1.000}
				&  {1.000} \\
				& \multirow[c]{2}{*}{$\widetilde T_{\mathrm{SRMR}}$}
				& N & 0.043 & 0.305 & 0.966 & 1.000 & 1.000 \\
				& &  {B}
				&  {0.052}
				&  {0.338}
				&  {0.968}
				&  {1.000}
				&  {1.000} \\
				& \multirow[c]{2}{*}{$T_{\mathrm{ZH}}$}
				& N & 0.036 & 0.075 & 0.298 & 0.711 & 0.944 \\
				& & B & 0.060 & 0.117 & 0.397 & 0.786 & 0.966 \\
				& $T_{\mathrm{MMR}}$
				& B & 0.064 & 0.065 & 0.074 & 0.085 & 0.093 \\
				& $T_{\mathrm{STU}}$
				& B & 0.063 & 0.064 & 0.072 & 0.087 & 0.094 \\
				\midrule
				\multirow[c]{8}{*}{Example~\ref{example: simulation chebyshev}}
				& \multirow[c]{2}{*}{$T_{\mathrm{SRMR}}$}
				& N & 0.041 & 0.389 & 0.996 & 1.000 & 1.000 \\
				& &  {B}
				&  {0.055}
				&  {0.425}
				&  {0.996}
				&  {1.000}
				&  {1.000} \\
				& \multirow[c]{2}{*}{$\widetilde T_{\mathrm{SRMR}}$}
				& N & 0.046 & 0.376 & 0.989 & 1.000 & 1.000 \\
				& &  {B}
				&  {0.053}
				&  {0.406}
				&  {0.989}
				&  {1.000}
				&  {1.000} \\
				& \multirow[c]{2}{*}{$T_{\mathrm{ZH}}$}
				& N & 0.027 & 0.477 & 1.000 & 1.000 & 1.000 \\
				& & B & 0.055 & 0.580 & 1.000 & 1.000 & 1.000 \\
				& $T_{\mathrm{MMR}}$
				& B & 0.045 & 0.082 & 0.296 & 0.916 & 1.000 \\
				& $T_{\mathrm{STU}}$
				& B & 0.051 & 0.084 & 0.302 & 0.931 & 1.000 \\
				\midrule
				\multirow[c]{8}{*}{Example~\ref{example: simulation normal cdf}}
				& \multirow[c]{2}{*}{$T_{\mathrm{SRMR}}$}
				& N & 0.041 & 0.348 & 0.990 & 1.000 & 1.000 \\
				& &  {B}
				&  {0.051}
				&  {0.377}
				&  {0.991}
				&  {1.000}
				&  {1.000} \\
				& \multirow[c]{2}{*}{$\widetilde T_{\mathrm{SRMR}}$}
				& N & 0.045 & 0.358 & 0.984 & 1.000 & 1.000 \\
				& &  {B}
				&  {0.051}
				&  {0.384}
				&  {0.985}
				&  {1.000}
				&  {1.000} \\
				& \multirow[c]{2}{*}{$T_{\mathrm{ZH}}$}
				& N & 0.048 & 0.448 & 0.998 & 1.000 & 1.000 \\
				& & B & 0.058 & 0.472 & 0.998 & 1.000 & 1.000 \\
				& $T_{\mathrm{MMR}}$
				& B & 0.050 & 0.056 & 0.073 & 0.110 & 0.152 \\
				& $T_{\mathrm{STU}}$
				& B & 0.051 & 0.056 & 0.074 & 0.110 & 0.151 \\
				\bottomrule
			\end{tabular}
			\endgroup
		\end{table}
		
		
		\begin{table}[!htb]
			\centering
			\caption{Empirical sizes and powers for $p=2$,
				with $n=N=500$, $q=10$, and $\lambda=10^{-4}$.}
			\label{tab:comparison_p2}
			\begingroup
			\small
			\renewcommand{\arraystretch}{1.1}
			\setlength{\tabcolsep}{4pt}
			\begin{tabular}{@{}cccccccc@{}}
				\toprule
				\multirow[c]{2}{*}{Example}
				& \multirow[c]{2}{*}{Methods}
				& \multirow[c]{2}{*}{Calibration}
				& \multicolumn{5}{c}{$a$}
				\\
				\cmidrule(lr){4-8}
				& & & 0.0 & 0.2 & 0.4 & 0.6 & 0.8
				\\
				\midrule
				\multirow[c]{8}{*}{Example~\ref{example: simulation bridge}}
				& \multirow[c]{2}{*}{$T_{\mathrm{SRMR}}$}
				& N & 0.045 & 0.232 & 0.916 & 1.000 & 1.000 \\
				& &  {B}
				&  {0.055}
				&  {0.258}
				&  {0.925}
				&  {1.000}
				&  {1.000} \\
				& \multirow[c]{2}{*}{$\widetilde T_{\mathrm{SRMR}}$}
				& N & 0.037 & 0.211 & 0.884 & 1.000 & 1.000 \\
				& &  {B}
				&  {0.048}
				&  {0.239}
				&  {0.898}
				&  {1.000}
				&  {1.000} \\
				& \multirow[c]{2}{*}{$T_{\mathrm{ZH}}$}
				& N & 0.022 & 0.022 & 0.019 & 0.016 & 0.016 \\
				& & B & 0.048 & 0.050 & 0.050 & 0.048 & 0.046 \\
				& $T_{\mathrm{MMR}}$
				& B & 0.053 & 0.059 & 0.060 & 0.066 & 0.070 \\
				& $T_{\mathrm{STU}}$
				& B & 0.049 & 0.051 & 0.054 & 0.061 & 0.068 \\
				\midrule
				\multirow[c]{8}{*}{Example~\ref{example: simulation chebyshev}}
				& \multirow[c]{2}{*}{$T_{\mathrm{SRMR}}$}
				& N & 0.037 & 0.183 & 0.842 & 0.999 & 1.000 \\
				& &  {B}
				&  {0.042}
				&  {0.204}
				&  {0.862}
				&  {0.999}
				&  {1.000} \\
				& \multirow[c]{2}{*}{$\widetilde T_{\mathrm{SRMR}}$}
				& N & 0.038 & 0.195 & 0.803 & 0.997 & 1.000 \\
				& &  {B}
				&  {0.045}
				&  {0.210}
				&  {0.820}
				&  {0.998}
				&  {1.000} \\
				& \multirow[c]{2}{*}{$T_{\mathrm{ZH}}$}
				& N & 0.025 & 0.079 & 0.526 & 0.951 & 0.999 \\
				& & B & 0.050 & 0.154 & 0.641 & 0.978 & 1.000 \\
				& $T_{\mathrm{MMR}}$
				& B & 0.057 & 0.062 & 0.082 & 0.119 & 0.183 \\
				& $T_{\mathrm{STU}}$
				& B & 0.055 & 0.067 & 0.137 & 0.303 & 0.595 \\
				\midrule
				\multirow[c]{8}{*}{Example~\ref{example: simulation normal cdf}}
				& \multirow[c]{2}{*}{$T_{\mathrm{SRMR}}$}
				& N & 0.040 & 0.190 & 0.815 & 1.000 & 1.000 \\
				& &  {B}
				&  {0.052}
				&  {0.215}
				&  {0.835}
				&  {1.000}
				&  {1.000} \\
				& \multirow[c]{2}{*}{$\widetilde T_{\mathrm{SRMR}}$}
				& N & 0.039 & 0.173 & 0.748 & 0.991 & 1.000 \\
				& &  {B}
				&  {0.047}
				&  {0.199}
				&  {0.773}
				&  {0.991}
				&  {1.000} \\
				& \multirow[c]{2}{*}{$T_{\mathrm{ZH}}$}
				& N & 0.039 & 0.071 & 0.187 & 0.482 & 0.789 \\
				& & B & 0.049 & 0.082 & 0.218 & 0.516 & 0.809 \\
				& $T_{\mathrm{MMR}}$
				& B & 0.051 & 0.051 & 0.052 & 0.063 & 0.070 \\
				& $T_{\mathrm{STU}}$
				& B & 0.045 & 0.047 & 0.062 & 0.073 & 0.086 \\
				\bottomrule
			\end{tabular}
			\endgroup
		\end{table}
		
		
		\begin{table}[!htb]
			\centering
			\caption{Empirical sizes and powers for $p=4$,
				with $n=N=500$, $q=10$, and $\lambda=10^{-4}$.}
			\label{tab:comparison_p4}
			\begingroup
			\small
			\renewcommand{\arraystretch}{1.1}
			\setlength{\tabcolsep}{4pt}
			\begin{tabular}{@{}cccccccc@{}}
				\toprule
				\multirow[c]{2}{*}{Example}
				& \multirow[c]{2}{*}{Methods}
				& \multirow[c]{2}{*}{Calibration}
				& \multicolumn{5}{c}{$a$}
				\\
				\cmidrule(lr){4-8}
				& & & 0.0 & 0.4 & 0.8 & 1.2 & 1.6
				\\
				\midrule
				\multirow[c]{8}{*}{Example~\ref{example: simulation bridge}}
				& \multirow[c]{2}{*}{$T_{\mathrm{SRMR}}$}
				& N & 0.051 & 0.550 & 0.969 & 0.993 & 0.998 \\
				& &  {B}
				&  {0.060}
				&  {0.582}
				&  {0.970}
				&  {0.993}
				&  {0.998} \\
				& \multirow[c]{2}{*}{$\widetilde T_{\mathrm{SRMR}}$}
				& N & 0.050 & 0.516 & 0.957 & 0.991 & 0.997 \\
				& &  {B}
				&  {0.055}
				&  {0.541}
				&  {0.963}
				&  {0.992}
				&  {0.996} \\
				& \multirow[c]{2}{*}{$T_{\mathrm{ZH}}$}
				& N & 0.005 & 0.002 & 0.004 & 0.006 & 0.006 \\
				& & B & 0.045 & 0.046 & 0.052 & 0.048 & 0.049 \\
				& $T_{\mathrm{MMR}}$
				& B & 0.042 & 0.044 & 0.057 & 0.056 & 0.054 \\
				& $T_{\mathrm{STU}}$
				& B & 0.037 & 0.046 & 0.042 & 0.042 & 0.048 \\
				\midrule
				\multirow[c]{8}{*}{Example~\ref{example: simulation chebyshev}}
				& \multirow[c]{2}{*}{$T_{\mathrm{SRMR}}$}
				& N & 0.042 & 0.301 & 0.936 & 1.000 & 1.000 \\
				& &  {B}
				&  {0.056}
				&  {0.337}
				&  {0.951}
				&  {1.000}
				&  {1.000} \\
				& \multirow[c]{2}{*}{$\widetilde T_{\mathrm{SRMR}}$}
				& N & 0.038 & 0.226 & 0.852 & 0.985 & 0.997 \\
				& &  {B}
				&  {0.045}
				&  {0.267}
				&  {0.876}
				&  {0.986}
				&  {0.996} \\
				& \multirow[c]{2}{*}{$T_{\mathrm{ZH}}$}
				& N & 0.034 & 0.035 & 0.042 & 0.055 & 0.057 \\
				& & B & 0.059 & 0.063 & 0.081 & 0.097 & 0.105 \\
				& $T_{\mathrm{MMR}}$
				& B & 0.062 & 0.067 & 0.091 & 0.111 & 0.127 \\
				& $T_{\mathrm{STU}}$
				& B & 0.047 & 0.064 & 0.103 & 0.154 & 0.207 \\
				\midrule
				\multirow[c]{8}{*}{Example~\ref{example: simulation normal cdf}}
				& \multirow[c]{2}{*}{$T_{\mathrm{SRMR}}$}
				& N & 0.043 & 0.151 & 0.610 & 0.902 & 0.969 \\
				& &  {B}
				&  {0.055}
				&  {0.185}
				&  {0.656}
				&  {0.921}
				&  {0.975} \\
				& \multirow[c]{2}{*}{$\widetilde T_{\mathrm{SRMR}}$}
				& N & 0.047 & 0.118 & 0.445 & 0.755 & 0.875 \\
				& &  {B}
				&  {0.052}
				&  {0.146}
				&  {0.488}
				&  {0.784}
				&  {0.898} \\
				& \multirow[c]{2}{*}{$T_{\mathrm{ZH}}$}
				& N & 0.050 & 0.047 & 0.056 & 0.066 & 0.074 \\
				& & B & 0.066 & 0.062 & 0.068 & 0.083 & 0.086 \\
				& $T_{\mathrm{MMR}}$
				& B & 0.044 & 0.054 & 0.061 & 0.061 & 0.066 \\
				& $T_{\mathrm{STU}}$
				& B & 0.046 & 0.052 & 0.058 & 0.063 & 0.064 \\
				\bottomrule
			\end{tabular}
			\endgroup
		\end{table}

        \medskip
		\noindent
		\textbf{General comparisons.}	 
	The  preceding simulations  show that the power improvement of the proposed tests over
	the MMR benchmark depends on the regularization parameter and
	the model structure. On the one hand,
	the best choice of regularization parameter depends on the unknown parameters,
	as discussed in Remark~\ref{remark: parameter selection}. On the other hand, it is known that   kernel selection can also 
	affect testing performance.
	To improve robustness of the proposed tests,
	we consider an  aggregation strategy based on an  idea similar to that used
	in~\citet{hagrass-two-sample-test,hagrass-gof-test,hagrass2026ksd}. 
	To illustrate this approach, we consider the Gaussian kernel
	$K_h^{\mathrm G}(x,x')
	=\exp\{-\|x-x'\|_2^2/h\}$
	and the inverse multiquadric kernel
	$K_h^{\mathrm{IMQ}}(x,x')
	=(1+\|x-x'\|_2^2/h)^{-1/2}$. Given  prescribed grids $\Lambda$ and $H$ of
	regularization parameters and bandwidths, we compute the studentized SRMR
	statistic for each $(\lambda , h) \in \Lambda \times H $, denoted as $\widehat S_{\lambda,h}$, and let $\widehat q_{1-\alpha/(|\Lambda||H|)}^{\,\lambda,h}$
	denote the corresponding empirical wild-bootstrap quantile.
	With a Bonferroni allocation of the significance level, the
	aggregated procedure rejects the null hypothesis whenever
	$\widehat S_{\lambda,h}>
	\widehat q_{1-\alpha/(|\Lambda||H|)}^{\,\lambda,h}$
	for at least one $(\lambda,h)\in\Lambda\times H$.

	We evaluate this procedure under  four regression models. Specifically,  data are generated from
	\begin{equation}
		\label{equ: single index alternatives}
		\begin{aligned}
			H_{11}:\quad
			Y
			&=
			\theta^\top X
			+a\cos(0.6\pi\theta^\top X)+e,\\
			H_{12}:\quad
			Y
			&=
			\theta^\top X
			+\frac{a}{\sqrt p}
			\sum_{\ell=1}^{p}
			\left[
			\cos(0.6\pi X_\ell)
			-\exp\left\{-\frac{(0.6\pi)^2}{2}\right\}
			\right]+e,\\
			H_{13}:\quad
			Y
			&=
			\theta^\top X
			+\frac{a}{\sqrt p}
			\sum_{\ell=1}^{p}
			\left\{
			\exp(-X_\ell^2/2)-\frac{1}{\sqrt2}
			\right\}+e,\\
			H_{14}:\quad
			Y
			&=
			\theta^\top X
			+a\left[
			\mathds{1}\{|\theta^\top X|>1\}
			-2\{1-\Phi(1)\}
			\right]+e,
		\end{aligned}
	\end{equation}
	where $\theta=\mathbf1_p/\sqrt p$, $X\sim N(0,I_p)$,
	$e\sim N(0,1)$ is independent of $X$, and $\Phi$ denotes
	the standard normal distribution function.
	Their common null hypothesis
	corresponds to $a=0$, when the regression function belongs to
	$\mathcal G(\Theta):=\{x\mapsto\theta^\top x:
	\theta\in\mathbb R^p\}$.
	Larger values of $|a|$ produce stronger departures from this
	linear family.
	
	In the following simulations, we set $n=N=200$ and $p\in\{4,8\}$. We use the regularization grid
	$\Lambda=\{10^k:k=-6,\ldots,-1\}$ and the bandwidth grid
	$H=\{h_m\cdot 10^{j/2}:j=-4,\ldots,4\}$, where $	h_m
	=
	\operatorname{median}
	\{\|X_i-X_j\|_2^2:1\leq i<j\leq n\}$
	is determined from the labeled covariates by the median
	heuristic~\citep{gretton2012kernel}. 
	We denote the resulting procedures by
	$T_{\mathrm{SRMR}}^{\mathrm{Bon}}$ and
	$\widetilde T_{\mathrm{SRMR}}^{\mathrm{Bon}}$, respectively.  As discussed by \citet{hagrass-two-sample-test}, the number of
	resampling replications $B$ can affect the calibration of
	aggregated tests. We observe a similar phenomenon, as shown in
	Table~\ref{tab: effect_B}, where a small \(B\) may lead to
	anti-conservative behavior, whereas a large one may make the test
	conservative.  Based on these results, we set $B=200$
	for the SRMR-based tests, which provides reasonable
	empirical Type~I error control in the reported settings.
	For comparison, $T_{\mathrm{MMR}}$ uses the same kernel as
	SRMR with bandwidth $h=h_m$, while $T_{\mathrm{ZH}}$ and
	$T_{\mathrm{STU}}$ retain the statistics and tuning choices
	used in the preceding experiments. All three competing tests
	use wild-bootstrap calibration with $B=500$ replications.
	\begin{table}[!htp]
		\centering
		\caption{Empirical sizes for different numbers of bootstrap
			replications, with $n=N=200$. }
		\label{tab: effect_B}
		\begingroup
		\small
		\renewcommand{\arraystretch}{1.25}
		\setlength{\tabcolsep}{6pt}
		\begin{tabular}{@{}cccccccc@{}}
			\toprule
			\multirow[c]{2}{*}{Kernel}
			& \multirow[c]{2}{*}{$p$}
			& \multirow[c]{2}{*}{Methods}
			& \multicolumn{5}{c}{$B$} \\
			\cmidrule(lr){4-8}
			& & & 100 & 150 & 200 & 250 & 300 \\
			\midrule
			\multirow[c]{4}{*}{Gaussian}
			& \multirow[c]{2}{*}{4}
			& $T_{\mathrm{SRMR}}^{\mathrm{Bon}}$
			& 0.096 & 0.070 & 0.054 & 0.045 & 0.039 \\
			& & $\widetilde T_{\mathrm{SRMR}}^{\mathrm{Bon}}$
			& 0.087 & 0.061 & 0.048 & 0.040 & 0.036 \\
			& \multirow[c]{2}{*}{8}
			& $T_{\mathrm{SRMR}}^{\mathrm{Bon}}$
			& 0.081 & 0.056 & 0.043 & 0.036 & 0.032 \\
			& & $\widetilde T_{\mathrm{SRMR}}^{\mathrm{Bon}}$
			& 0.084 & 0.060 & 0.049 & 0.041 & 0.036 \\
			\midrule
			\multirow[c]{4}{*}{IMQ}
			& \multirow[c]{2}{*}{4}
			& $T_{\mathrm{SRMR}}^{\mathrm{Bon}}$
			& 0.072 & 0.050 & 0.041 & 0.032 & 0.028 \\
			& & $\widetilde T_{\mathrm{SRMR}}^{\mathrm{Bon}}$
			& 0.065 & 0.044 & 0.035 & 0.027 & 0.024 \\
			& \multirow[c]{2}{*}{8}
			& $T_{\mathrm{SRMR}}^{\mathrm{Bon}}$
			& 0.056 & 0.046 & 0.032 & 0.029 & 0.024 \\
			& & $\widetilde T_{\mathrm{SRMR}}^{\mathrm{Bon}}$
			& 0.055 & 0.042 & 0.030 & 0.027 & 0.025 \\
			\bottomrule
		\end{tabular}
		\endgroup
	\end{table}
	
\begin{figure}[!htp]
		\centering
		\captionsetup[subfigure]{font=small,skip=2pt}
		
		\begin{subfigure}[t]{0.46\textwidth}
			\centering
			\includegraphics[width=\linewidth]
			{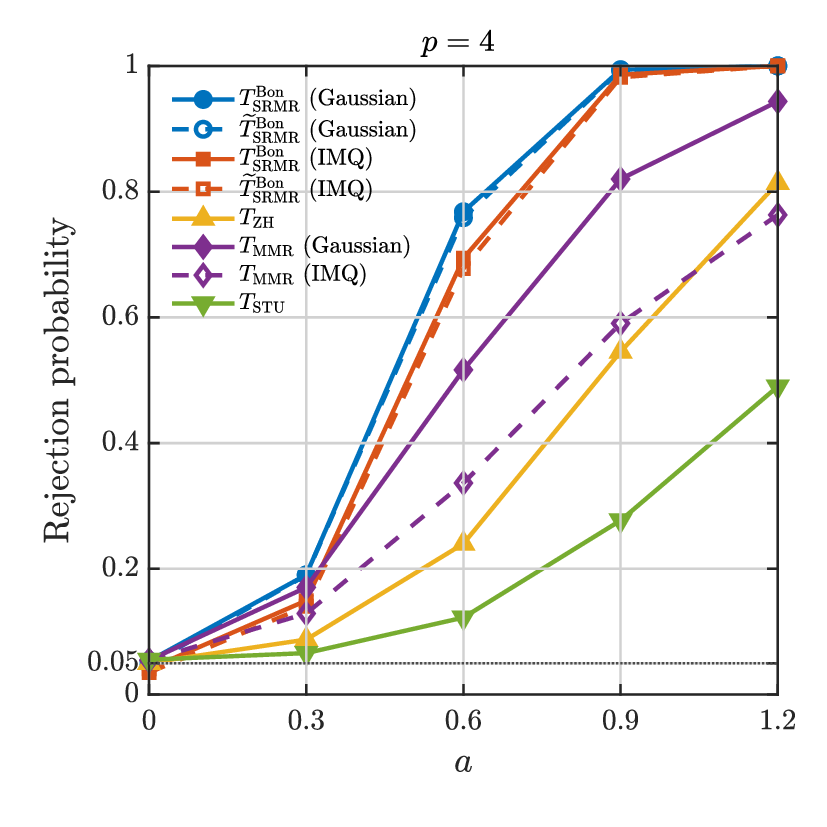}
			\label{fig:power_comparison_m1p4}
		\end{subfigure}
		\hfill
		\begin{subfigure}[t]{0.46\textwidth}
			\centering
			\includegraphics[width=\linewidth]
			{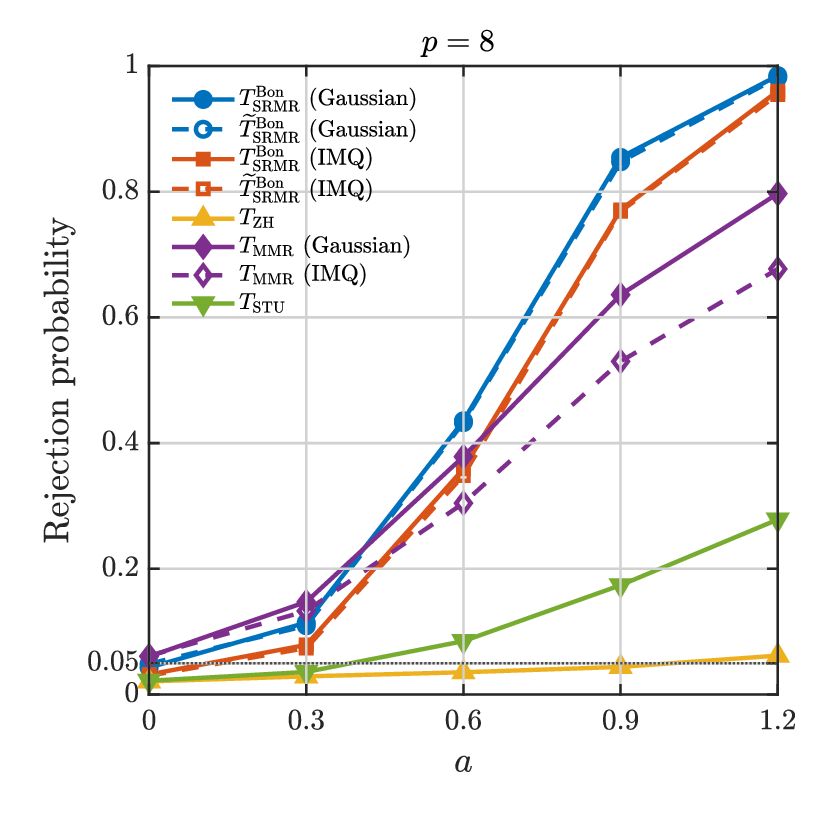}
			\label{fig:power_comparison_m1p8}
		\end{subfigure}
		
		\caption{Power comparisons under $H_{11}$ using the Gaussian
			and IMQ kernels, with $n=N=200$.}
		\label{fig:power_comparison_model1}
	\end{figure}

\begin{figure}[!htp]
		\centering
		\captionsetup[subfigure]{font=small,skip=2pt}
		
		\begin{subfigure}[t]{0.46\textwidth}
			\centering
			\includegraphics[width=\linewidth]
			{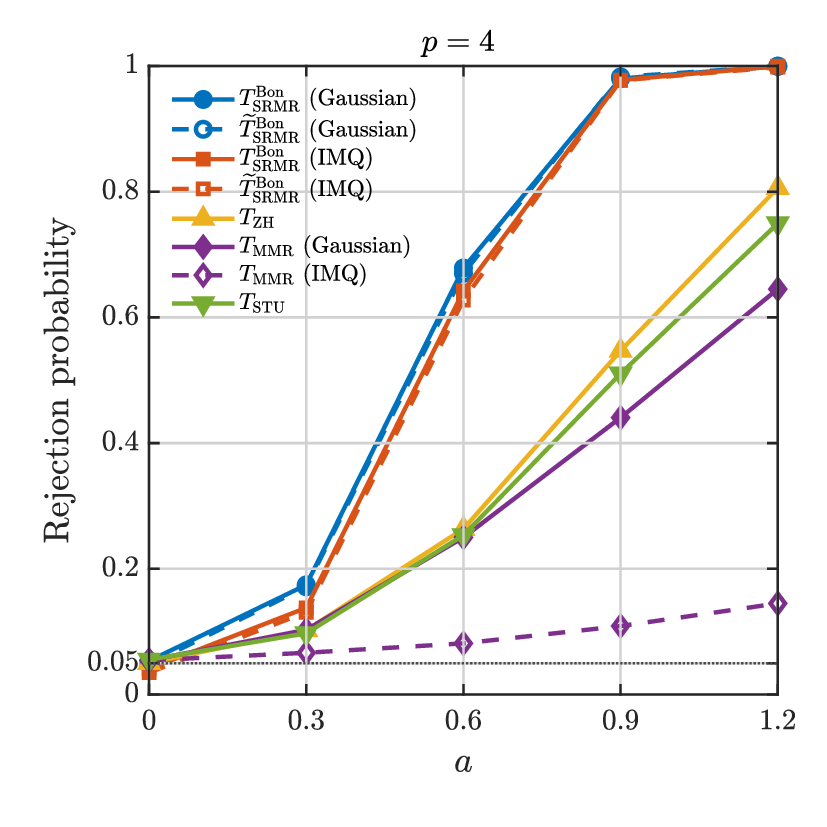}
		\end{subfigure}
		\hfill
		\begin{subfigure}[t]{0.46\textwidth}
			\centering
			\includegraphics[width=\linewidth]
			{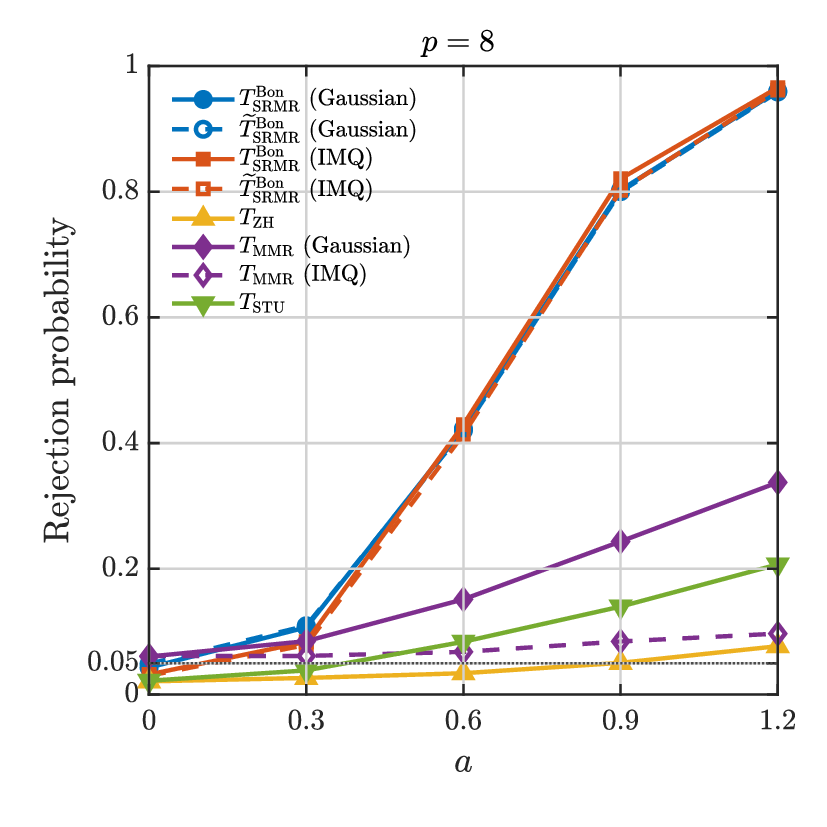}
		\end{subfigure}
		
		\caption{Power comparisons under $H_{12}$ using the Gaussian
			and IMQ kernels, with $n=N=200$.
			The horizontal dotted lines indicate the nominal
			significance level $0.05$.}
		\label{fig:power_comparison_model2}
	\end{figure}

\begin{figure}[!htp]
		\centering
		\captionsetup[subfigure]{font=small,skip=2pt}
		
		\begin{subfigure}[t]{0.46\textwidth}
			\centering
			\includegraphics[width=\linewidth]
			{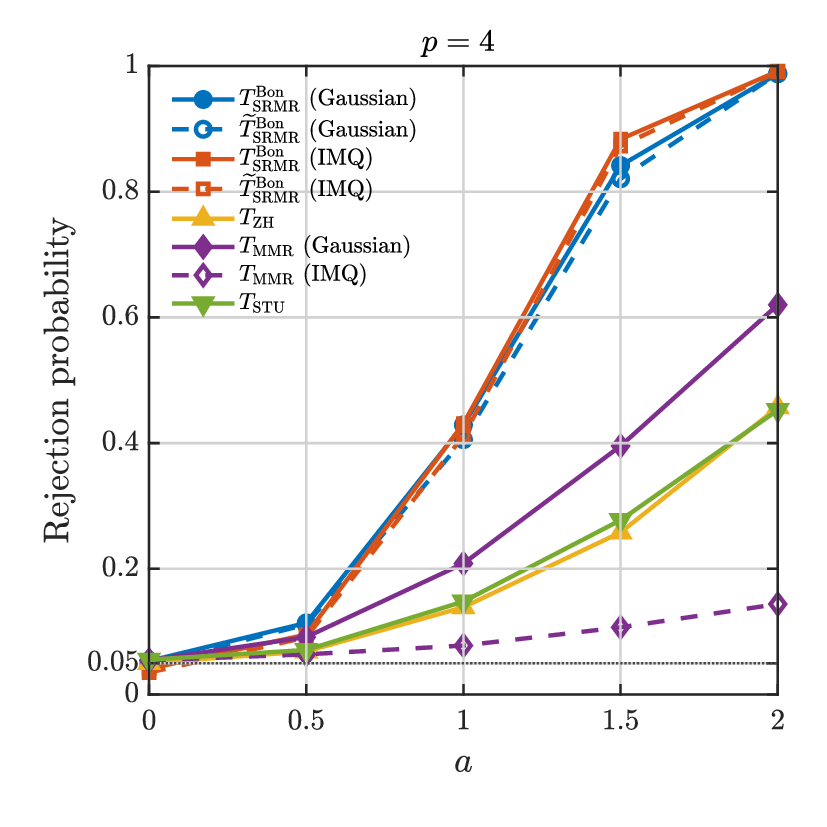}
		\end{subfigure}
		\hfill
		\begin{subfigure}[t]{0.46\textwidth}
			\centering
			\includegraphics[width=\linewidth]
			{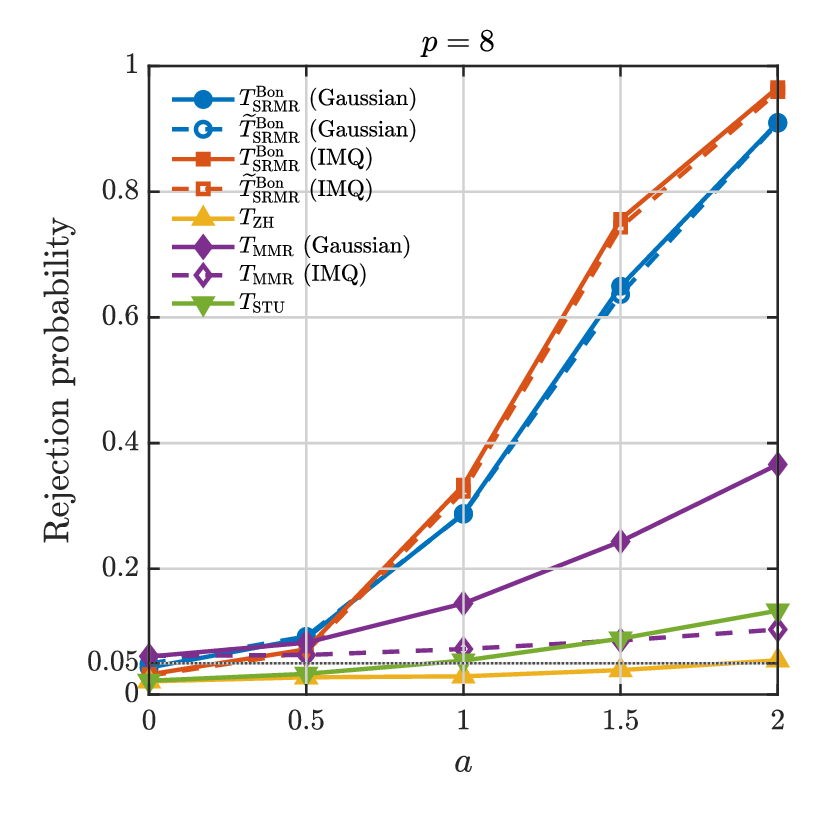}
		\end{subfigure}
		
		\caption{Power comparisons under $H_{13}$ using the Gaussian
			and IMQ kernels, with $n=N=200$.
			The horizontal dotted lines indicate the nominal
			significance level $0.05$.}
		\label{fig:power_comparison_model3}
	\end{figure}

\begin{figure}[!htp]
		\centering
		\captionsetup[subfigure]{font=small,skip=2pt}
		
		\begin{subfigure}[t]{0.46\textwidth}
			\centering
			\includegraphics[width=\linewidth]
			{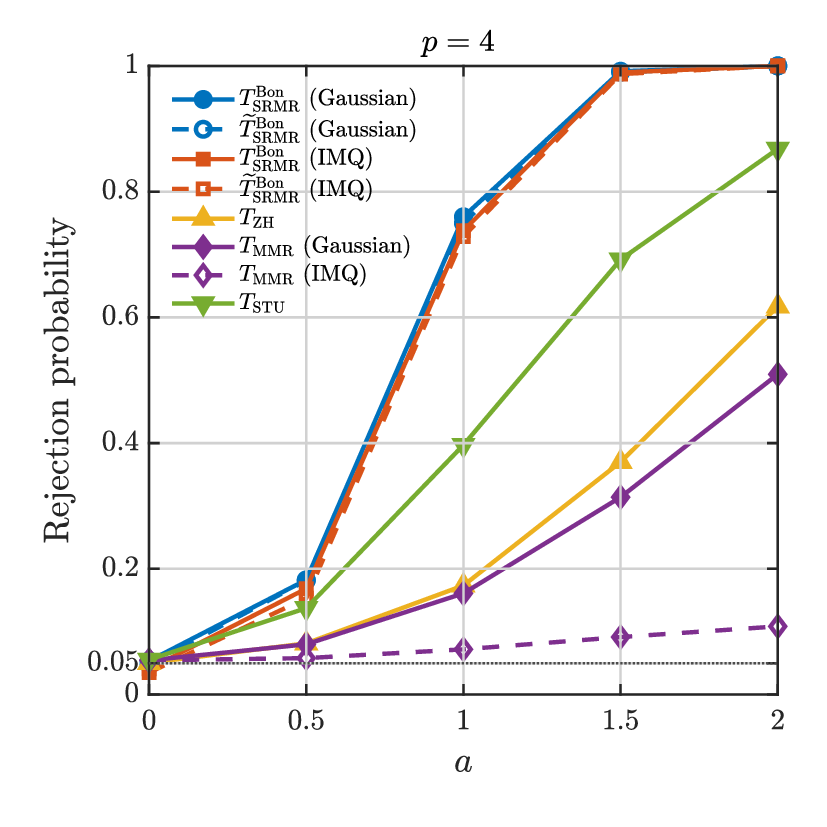}
		\end{subfigure}
		\hfill
		\begin{subfigure}[t]{0.46\textwidth}
			\centering
			\includegraphics[width=\linewidth]
			{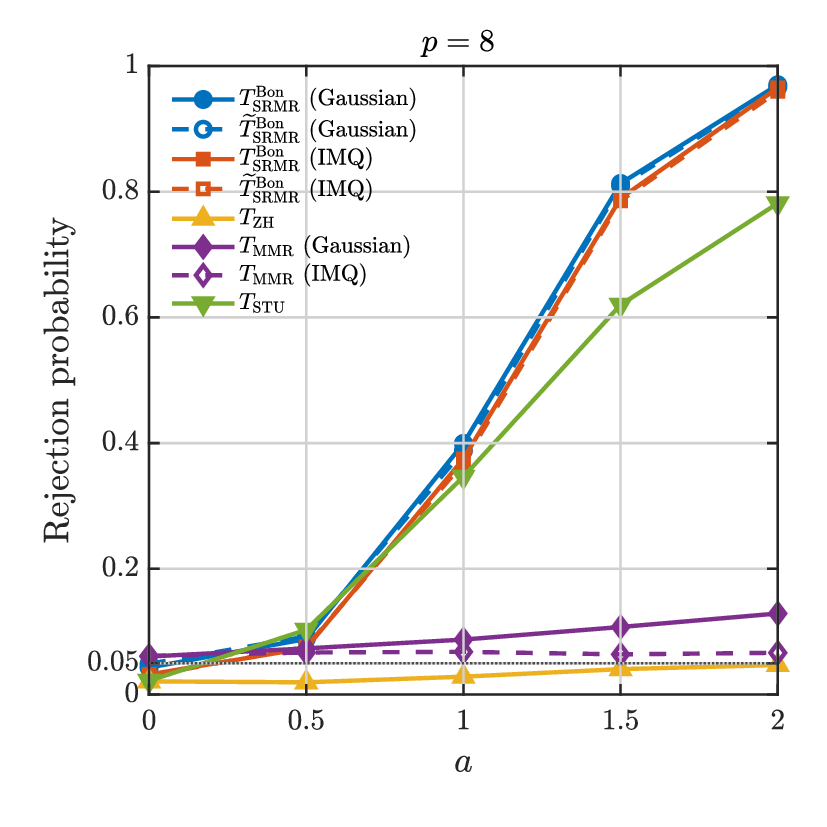}
		\end{subfigure}
		
		\caption{Power comparisons under $H_{14}$ using the Gaussian
			and IMQ kernels, with $n=N=200$.
			The horizontal dotted lines indicate the nominal
			significance level $0.05$.}
		\label{fig:power_comparison_model4}
	\end{figure}

	Figures~\ref{fig:power_comparison_model1}--\ref{fig:power_comparison_model4}
	show that the SRMR-based tests, combined with the Bonferroni
	aggregation strategy, achieve competitive empirical power
	across the four models.
	For example, the SRMR-based tests with the Gaussian kernel
	have a clear advantage over $T_{\mathrm{ZH}}$,
	$T_{\mathrm{MMR}}$, and $T_{\mathrm{STU}}$ under the additive
	departures $H_{12}$ and $H_{13}$ in
	Figures~\ref{fig:power_comparison_model2}
	and~\ref{fig:power_comparison_model3}, particularly at moderate
	and strong signal levels.
	Although our tests do not have the highest empirical power in every reported setting, for example under weak signals in $H_{11}$ and
	$H_{14}$, they outperform the competing methods in most of the reported settings.
	Similar advantages are observed with the IMQ kernel.
	We also observe that kernel choice and the presence of
	covariate shift affect testing performance.
	Nevertheless, the results indicate that the proposed tests
	with aggregation can be effective across different  alternatives.
	
	%
	%

		\subsection{A Real Data Example}
		\label{sec: real data}
		We apply the proposed test to a publicly available dataset
		from a survey conducted by the Los Angeles Homeless Services
		Authority (LAHSA) during 2004--2005.
		Among the 2,054 census tracts in Los Angeles County, 244 areas believed to have large homeless populations were designated as ``hot tracts" and surveyed in full. A further 265 tracts were selected through stratified random sampling from the remaining 1,810 non-hot tracts, leaving 1,545 tracts unvisited.  Following the analysis in~\citet{jiang2026semi}, we focus  on the non-hot tracts.  The response variable is the street homeless count in each tract. The  covariates   are \texttt{PctVacant} (per
		centage of unoccupied housing units), \texttt{PctOwnerOcc} (percentage of owner-occupied housing
		units), \texttt{PctMinority} (percentage of non-Caucasian population), and \texttt{MHI} (median household
		income). These four covariates were found to be significant
		by~\citet{Brian2010}.  The resulting dataset contains \(n=265\) labeled observations, and \(N=1{,}545\) unlabeled observations.
		
		To examine whether a linear specification adequately describes
		the conditional mean, we compare a linear model with an additive
		quadratic model:  
		\begin{equation}
			\label{equ: homeless models}
			\begin{aligned}
				\mathrm{LM1}:\quad
				\mathbb E(Y\mid X)
				&= \theta_1
				+ \sum_{\ell=1}^{p}\theta_{\ell+1}X_\ell,\\
				\mathrm{LM2}:\quad
				\mathbb E(Y\mid X)
				&= \theta_1
				+ \sum_{\ell=1}^{p}\theta_{\ell+1}X_\ell
				+ \sum_{\ell=1}^{p}\theta_{p+\ell+1}X_\ell^2.
			\end{aligned}
		\end{equation}
		Here, the covariates $X_i$'s are  standardized using the labeled-sample means and
		standard deviations, with the same transformation applied to the
		unlabeled observations,
		while the response $Y$ is retained on its original scale.
		We apply the normal-calibrated test $T_{\mathrm{SRMR}}$
		using the kernel in Example~\ref{example: simulation normal cdf},
		auxiliary samples of sizes $N\in\{100,200,500\}$ randomly
		drawn from the unlabeled observations, and regularization
		parameters $\lambda\in\{10^{-6},10^{-5},10^{-4}\}$.
		
\begin{figure}[t]
			\centering
			\begin{subfigure}[t]{0.47\textwidth}
				\centering
				\includegraphics[width=\linewidth]{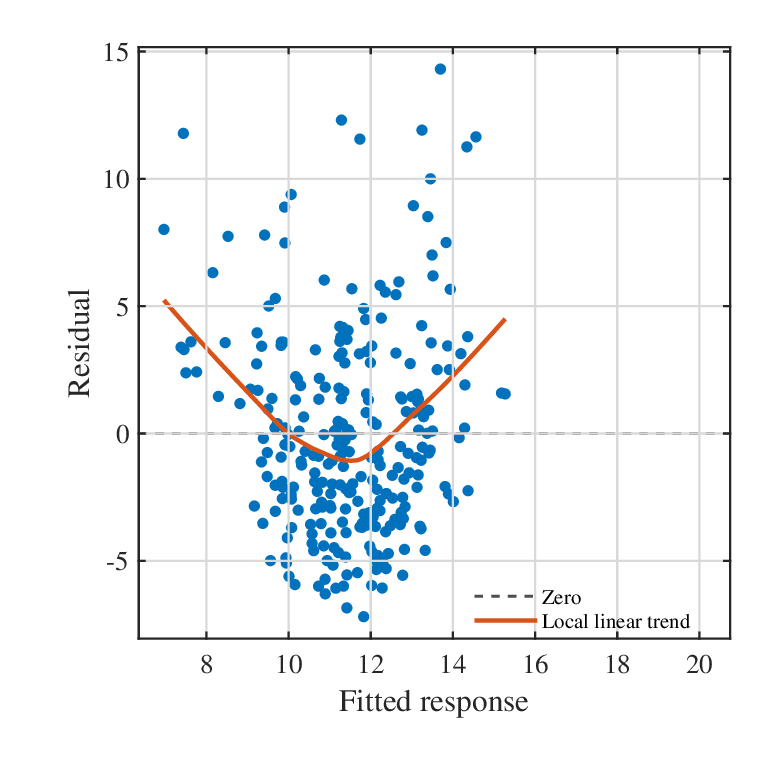}
				\caption{LM1: linear model}
				\label{fig:homeless_residual_linear}
			\end{subfigure} 
			\begin{subfigure}[t]{0.47\textwidth}
				\centering
				\includegraphics[width=\linewidth]{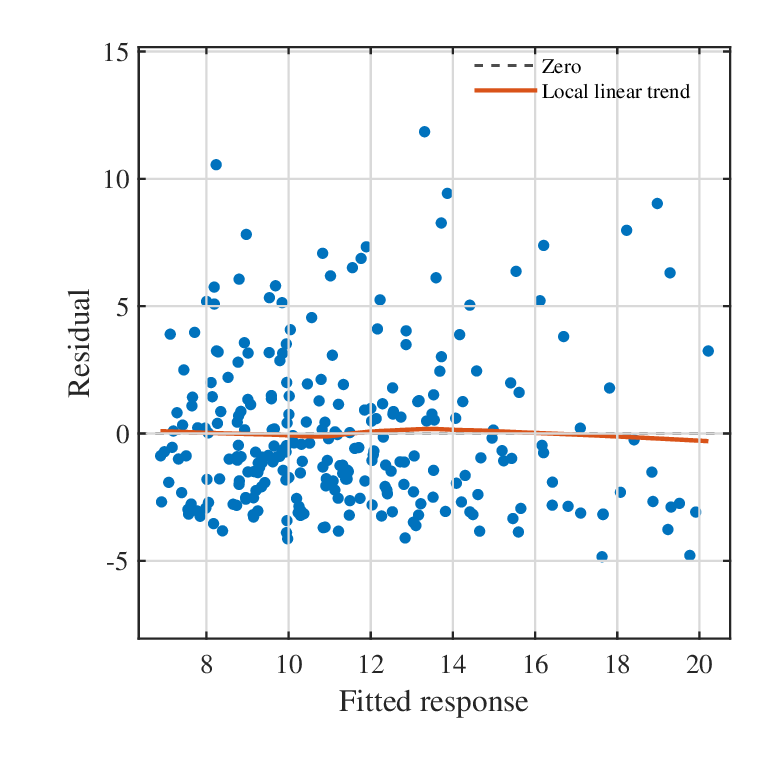}
				\caption{LM2: additive quadratic model}
				\label{fig:homeless_residual_quadratic}
			\end{subfigure}
			\caption{Residuals plotted against fitted responses for
				the two specifications.}
			\label{fig:homeless_residuals}
		\end{figure}

		Table~\ref{tab:homeless_pvalues} reports the resulting $p$-values.
		At the $5\%$ significance level, LM1 is rejected for every combination of $N$ and $\lambda$,
		with $p$-values ranging from $0.0082$ to $0.0269$.
		In contrast, LM2 is not rejected in any of the reported settings,
		with $p$-values between $0.1530$ and $0.5769$. These results consistently support the  inclusion of quadratic terms.  The residual plots in Figure~\ref{fig:homeless_residuals}
		also agree with these test results.
		The residuals from LM1 display a curved pattern, suggesting that
		the linear specification does not adequately capture the conditional
		mean structure. This pattern is less pronounced under LM2, providing further support for the additive quadratic specification.  
		
		\begin{table}[t]
			\centering
			\caption{$p$-values for LM1 and LM2.}
			\label{tab:homeless_pvalues}
			\begingroup
			\small
			\renewcommand{\arraystretch}{1.15}
			\setlength{\tabcolsep}{7pt}
			\begin{tabular}{@{}ccccccc@{}}
				\toprule
				& \multicolumn{3}{c}{LM1}
				& \multicolumn{3}{c}{LM2}\\
				\cmidrule(lr){2-4}
				\cmidrule(lr){5-7}
				$N$
				& $\lambda=10^{-6}$
				& $\lambda=10^{-5}$
				& $\lambda=10^{-4}$
				& $\lambda=10^{-6}$
				& $\lambda=10^{-5}$
				& $\lambda=10^{-4}$\\
				\midrule
				100 & 0.0126 & 0.0126 & 0.0123
				& 0.1530 & 0.1530 & 0.1538\\
				200 & 0.0269 & 0.0268 & 0.0257
				& 0.5753 & 0.5755 & 0.5769\\
				500 & 0.0105 & 0.0103 & 0.0082
				& 0.4117 & 0.4141 & 0.4336\\
				\bottomrule
			\end{tabular}
			\endgroup
		\end{table}
		
		\section{Concluding Remarks}
		\label{sec: conclusion}
		This paper develops a spectral regularized kernel-embedding approach
		that incorporates auxiliary covariates into regression
		specification testing, and establishes its calibration and detection properties. Our work also
		points to several interesting questions for further investigation. 
		
		Our analysis assumes that the density ratio between the
		auxiliary and labeled covariate distributions is bounded
		above and away from zero.
		Replacing these bounds with weaker 	conditions could accommodate auxiliary data from a broader
		range of sources~\citep{ma2023optimally,gogolashvili2023importance,fan2025spectral}.
		An important question is how such distributional differences
		affect test calibration and the resulting detection properties.

	In practice, our method involves choosing the regularization
	parameter, the kernel, and the auxiliary sample size.
	Existing studies provide useful guidance for parameter and
	kernel selection~\citep{hagrass-gof-test,hagrass-two-sample-test},
	and further theoretical analysis could help guide these
	choices in our setting.
	Our results also provide sufficient conditions on the auxiliary sample size $N$, which raises the computational issue. 
	Random Fourier features and Nystr\"om
	approximations have been studied for kernel distribution tests~\citep{mukherjee2025randomfeatures,chatalic2025scalable}, offering possible ways to reduce this cost.
	Their incorporation into our methods, together with the resulting
	effects on testing performance, would be worth studying.
	
	More generally, Tikhonov regularization is a special case
	of spectral filtering.
	The use of general spectral regularizers in distribution
	testing~\citep{hagrass-gof-test,hagrass-two-sample-test}
	suggests a further direction for extending SRMR beyond
	the particular regularization scheme considered here.
		\bibliography{reference}
		
		\appendix
		\appendix

\section{Error Analysis}
	\label{sec: error analysis}
    
	In this section,  we analyze the stochastic errors of the proposed SRMR statistic relative to the target signal, namely, the squared $L^2(P_X)$ norm of the conditional residual function in \eqref{equ: conditional residual function}.
	This analysis is used in the proofs of Theorems~\ref{theorem: asymptotic null distribution} and \ref{theorem: uniform power}.

For the Tikhonov filter $g_\lambda(t):=(t+\lambda)^{-1}$,
	$t\geq0$, we use the abbreviations
	\begin{equation}
		\label{equ: TLTDL}
		T_\lambda
		:=g_\lambda(L_{K,0})
		=(L_{K,0}+\lambda I)^{-1},
		\qquad
		T_{D,\lambda}
		:=g_\lambda(L_{K,D})
		=(L_{K,D}+\lambda I)^{-1}.
	\end{equation}
	We use the residual shorthand $	\epsilon:=\epsilon(Z,\theta^\star)$, $
	\epsilon_i:=\epsilon(Z_i,\theta^\star)$, $
	\widehat\epsilon:=\epsilon(Z,\widehat\theta)$, and $
	\widehat\epsilon_i:=\epsilon(Z_i,\widehat\theta)$.
	Define $	 
	\mu^\star
	:=\mathbb E(K_X\epsilon)
	=I_{K,X}^{\ast}\mathbf f$, $
	\omega:=K_X\epsilon-\mu^\star$, and $ 
	\omega_i:=K_{X_i}\epsilon_i-\mu^\star$. 
	Their fitted counterparts are
	$\widehat\omega:=K_X\widehat\epsilon-\mu^\star$
	and
	$\widehat\omega_i:=K_{X_i}\widehat\epsilon_i-\mu^\star$.
    
	For the fixed pseudo-true parameter
	$\theta^\star$, we define the semi-population quantity
	\begin{equation}
		\label{equ: semi population}
		\eta_\lambda^{s}(\theta^\star)
		:=
		\mathbb E\left\{\widehat\eta_\lambda(\theta^\star)\mid D\right\}
		=	\left\|
		T_{D,\lambda}^{1/2} \mu^{\star}
		\right\|_K^2=
		\left\|
		T_{D,\lambda}^{1/2}I_{K,X}^{\ast}\mathbf f
		\right\|_K^2.
	\end{equation}
	The discrepancy of the statistic from the target signal can then be decomposed as
	\begin{equation}
		\label{equ: error decomposition}
		\begin{aligned}
			\widehat\eta_\lambda(\widehat\theta)
			\ -\ \|\mathbf f\|_{L^2(P_X)}^2
			\	=  \
			\underbrace{
				\widehat\eta_\lambda(\widehat\theta)
				\	-\ \widehat\eta_\lambda(\theta^\star)
			}_{\textit{Parametric  error}}\ +\
			\underbrace{
				\widehat\eta_\lambda(\theta^\star)
				\ -\ \eta_\lambda^s(\theta^\star)
			}_{\textit{Sample  error}}\ +\
			\underbrace{
				\eta_\lambda^s(\theta^\star)
				\ -\ \|\mathbf f\|_{L^2(P_X)}^2
			}_{\textit{Approximation  error}}.
		\end{aligned}
	\end{equation}

	The \emph{parametric error} arises from replacing
	$\theta^\star$ with its estimator $\widehat\theta$.
	This error is common in regression specification
	testing~\citep{zheng1996consistent,hitomi2023optimal,
		guerre2002optimal}, but does not arise in related
	distribution-testing analyses, where no  parameter
	is fitted~\citep{balasubramanian2021optimality,
		hagrass-two-sample-test,hagrass2026ksd,hagrass-gof-test}.
	Under Assumption~\ref{assumption: regularities on parametric families},
	Taylor's formula separates the parameter increment
	$\widehat\theta-\theta^\star$ from random terms involving
	the derivatives of $g(x,\theta)$ and the empirical integral
	operator. 
	Lemma~\ref{lemma: parametric error lemma} provides the
	required conditional moment bounds for these terms.
	Combining them with the parameter-estimation rates in
	Assumptions~\ref{assumption: identibility and parametric estimation under H0}
	and~\ref{assumption: uniform parametric estimation under H1}, we can
	control the parametric error.
	
	The \emph{sample error} arises from using the labeled
	observations to estimate the population residual embedding.
	It can be decomposed into a centered quadratic term and
	a linear term.
	Lemma~\ref{lemma: sample error lemma} provides conditional
	second-moment bounds for both terms, which yield a stochastic
	bound for the sample error.
	
	The \emph{approximation error} includes the effects of
	Tikhonov regularization and estimation of the integral operator.
	For the power analysis, we only need
	$\eta_\lambda^s(\theta^\star)$ to remain sufficiently large
	when the target signal $\|\mathbf{f}\|_{L^2(P_X)}^2$ is large.
	We therefore use operator inequalities and the source
	condition to derive a lower bound for
	$\eta_\lambda^s(\theta^\star)$, rather than require it to
	approximate $\|\mathbf f\|_{L^2(P_X)}^2$ accurately.
	This weaker requirement avoids the saturation restriction of Tikhonov regularization; see
	\citet[Remark~4.3]{hagrass-two-sample-test}.

	\begin{lemma}
		\label{lemma: parametric error lemma}
		Suppose that
		Assumptions~\ref{assumption: auxiliary covariate distribution}, \ref{assumption: conditional variance}
		and~\ref{assumption: eigenvalues and eigenfunctions} hold, and $\|\mathbf f\|_{L^\infty(P_X)}\leq B$ 	for some fixed $B>0$. Recall that
		$\mu^\star=I_{K,X}^{\ast}\mathbf f$,
		$\omega_i=K_{X_i}\epsilon_i-\mu^\star$, and let $T_{D,\lambda}$ be as in \eqref{equ: TLTDL}. For real-valued coefficients
		$\mathbf h=(h_1,\ldots,h_n)$, which may depend on the labeled observations, define
		\[
		\begin{aligned}
			U_n(\mathbf h)
			&:=
			\frac1{n(n-1)}
			\sum_{i\ne j}
			\left\langle
			T_{D,\lambda}^{1/2}K_{X_i}h_i,
			T_{D,\lambda}^{1/2}\omega_j
			\right\rangle_K,
			\\
			V_n(\mathbf h)
			&:=
			\frac1n\sum_{i=1}^n
			\left\langle
			T_{D,\lambda}^{1/2}K_{X_i}h_i,
			T_{D,\lambda}^{1/2}\mu^\star
			\right\rangle_K.
		\end{aligned}
		\]
		When $h_i=h(X_i)$ for a deterministic function $h$,
		write $U_n(h)$ for $U_n(\mathbf h)$.  
		\begin{enumerate}
			\item[(i)] 
			For deterministic $h\in L^2(P_X)$,
			\begin{equation}
				\label{equ: conditional U fixed shift}
				\begin{aligned} 
					\mathbb E\left\{U_n^2(h) \mid D\right\}
					&\lesssim
					\mathscr P_{D,\lambda}^4
					\|h\|_{L^2(P_X)}^2
					\left(n^{-1}+n^{-2}\lambda^{-s}\right).
				\end{aligned}
			\end{equation}
			
			\item[(ii)]
			If $|h_i|\leq M_h(X_i)$ almost surely for a deterministic  $M_h\in L^1(P_X)$, then
			\begin{equation}
				\label{equ: conditional U adaptive shift}
				\mathbb E\{|U_n(\mathbf h)|\mid D\}
				\lesssim
				\mathscr P_{D,\lambda}^2
				\|M_h\|_{L^1(P_X)}
				n^{-1/2}\lambda^{-s/2}.
			\end{equation}
			
			\item[(iii)] 
			If $|h_i|\leq M_h(X_i)$ almost surely for a deterministic  $M_h\in L^2(P_X)$, then
			\begin{equation}
				\label{equ: conditional V adaptive shift}
				\mathbb E\{|V_n(\mathbf h)|\mid D\}
				\lesssim \mathscr P_{D,\lambda}^2
				\|M_h\|_{L^2(P_X)}
				\|\mathbf f\|_{L^2(P_X)}.
			\end{equation}
		\end{enumerate} 
		Here, 
		\begin{equation}
			\label{equ: PDL}
			\mathscr{P}_{D,\lambda}  = \left\| T_{D,\lambda}^{1/2} \,T_{\lambda}^{-1/2}\right\|,
		\end{equation} 
		where $
		T_\lambda  $ and $  
		T_{D,\lambda} $ are defined in \eqref{equ: TLTDL}. 
	\end{lemma}
	
	\begin{proof}
		\emph{(i).}
		Conditional on $D$, define the symmetric kernel
		\[
		\begin{aligned}
			H(Z_1,Z_2) = 
			\frac12
			\left\langle
			T_{D,\lambda}^{1/2}K_{X_1}h(X_1),
			T_{D,\lambda}^{1/2}\omega_2
			\right\rangle_K +
			\frac12
			\left\langle
			T_{D,\lambda}^{1/2}K_{X_2}h(X_2),
			T_{D,\lambda}^{1/2}\omega_1
			\right\rangle_K.
		\end{aligned}
		\]
		Then
		\[
		U_n(h)=\binom n2^{-1}\sum_{i<j}H(Z_i,Z_j), 
		\quad \mathrm{and} \quad  \mathbb E\left\{H(Z_1,Z_2)\mid D\right\}=0. 
		\]
		This implies  \(\mathbb E\{U_n(h)\mid D\}=0\).
		The first conditional projection of $H$ is
		\[
		\psi_1(Z_1)
		:=
		\mathbb E\{H(Z_1,Z_2)\mid D,Z_1\}
		=
		\frac12
		\left\langle
		T_{D,\lambda}^{1/2}\omega_1,
		T_{D,\lambda}^{1/2}I_{K,X}^{\ast}h
		\right\rangle_K.
		\]
		Then  by the independence
		of $D$ and the labeled observations and Lemma~\ref{lemma: Balazs2005},
		\[
		\begin{aligned}
			4\mathbb E\{\psi_1^2(Z_1)\mid D\}
			&=
			\left\langle
			T_{D,\lambda}^{1/2}I_{K,X}^\ast h,
			\left\{T_{D,\lambda}^{1/2}\mathbb E(\omega\otimes\omega)
			T_{D,\lambda}^{1/2}\right\}
			T_{D,\lambda}^{1/2}I_{K,X}^\ast h
			\right\rangle_K\\
			&\leq
			\left\|T_{D,\lambda}^{1/2}\mathbb E(\omega\otimes \omega)
			T_{D,\lambda}^{1/2}\right\|
			\left\|T_{D,\lambda}^{1/2}I_{K,X}^\ast h\right\|_K^2\\
			&\lesssim
			\mathscr P_{D,\lambda}^4
			\|h\|_{L^2(P_X)}^2.
		\end{aligned}
		\] 
		where the last step follows from  \eqref{equ: covariance empirical shift}
		and~\eqref{equ: empirical feature bound shift}.  The second conditional projection of $H$ is $\psi_2(Z_1,Z_2) = H(Z_1,Z_2) - \psi_1(Z_1) - \psi_1(Z_2)
		$. By the symmetry of $H$ and the inequality $(a+b)^2\leq2(a^2+b^2)$,
		\[
		\begin{aligned}
			\mathbb E\{\psi_2^2(Z_1,Z_2)\mid D\}& = \mathbb 
			E \left\{H^2(Z_1,Z_2) \mid D\right\} - 2\mathbb{E}\left\{\psi_1^2(Z)\mid D\right\}\\
			&\leq
			\mathbb E\left\{
			\left\langle
			T_{D,\lambda}^{1/2}K_{X_1}h(X_1),
			T_{D,\lambda}^{1/2}\omega_2
			\right\rangle_K^2
			\mid D\right\}
			\\
			&\leq
			\left\|
			T_{D,\lambda}^{1/2}
			\mathbb E(\omega\otimes\omega)
			T_{D,\lambda}^{1/2}
			\right\| 
			\mathbb E\left\{
			\left\|T_{D,\lambda}^{1/2}K_X\right\|_K^2
			h^2(X) \mid D\right\}
			\\
			&\lesssim
			\mathscr P_{D,\lambda}^4\lambda^{-s}
			\|h\|_{L^2(P_X)}^2,
		\end{aligned}
		\]
		where the last step also follows from   \eqref{equ: covariance empirical shift}
		and~\eqref{equ: empirical feature bound shift}.  
		Hoeffding's variance identity in Lemma~\ref{lemma: conditional Hoeffding decomposition}-\eqref{equ: conditional variance projection} therefore yields
		\[
		\begin{aligned}
			\mathbb E\left\{U_n^2(h)\mid D\right\}
			= &
			\frac{4}{n}
			\mathbb E\{\psi_1^2(Z_1)\mid D\}
			+
			\frac{2}{n(n-1)}
			\mathbb E\{\psi_2^2(Z_1,Z_2)\mid D\}
			\\[2mm]
			\lesssim &
			\mathscr P_{D,\lambda}^4
			\|h\|_{L^2(P_X)}^2
			\left(n^{-1}+n^{-2}\lambda^{-s}\right).
		\end{aligned}
		\]
		This proves \eqref{equ: conditional U fixed shift}.
		
		\medskip
		\noindent 	\emph{(ii).}
		Write
		\[
		S_i
		:=
		\frac1{n-1}\sum_{j\ne i}
		\left\langle
		T_{D,\lambda}^{1/2}K_{X_i},
		T_{D,\lambda}^{1/2}\omega_j
		\right\rangle_K.
		\]
		Conditional on $D$ and $X_i$, the summands are independent
		and centered. Then by Lemma~\ref{lemma: Balazs2005},
		\[
		\begin{aligned}
			(n-1)\mathbb E(S_i^2\mid D,X_i)
			&=
			\left\langle
			T_{D,\lambda}^{1/2}K_{X_i},
			\left\{
			T_{D,\lambda}^{1/2}
			\mathbb E(\omega\otimes\omega)
			T_{D,\lambda}^{1/2}
			\right\}
			T_{D,\lambda}^{1/2}K_{X_i}
			\right\rangle_K
			\\
			&\leq
			\left\|
			T_{D,\lambda}^{1/2}
			\mathbb E(\omega\otimes\omega)
			T_{D,\lambda}^{1/2}
			\right\|
			\left\|T_{D,\lambda}^{1/2}K_{X_i}\right\|_K^2
			\\
			&\lesssim
			\mathscr P_{D,\lambda}^4\lambda^{-s},
		\end{aligned}
		\]
		where the last step follows from \eqref{equ: covariance empirical shift}
		and~\eqref{equ: empirical feature bound shift}.
		Since $|h_i| \leq M_h(X_i)$ almost surely, triangle inequality yields
		\[
		|U_n(\mathbf h)|
		=
		\left|\frac1n\sum_{i=1}^n h_iS_i\right|
		\leq
		\frac1n\sum_{i=1}^n M_h(X_i)|S_i|.
		\]
		Hence,	conditional Cauchy--Schwarz inequality gives
		\[
		\begin{aligned}
			\mathbb E\{|U_n(\mathbf h)|\mid D\}
			&\leq
			\frac1n\sum_{i=1}^n
			\mathbb E\left[
			M_h(X_i)
			\{\mathbb E(S_i^2\mid D,X_i)\}^{1/2}
			\mid D\right]
			\\
			&\lesssim
			\mathscr P_{D,\lambda}^2
			n^{-1/2}\lambda^{-s/2}
			\|M_h\|_{L^1(P_X)}.
		\end{aligned}
		\]
		Here we used  the independence
		of $D$ and the labeled observations.
		This proves \eqref{equ: conditional U adaptive shift}.
		
		\medskip
		\noindent 	\emph{(iii).}
		By the self-adjointness of $T_{D,\lambda}$ and  the  reproducing property,
		$$
		\left\langle
		T_{D,\lambda}^{1/2}K_{x},
		T_{D,\lambda}^{1/2}\mu^\star
		\right\rangle_K = 	\left\langle
		K_{x},
		T_{D,\lambda} I_{K,X}^{\ast}\mathbf
		f
		\right\rangle_K =\{I_{K,X}T_{D,\lambda}I_{K,X}^\ast\mathbf f\}(x).
		$$
		Since $|h_i| \leq M_h(X_i)$ almost surely, the triangle inequality yields
		\[
		\begin{aligned}
			\mathbb E\{|V_n(\mathbf h)|\mid D\}
			&\leq
			\frac1n\sum_{i=1}^n
			\mathbb E\left[
			M_h(X_i)
			\left|
			\{I_{K,X}T_{D,\lambda}I_{K,X}^{\ast}\mathbf f\}(X_i)
			\right|
			\mid D\right]
			\\
			&\leq
			\|M_h\|_{L^2(P_X)}
			\left\|
			I_{K,X}T_{D,\lambda}I_{K,X}^{\ast}\mathbf f
			\right\|_{L^2(P_X)}
			\\
			&\leq
			\|M_h\|_{L^2(P_X)}
			\left\|T_{D,\lambda}^{1/2}I_{K,X}^{\ast}\right\|^2
			\|\mathbf f\|_{L^2(P_X)}
			\\
			&\lesssim \mathscr P_{D,\lambda}^2
			\|M_h\|_{L^2(P_X)}
			\|\mathbf f\|_{L^2(P_X)}.
		\end{aligned}
		\]
		The second step uses independence of $D$ and the labeled
		covariates and Cauchy--Schwarz inequality.
		The third step follows since $\|A^{\ast}A\| = \|A\|^2$ with $A  = T_{D,\lambda}^{1/2}I_{K,X}^{\ast}$. The last step uses
		\eqref{equ: empirical feature bound shift}.
		This proves \eqref{equ: conditional V adaptive shift}.
	\end{proof}
	
	\begin{lemma}
		\label{lemma: sample error lemma}
		Suppose that
		Assumptions~\ref{assumption: auxiliary covariate distribution},
		\ref{assumption: conditional variance},
		and~\ref{assumption: eigenvalues and eigenfunctions} hold,  and
		$\|\mathbf f\|_{L^\infty(P_X)}\leq B$
		for some fixed $B>0$. Recall that
		$\mu^\star=I_{K,X}^{\ast}\mathbf f$,
		$\omega_i=K_{X_i}\epsilon_i-\mu^\star$, and $T_{D,\lambda}$ in \eqref{equ: TLTDL}. 
		Define
		\[
		\begin{aligned}
			W_{n1}
			&:=
			\frac1{n(n-1)}
			\sum_{i\ne j}
			\left\langle
			T_{D,\lambda}^{1/2}\omega_i,
			T_{D,\lambda}^{1/2}\omega_j
			\right\rangle_K,\qquad 
			W_{n2}
			&:=
			\frac1n\sum_{i=1}^n
			\left\langle
			T_{D,\lambda}^{1/2}\omega_i,
			T_{D,\lambda}^{1/2}\mu^\star
			\right\rangle_K.
		\end{aligned}
		\]
		Then, 
		\begin{align}
			\mathbb E( W_{n1} ^2\mid D)
			&\lesssim
			n^{-2}\mathscr P_{D,\lambda}^4\lambda^{-s},
			\label{equ: centered pair variance alternatives shift}
			\\
			\mathbb E( W_{n2} ^2\mid D)
			&\lesssim
			n^{-1}\mathscr P_{D,\lambda}^4
			\|\mathbf f\|_{L^2(P_X)}^2,
			\label{equ: linear sample variance alternatives shift}
		\end{align} 
		where $\mathscr{P}_{D,\lambda}$ is defined in \eqref{equ: PDL}.
	\end{lemma}
	
	\begin{proof} 
		For the quadratic term, the kernel forming $W_{n1}$ is conditionally
		degenerate, that is,
		\[
		\mathbb E\left\{
		\left\langle
		T_{D,\lambda}^{1/2}\omega_i,
		T_{D,\lambda}^{1/2}\omega_j
		\right\rangle_K
		\mid D,Z_i\right\}=0,
		\qquad i\ne j.
		\]
		By Lemma~\ref{lemma: conditional Hoeffding decomposition},
		\[
		\begin{aligned}
			n(n-1)/2 \cdot\mathbb E(W_{n1}^2\mid D)
			&= 
			\mathbb E\left\{
			\left\langle
			T_{D,\lambda}^{1/2}\omega_1,
			T_{D,\lambda}^{1/2}\omega_2
			\right\rangle_K^2
			\mid D\right\} \\
			&= 
			\mathbb E\left\{
			\left\langle
			T_{D,\lambda}^{1/2} \omega_1\otimes \omega_1  T_{D,\lambda}^{1/2},
			T_{D,\lambda}^{1/2} \omega_2\otimes \omega_2 T_{D,\lambda}^{1/2}
			\right\rangle_{HS}
			\mid D\right\}  	\\
			&= 
			\operatorname{Tr}\left[
			\left\{
			T_{D,\lambda}^{1/2}
			\mathbb E(\omega\otimes\omega)
			T_{D,\lambda}^{1/2}
			\right\}^2
			\right]
			\\
			&\leq 
			\left\|
			T_{D,\lambda}^{1/2}
			\mathbb E(\omega\otimes\omega)
			T_{D,\lambda}^{1/2}
			\right\| 
			\operatorname{Tr}\left\{
			T_{D,\lambda}^{1/2}
			\mathbb E(\omega\otimes\omega)
			T_{D,\lambda}^{1/2}
			\right\}
			\\ 
			&\lesssim \mathscr P_{D,\lambda}^4\lambda^{-s}.
		\end{aligned}
		\]
		The second and third steps follow from Lemma~\ref{lemma: Balazs2005} and the independence of $D$ and the labeled observations. 
		The fourth step uses Lemma~\ref{lemma: trace inequality}.
		The final  bounds follow from
		\eqref{equ: covariance empirical shift}.
		This proves
		\eqref{equ: centered pair variance alternatives shift}.
		
		For the linear term, the independence of $D$ and labeled observations, together with Lemma~\ref{lemma: Balazs2005}, yields 
		\[
		\begin{aligned}
			n\, \mathbb E(W_{n2}^2\mid D)
			&=
			\mathbb E\left\{
			\left\langle
			T_{D,\lambda}^{1/2}\omega,
			T_{D,\lambda}^{1/2}\mu^\star
			\right\rangle_K^2
			\mid D\right\}
			\\
			&=
			\left\langle
			T_{D,\lambda}^{1/2}\mu^\star,
			\left\{
			T_{D,\lambda}^{1/2}
			\mathbb E(\omega\otimes\omega)
			T_{D,\lambda}^{1/2}
			\right\}
			T_{D,\lambda}^{1/2}\mu^\star
			\right\rangle_K
			\\
			&\leq
			\left\|
			T_{D,\lambda}^{1/2}
			\mathbb E(\omega\otimes\omega)
			T_{D,\lambda}^{1/2}
			\right\|
			\left\|T_{D,\lambda}^{1/2}I_{K,X}^{\ast}\mathbf f\right\|_K^2
			\\
			&\lesssim
			\mathscr P_{D,\lambda}^4
			\|\mathbf f\|_{L^2(P_X)}^2,
		\end{aligned}
		\]
		where  the last inequality follows from 
		\eqref{equ: covariance empirical shift},
		and~\eqref{equ: empirical feature bound shift}.
		This proves
		\eqref{equ: linear sample variance alternatives shift}.
	\end{proof} 
	
	The following proposition characterizes the three error
	components in~\eqref{equ: error decomposition}.
	\begin{prop}
		\label{prop: error decomposition}
		Suppose that
		Assumptions~\ref{assumption: auxiliary covariate distribution},
		\ref{assumption: regularities on parametric families},
		\ref{assumption: conditional variance},
		\ref{assumption: identibility and parametric estimation under H0},
		\ref{assumption: eigenvalues and eigenfunctions}, and
		\ref{assumption: uniform parametric estimation under H1}
		hold, and the conditional residual function 
		\(\mathbf f\) in \eqref{equ: conditional residual function} belongs to the source unit ball $\mathcal
		F_{r,B}$. 
		The following statements hold:
		\begin{enumerate}
			\item [(i)] The parametric error in \eqref{equ: error decomposition} satisfies
			\begin{equation}
				\label{equ: total parametric rate shift}
				\widehat\eta_\lambda(\widehat\theta)
				-\widehat\eta_\lambda(\theta^\star)
				=O_p\!\left[
				\mathbb E(\mathscr P_{D,\lambda}^2)
				\left\{
				\frac1n+\frac{\lambda^{-s}}{n^2}
				+\frac{\|\mathbf f\|_{L^2(P_X)}}{\sqrt n}
				\right\}\right].
			\end{equation}
			
			\item [(ii)] The sample error in \eqref{equ: error decomposition} satisfies
			\begin{equation}
				\label{equ: sample error rate alternatives shift}
				\widehat\eta_\lambda(\theta^\star)-\eta_\lambda^s(\theta^\star)
				=O_p\!\left[
				\mathbb E\left(\mathscr P^2_{D,\lambda} \right)
				\left\{\frac{\lambda^{-s/2}}n+
				\frac{\|\mathbf f\|_{L^2(P_X)}}{\sqrt n}\right\}\right].
			\end{equation} 
			\item[(iii)]
			The   approximation error is controlled through the following
			lower bound:
			\begin{equation}
				\label{equ: approximation error}
				\eta_\lambda^s(\theta^\star)
				\geq
				\frac{1}{2\overline c(1+\mathscr Q_{D,\lambda})}
				\left\{
				\|\mathbf f\|_{L^2(P_X)}^2-\lambda^{2r}
				\right\}.
			\end{equation}
		\end{enumerate}
		Here, $\mathscr{P}_{D,\lambda}$ is defined in \eqref{equ: PDL}, and  
		\begin{align}
			\mathscr{Q}_{D,\lambda}  := \left\| T_{\lambda}^{1/2} (L_{K,0} -L_{K,D})T_{\lambda}^{1/2}\right\|,
			\label{equ: QDL}
		\end{align}
		where   $
		T_\lambda  $ and $  
		T_{D,\lambda} $ are defined in \eqref{equ: TLTDL}. 
	\end{prop}
	
	\begin{proof}

		\emph{(i) Parametric error.}
		The parametric error admits the decomposition
		\begin{equation}
			\label{equ: parametric decomposition shift}
			\begin{aligned}
				\widehat\eta_\lambda(\widehat\theta)
				-\widehat\eta_\lambda(\theta^\star)
				={}&
				\frac{2}{n(n-1)}
				\sum_{i\ne j}
				\left\langle
				T_{D,\lambda}^{1/2}K_{X_i}
				(\widehat\epsilon_i-\epsilon_i),
				T_{D,\lambda}^{1/2}\omega_j
				\right\rangle_K
				\\
				&+
				\frac{2}{n}
				\sum_{i=1}^n
				\left\langle
				T_{D,\lambda}^{1/2}K_{X_i}
				(\widehat\epsilon_i-\epsilon_i),
				T_{D,\lambda}^{1/2}\mu^\star
				\right\rangle_K
				\\
				&+
				\frac{1}{n(n-1)}
				\sum_{i\ne j}
				\left\langle
				T_{D,\lambda}^{1/2}K_{X_i}
				(\widehat\epsilon_i-\epsilon_i),
				T_{D,\lambda}^{1/2}K_{X_j}
				(\widehat\epsilon_j-\epsilon_j)
				\right\rangle_K
				\\
				=:{}&P_1+P_2+P_3.
			\end{aligned}
		\end{equation}
		For notational convenience, we denote
		\begin{equation}
			\label{equ: JH}
			\overline J_i
			:=
			\int_0^1
			J_g\{X_i,\theta^\star+t(\widehat\theta-\theta^\star)\}\,dt,
			\qquad
			\overline H_i
			:=
			\int_0^1(1-t)
			H_g\{X_i,\theta^\star+t(\widehat\theta-\theta^\star)\}\,dt.
		\end{equation}
		Define
		\begin{align}
			G_1
			&:=
			-\frac{2}{n(n-1)}
			\sum_{i\ne j}
			\left\langle
			T_{D,\lambda}^{1/2}K_{X_i},
			T_{D,\lambda}^{1/2}\omega_j
			\right\rangle_K
			J_g(X_i,\theta^\star),
			\label{equ: G1 shift}
			\\
			G_2
			&:=
			-\frac{2}{n(n-1)}
			\sum_{i\ne j}
			\left\langle
			T_{D,\lambda}^{1/2}K_{X_i},
			T_{D,\lambda}^{1/2}\omega_j
			\right\rangle_K
			\overline H_i,
			\label{equ: G2 shift}
			\\
			G_3
			&:=
			-\frac{2}{n}
			\sum_{i=1}^n
			\left\langle
			T_{D,\lambda}^{1/2}K_{X_i},
			T_{D,\lambda}^{1/2}\mu^\star
			\right\rangle_K
			\overline J_i.
			\label{equ: G3 shift}
		\end{align}
		Here $G_1,G_3\in\mathbb R^d$ and
		$G_2\in\mathbb R^{d\times d}$.
		Define the linear operator
		$G_4:\mathbb R^d\to\mathcal H_K$ by
		\begin{equation}
			\label{equ: G4 shift}
			G_4(\mathbf a)
			:=
			\frac1n\sum_{i=1}^n
			T_{D,\lambda}^{1/2}K_{X_i}
			J_g(X_i,\theta^\star)^\top\mathbf a,
			\qquad \mathbf a\in\mathbb R^d,
		\end{equation}
		and the bilinear operator
		$G_5:\mathbb R^d\times\mathbb R^d\to\mathcal H_K$ by
		\begin{equation}
			\label{equ: G5 shift}
			G_5(\mathbf a,\mathbf b)
			:=
			\frac1n\sum_{i=1}^n
			T_{D,\lambda}^{1/2}K_{X_i}\,
			\mathbf a^\top\overline H_i\mathbf b,
			\qquad \mathbf a,\mathbf b\in\mathbb R^d.
		\end{equation}
		Finally, define  
		$G_6\in \mathbb R^{d\times d}$ by
		\begin{equation}
			\label{equ: G6 shift}
			G_6 
			:=
			\frac1n\sum_{i=1}^n
			\left\|T_{D,\lambda}^{1/2}K_{X_i}\right\|_K^2
			\overline J_i\overline J_i^\top.
		\end{equation}  
		
		Under Assumption~\ref{assumption: regularities on parametric families}, the integral forms of the first- and second-order
		Taylor expansions give
		\begin{equation}
			\label{equ: residual Taylor shift}
			\begin{aligned}
				\widehat\epsilon_i-\epsilon_i
				=
				-\overline J_i^\top(\widehat\theta-\theta^\star) =
				-J_g(X_i,\theta^\star)^\top
				(\widehat\theta-\theta^\star)
				-(\widehat\theta-\theta^\star)^\top
				\overline H_i(\widehat\theta-\theta^\star).
			\end{aligned}
		\end{equation}
		Consequently,
		\[
		\begin{aligned}
			P_1 =
			G_1^\top(\widehat\theta-\theta^\star)
			+(\widehat\theta-\theta^\star)^\top
			G_2(\widehat\theta-\theta^\star),	\qquad P_2 =
			G_3^\top(\widehat\theta-\theta^\star),
		\end{aligned}
		\]
		and hence
		\begin{equation}
			\label{equ: P1 and P2}
			\begin{aligned}
				|P_1|
				\leq
				\|G_1\|_2\|\widehat\theta-\theta^\star\|_2
				+\|G_2\| 
				\|\widehat\theta-\theta^\star\|_2^2,\qquad  |P_2|
				\leq
				\|G_3\|_2\|\widehat\theta-\theta^\star\|_2.
			\end{aligned}
		\end{equation} 
		Furthermore, we have
		\[
		\begin{aligned}
			P_3
			={}&
			\frac{n}{n-1}
			\left\|
			\frac1n\sum_{i=1}^n
			T_{D,\lambda}^{1/2}K_{X_i}
			(\widehat\epsilon_i-\epsilon_i)
			\right\|_K^2 -
			\frac{1}{n(n-1)}
			\sum_{i=1}^n
			\left\|T_{D,\lambda}^{1/2}K_{X_i}\right\|_K^2
			(\widehat\epsilon_i-\epsilon_i)^2.
		\end{aligned}
		\]
		By the second-order expansion in \eqref{equ: residual Taylor shift} and the definitions
		of $G_4$ and $G_5$,
		\[
		\begin{aligned}
			\frac1n\sum_{i=1}^n
			T_{D,\lambda}^{1/2}K_{X_i}
			(\widehat\epsilon_i-\epsilon_i)
			=
			-G_4(\widehat\theta-\theta^\star)-
			G_5(\widehat\theta-\theta^\star,
			\widehat\theta-\theta^\star).
		\end{aligned}
		\]
		By the first-order expansion in \eqref{equ: residual Taylor shift} and the definition
		of $G_6$,
		\[
		\begin{aligned}
			\frac1n\sum_{i=1}^n
			\left\|T_{D,\lambda}^{1/2}K_{X_i}\right\|_K^2
			(\widehat\epsilon_i-\epsilon_i)^2 =
			\frac1n\sum_{i=1}^n
			\left\|T_{D,\lambda}^{1/2}K_{X_i}\right\|_K^2
			\left\{
			\overline J_i^\top(\widehat\theta-\theta^\star)
			\right\}^2 =
			(\widehat\theta-\theta^\star)^\top
			G_6(\widehat\theta-\theta^\star).
		\end{aligned}
		\]
		Therefore,
		\[
		\begin{aligned}
			P_3 =
			\frac{n}{n-1}
			\left\|
			G_4(\widehat\theta-\theta^\star)
			+
			G_5(\widehat\theta-\theta^\star,
			\widehat\theta-\theta^\star)
			\right\|_K^2 -
			\frac{1}{n-1}
			(\widehat\theta-\theta^\star)^\top
			G_6(\widehat\theta-\theta^\star),
		\end{aligned}
		\]
		and hence
		\begin{equation}
			\label{equ: P3}
			\begin{aligned}
				|P_3| \leq 
				\frac{2n}{n-1}
				\|G_4\|^2
				\|\widehat\theta-\theta^\star\|_2^2 +
				\frac{2n}{n-1}
				\|G_5\|^2
				\|\widehat\theta-\theta^\star\|_2^4 &+
				\frac{1}{n-1}
				\|G_6\| 
				\|\widehat\theta-\theta^\star\|_2^2.
			\end{aligned}
		\end{equation} 
		
		We next bound $G_1,\ldots,G_6$.  By Assumption~\ref{assumption: regularities on parametric families},
		\begin{equation}
			\label{equ: JH bound}
			\begin{aligned}
				\|\overline H_i\| 
				&\leq
				\int_0^1(1-t)
				\left\|
				H_g\{X_i,\theta^\star+t(\widehat\theta-\theta^\star)\}
				\right\| \,dt
				\leq \frac12M_H(X_i),
				\\
				\|\overline J_i\|_2
				&\leq
				\int_0^1
				\left\|
				J_g\{X_i,\theta^\star+t(\widehat\theta-\theta^\star)\}
				\right\|_2\,dt
				\leq M_J(X_i).
			\end{aligned}
		\end{equation}
		Moreover,
		\begin{equation}
			\label{equ: Jg-coordinate-bound}
			\sum_{\ell=1}^d
			\|J_{g,\ell}(\cdot,\theta^\star)\|_{L^2(P_X)}^2
			=
			\mathbb E\|J_g(X,\theta^\star)\|_2^2
			\leq \|M_J\|_{L^2(P_X)}^2.
		\end{equation}
		
		For  $G_1$, apply
		\eqref{equ: conditional U fixed shift}
		with $h=J_{g,\ell}(\cdot,\theta^\star)$, the $\ell$-th coordinate of $J_{g}(\cdot, \theta^{\star})$.
		Summing the resulting conditional second-moment bounds gives
		\[
		\begin{aligned}
			\mathbb E(\|G_1\|_2^2\mid D)
			&\lesssim
			\mathscr P_{D,\lambda}^4
			\left(n^{-1}+n^{-2}\lambda^{-s}\right)
			\sum_{\ell=1}^d
			\|J_{g,\ell}(\cdot,\theta^\star)\|_{L^2(P_X)}^2
			\\
			&\leq \mathscr P_{D,\lambda}^4
			\|M_J\|_{L^2(P_X)}^2
			\left(n^{-1}+n^{-2}\lambda^{-s}\right).
		\end{aligned}
		\]

		For $G_2$, its $(a,b)$-th entry  is $-2U_n(\mathbf h)$ with
		$h_i=(\overline H_i)_{ab}$, and   $|h_i| \leq M_H(X_i)/2$ by \eqref{equ: JH bound}.
		Thus \eqref{equ: conditional U adaptive shift} implies
		\[
		\begin{aligned}
			\mathbb E(\|G_2\| \mid D)
			&\leq
			\sum_{a,b=1}^d
			\mathbb E\{|(G_2)_{ab}|\mid D\} \lesssim
			\mathscr P_{D,\lambda}^2
			\|M_H\|_{L^1(P_X)}
			n^{-1/2}\lambda^{-s/2}.
		\end{aligned}
		\]
		
		For $G_3$, the $\ell$-th  coordinate  is $-2V_n(\mathbf h)$
		with $h_i=(\overline J_i)_\ell$,  and  $|h_i| \leq M_J(X_i)$ by~\eqref{equ: JH bound}. Hence~\eqref{equ: conditional V adaptive shift} gives
		\[
		\begin{aligned}
			\mathbb E(\|G_3\|_2\mid D)
			&\leq
			\sum_{\ell=1}^d
			\mathbb E\{|(G_3)_\ell|\mid D\} \lesssim
			\mathscr P_{D,\lambda}^2
			\|M_J\|_{L^2(P_X)}
			\|\mathbf f\|_{L^2(P_X)}.
		\end{aligned}
		\]
		
		For $G_4$, Cauchy--Schwarz inequality gives that, for every
		$\mathbf a\in\mathbb R^d$ with $\|\mathbf a\|_2\leq1$,
		\[
		\begin{aligned}
			\|G_4\mathbf a\|_K^2
			&=
			\left\|
			\sum_{\ell=1}^d a_\ell
			\left\{
			\frac1n\sum_{i=1}^n
			T_{D,\lambda}^{1/2}K_{X_i}
			J_{g,\ell}(X_i,\theta^\star)
			\right\}
			\right\|_K^2 \leq
			\sum_{\ell=1}^d
			\left\|
			\frac1n\sum_{i=1}^n
			T_{D,\lambda}^{1/2}K_{X_i}
			J_{g,\ell}(X_i,\theta^\star)
			\right\|_K^2.
		\end{aligned}
		\]
		Taking the supremum over $\mathbf a$ and expanding
		the conditional second moments yields
		\[
		\begin{aligned}
			\mathbb E(\|G_4\|^2\mid D)
			&\leq
			\sum_{\ell=1}^d
			\mathbb E\left[
			\left\|
			\frac1n\sum_{i=1}^n
			T_{D,\lambda}^{1/2}K_{X_i}
			J_{g,\ell}(X_i,\theta^\star)
			\right\|_K^2
			\Bigm|D\right]
			\\
			&=
			\frac1n
			\mathbb E\left\{
			\left\|T_{D,\lambda}^{1/2}K_X\right\|_K^2
			\|J_g(X,\theta^\star)\|_2^2
			\mid D\right\} +
			\frac{n-1}{n}
			\sum_{\ell=1}^d
			\left\|
			T_{D,\lambda}^{1/2}I_{K,X}^{\ast}
			J_{g,\ell}(\cdot,\theta^\star)
			\right\|_K^2
			\\
			&\lesssim
			\mathscr P_{D,\lambda}^2
			\|M_J\|_{L^2(P_X)}^2
			\left(1+n^{-1}\lambda^{-s}\right).
		\end{aligned}
		\]
		The equality follows from  independence of
		the labeled observations.
		The final inequality uses the independence of $D$ and the labeled observations, and bounds in
		\eqref{equ: empirical feature bound shift} and \eqref{equ: Jg-coordinate-bound}.
		
		For  $G_5$,
		the triangle inequality gives
		\[
		\begin{aligned}
			\|G_5\| 
			&\leq
			\frac1n\sum_{i=1}^n
			\left\|T_{D,\lambda}^{1/2}K_{X_i}\right\|_K
			\|\overline H_i\|  \leq
			\frac1{2n}\sum_{i=1}^n
			\left\|T_{D,\lambda}^{1/2}K_{X_i}\right\|_K
			M_H(X_i)
			\\
			&\lesssim
			\mathscr P_{D,\lambda}\lambda^{-s/2}\cdot
			\frac1n\sum_{i=1}^n M_H(X_i),
		\end{aligned}
		\]
		where the second step follows from~\eqref{equ: JH bound}, and the last step follows from
		\eqref{equ: empirical feature bound shift}.
		Since $D$ is independent of the labeled observations,
		\[
		\mathbb E(\|G_5\| \mid D)
		\lesssim
		\mathscr P_{D,\lambda}\lambda^{-s/2}
		\|M_H\|_{L^1(P_X)}.
		\]
		
		For $G_6$, again by \eqref{equ: empirical feature bound shift} and the envelope bound
		for $\|\overline J_i\|_2 \leq M_J(X_i)$ in \eqref{equ: JH bound}, 
		\[
		\begin{aligned}
			\|G_6\|  \leq
			\frac1n\sum_{i=1}^n
			\left\|T_{D,\lambda}^{1/2}K_{X_i}\right\|_K^2
			\|\overline J_i\|_2^2 \lesssim
			\mathscr P_{D,\lambda}^2\lambda^{-s}\cdot
			\frac1n\sum_{i=1}^n M_J(X_i)^2.
		\end{aligned}
		\] 
		Taking conditional expectations, and using the independence of $D$ and the labeled observations, gives
		\[
		\mathbb E(\|G_6\| \mid D)
		\lesssim
		\mathscr P_{D,\lambda}^2\lambda^{-s}
		\|M_J\|_{L^2(P_X)}^2.
		\]
		
		For $G_1$ and $G_4$, conditional Cauchy--Schwarz inequality,
		followed by integration over $D$, yields
		\[
		\begin{aligned}
			\mathbb E\|G_1\|_2
			&\leq
			\mathbb E\left[
			\{\mathbb E(\|G_1\|_2^2\mid D)\}^{1/2}
			\right] \lesssim
			\mathbb E(\mathscr P_{D,\lambda}^2)
			\|M_J\|_{L^2(P_X)}
			\left(n^{-1/2}+n^{-1}\lambda^{-s/2}\right),
			\\
			\mathbb E\|G_4\| 
			&\leq
			\mathbb E\left[
			\{\mathbb E(\|G_4\| ^2\mid D)\}^{1/2}
			\right] \lesssim
			\mathbb E(\mathscr P_{D,\lambda})
			\|M_J\|_{L^2(P_X)}
			\left(1+n^{-1/2}\lambda^{-s/2}\right).
		\end{aligned}
		\]
		Integrating the conditional first-moment bounds for
		$G_2,G_3,G_5,G_6$ and applying Markov's inequality
		therefore gives
		\begin{align}
			\|G_1\|_2
			&=
			O_p\!\left[
			\mathbb E(\mathscr P_{D,\lambda}^2)
			\left(n^{-1/2}+n^{-1}\lambda^{-s/2}\right)
			\right],
			\label{equ: G1 rate shift}
			\\
			\|G_2\| 
			&=
			O_p\!\left[
			\mathbb E(\mathscr P_{D,\lambda}^2)
			n^{-1/2}\lambda^{-s/2}
			\right],
			\label{equ: G2 rate shift}
			\\
			\|G_3\|_2
			&=
			O_p\!\left[
			\mathbb E(\mathscr P_{D,\lambda}^2)
			\|\mathbf f\|_{L^2(P_X)}
			\right],
			\label{equ: G3 rate shift}
			\\
			\|G_4\| 
			&=
			O_p\!\left[
			\mathbb E(\mathscr P_{D,\lambda})
			\left(1+n^{-1/2}\lambda^{-s/2}\right)
			\right],
			\label{equ: G4 rate shift}
			\\
			\|G_5\| 
			&=
			O_p\!\left[
			\mathbb E(\mathscr P_{D,\lambda})
			\lambda^{-s/2}
			\right],
			\label{equ: G5 rate shift}
			\\
			\|G_6\| 
			&=
			O_p\!\left[
			\mathbb E(\mathscr P_{D,\lambda}^2)
			\lambda^{-s}
			\right].
			\label{equ: G6 rate shift}
		\end{align}
		
		We now use the parameter-estimation bound in
		Assumption~\ref{assumption: identibility and parametric estimation under H0}
		under the null, or
		Assumption~\ref{assumption: uniform parametric estimation under H1}
		under alternatives.
		Combining \eqref{equ: P1 and P2} with
		\eqref{equ: G1 rate shift}, \eqref{equ: G2 rate shift},
		\eqref{equ: G3 rate shift} gives
		\begin{align}
			P_1
			&=
			O_p\!\left[
			\mathbb E(\mathscr P_{D,\lambda}^2)
			\left(n^{-1}+n^{-3/2}\lambda^{-s/2}\right)
			\right],
			\label{equ: P1 rate shift}
			\\
			P_2
			&=
			O_p\!\left[
			\mathbb E(\mathscr P_{D,\lambda}^2)
			n^{-1/2}\|\mathbf f\|_{L^2(P_X)}
			\right].
			\label{equ: P2 rate shift}
		\end{align}
		For $P_3$, using the inequality
		$\{\mathbb E(\mathscr P_{D,\lambda})\}^2
		\leq\mathbb E(\mathscr P_{D,\lambda}^2)$, we get from~\eqref{equ: G4 rate shift}, \eqref{equ: G5 rate shift},
		\eqref{equ: G6 rate shift}  that
		\begin{align*}
			&\frac{2n}{n-1}
			\|G_4\| ^2
			\|\widehat\theta-\theta^\star\|_2^2 =
			O_p\!\left[
			\mathbb E(\mathscr P_{D,\lambda}^2)
			\left(n^{-1}+n^{-2}\lambda^{-s}\right)
			\right], 
			\\
			&\frac{2n}{n-1}
			\|G_5\| ^2
			\|\widehat\theta-\theta^\star\|_2^4 =
			O_p\!\left[
			\mathbb E(\mathscr P_{D,\lambda}^2)
			n^{-2}\lambda^{-s}
			\right], 
			\\
			&\frac1{n-1}
			\|G_6\| 
			\|\widehat\theta-\theta^\star\|_2^2 =
			O_p\!\left[
			\mathbb E(\mathscr P_{D,\lambda}^2)
			n^{-2}\lambda^{-s}
			\right]. 
		\end{align*}
		Substituting these bounds into \eqref{equ: P3} yields
		\begin{equation}
			\label{equ: P3 rate shift}
			P_3
			=
			O_p\!\left[
			\mathbb E(\mathscr P_{D,\lambda}^2)
			\left(n^{-1}+n^{-2}\lambda^{-s}\right)
			\right].
		\end{equation}
		
		Finally, combining
		\eqref{equ: P1 rate shift},
		\eqref{equ: P2 rate shift},
		and~\eqref{equ: P3 rate shift},
		and using
		\[
		n^{-3/2}\lambda^{-s/2}
		\leq
		\frac12\left(n^{-1}+n^{-2}\lambda^{-s}\right),
		\]
		we obtain
		\[
		\widehat\eta_\lambda(\widehat\theta)
		-\widehat\eta_\lambda(\theta^\star)
		=
		O_p\!\left[
		\mathbb E(\mathscr P_{D,\lambda}^2)
		\left\{
		n^{-1}+n^{-2}\lambda^{-s}
		+n^{-1/2}\|\mathbf f\|_{L^2(P_X)}
		\right\}
		\right],
		\]
		which proves \eqref{equ: total parametric rate shift}.
		
		\medskip
		\noindent
		
		\emph{(ii) Sample error.}   
		The sample error admits the decomposition
		\[
		\begin{aligned}
			\widehat{\eta}_{\lambda}({\theta}^{\star})-{\eta}_{\lambda}^{s}({\theta}^{\star}) 
			&= \frac{1}{n(n-1)}\sum_{i\neq j}\left\langle  T_{D,\lambda}^{1/2}K_{X_i}\epsilon_i, T_{D,\lambda}^{1/2}K_{X_j}\epsilon_j \right\rangle_K-\left\langle T_{D,\lambda}^{1/2}I_{K,X}^{\ast}\mathbf f, T_{D,\lambda}^{1/2}I_{K,X}^{\ast}\mathbf f\right\rangle_{K}\\
			&=\frac{1}{n(n-1)}\sum_{i\neq j}\left\langle T_{D,\lambda}^{1/2}\omega_i, T_{D,\lambda}^{1/2}\omega_j\right\rangle_K+\frac{2}{n}\sum_{i=1}^n \left\langle T_{D,\lambda}^{1/2}\omega_i, T_{D,\lambda}^{1/2}I_{K,X}^{\ast}\mathbf f \right\rangle_K \\
			& =W_{n1}+2W_{n2}.
		\end{aligned}
		\]
		By 
		\eqref{equ: centered pair variance alternatives shift} and 
		\eqref{equ: linear sample variance alternatives shift}, the conditional Cauchy--Schwarz inequality   gives
		\[
		\begin{aligned}
			\mathbb E\left\{
			\left|
			\widehat\eta_\lambda(\theta^\star)
			-\eta_\lambda^s(\theta^\star)
			\right|
			\mid D\right\}
			&\leq
			\{\mathbb E(W_{n1}^2\mid D)\}^{1/2}
			+2\{\mathbb E(W_{n2}^2\mid D)\}^{1/2} \lesssim
			\mathscr P_{D,\lambda}^2
			\left\{
			\frac{\lambda^{-s/2}}n
			+\frac{\|\mathbf f\|_{L^2(P_X)}}{\sqrt n}
			\right\}.
		\end{aligned}
		\]
		Integrating over $D$ yields
		\[
		\mathbb E\left|
		\widehat\eta_\lambda(\theta^\star)
		-\eta_\lambda^s(\theta^\star)
		\right|
		\lesssim
		\mathbb E(\mathscr P_{D,\lambda}^2)
		\left\{
		\frac{\lambda^{-s/2}}n
		+\frac{\|\mathbf f\|_{L^2(P_X)}}{\sqrt n}
		\right\}.
		\]
		Markov's inequality now proves
		\eqref{equ: sample error rate alternatives shift}. 
		
		\medskip
		\noindent
		
		\emph{(iii) Approximation error.}	By the definition of $\mathscr{Q}_{D,\lambda}$ in \eqref{equ: QDL}, we have
		\[
		-\mathscr Q_{D,\lambda}I
		\preceq T_\lambda^{1/2}
		\left(L_{K,0}-L_{K,D}\right)T_\lambda^{1/2}
		\preceq\mathscr Q_{D,\lambda}I.
		\]
		Hence,
		\[
		\begin{aligned}
			T_\lambda^{1/2}(L_{K,D}+\lambda I)T_\lambda^{1/2}
			&=T_\lambda^{1/2}(L_{K,0}+\lambda I)T_\lambda^{1/2}
			-T_\lambda^{1/2}(L_{K,0}-L_{K,D})T_\lambda^{1/2}\\
			&=I-T_\lambda^{1/2}(L_{K,0}-L_{K,D})T_\lambda^{1/2}\\
			&\preceq(1+\mathscr Q_{D,\lambda})I.
		\end{aligned}
		\]
		Inverting this inequality
		and multiplying on both sides by \(T_\lambda^{1/2}\) gives
		$$
		T_{D,\lambda}  =T_{\lambda}^{1/2} T_{\lambda}^{-1/2} T_{D,\lambda}T_{\lambda}^{-1/2}T_{\lambda}^{1/2}\succeq
		(1+\mathscr Q_{D,\lambda})^{-1}T_\lambda.
		$$
		This  implies 
		\begin{equation}
			\label{equ: empirical population signal comparison shift}
			\eta_\lambda^s(\theta^\star)=	\left\|T_{D,\lambda}^{1/2}\mu^\star\right\|_K^2
			= \left\langle T_{D,\lambda} \mu^\star, \mu^\star \right\rangle_K\geq  \frac{1}{1+\mathscr Q_{D,\lambda}} \left\langle T_{\lambda} \mu^\star, \mu^\star \right\rangle_K=
			\frac{1}{1+\mathscr Q_{D,\lambda}}
			\left\|T_\lambda^{1/2}\mu^\star\right\|_K^2.
		\end{equation}
		The operator comparison in \eqref{equ: two operators are comparable}
		and \(\overline c\geq1\) imply that $ L_{K,0}+\lambda I
		\preceq\overline c(L_{K,X}+\lambda I)$, and thus
		$ T_\lambda\succeq
		\overline c^{-1}(L_{K,X}+\lambda I)^{-1}$. 
		Since $\mu^{\star} = I_{K,X}^{\ast}\mathbf{f}$, we have
		\[
		\begin{aligned}
			\left\|T_\lambda^{1/2}\mu^\star\right\|_K^2& =	\langle I_{K,X}^\ast\mathbf f,
			T_{\lambda}I_{K,X}^\ast\mathbf f\rangle_K 
			\geq\overline c^{-1}
			\langle I_{K,X}^\ast\mathbf f,
			(L_{K,X}+\lambda I)^{-1}I_{K,X}^\ast\mathbf f\rangle_K\\
			&=\overline c^{-1}
			\langle \mathbf f,  I_{K,X}
			(L_{K,X}+\lambda I)^{-1}I_{K,X}^\ast\mathbf f\rangle_{L^2(P_X)} \\
			&= \overline c^{-1}
			\langle \mathbf f,   
			(\mathcal L_{K,X}+\lambda I)^{-1}\mathcal L_{K,X} \mathbf f\rangle_{L^2(P_X)}\\
			&=\overline c^{-1}\sum_{k\geq1}
			\frac{\sigma_{X,k}}{\sigma_{X,k}+\lambda}
			\langle\mathbf f,\phi_{X,k}\rangle_{L^2(P_X)}^2\\
			&\geq\frac{1}{2\overline c}
			\sum_{\sigma_{X,k}\geq\lambda}
			\langle\mathbf f,\phi_{X,k}\rangle_{L^2(P_X)}^2.
		\end{aligned}
		\]
		The third step applies  the functional-calculus identity~\citep[Equation~(2.43)]{engl1996regularization}  
	\begin{equation}
		\label{equ: functional calculus identity}
			h(AA^\ast)A=Ah(A^\ast A)
	\end{equation}
		with $A=I_{K,X}^\ast$ and $h(t)=(t+\lambda)^{-1}$.
		The last step follows  since $x/(x+ \lambda) \geq 1/2$ if $x \geq \lambda$.
		Since $\mathbf{f}$ belongs to $\mathcal{F}_{r,B}$, we write \(\mathbf f=\mathcal L_{K,X}^{r}u\) for some $u \in L^2(P_X)$ and  $\|u\|_{L^2(P_X)} \leq 1$. Then
		\[
		\begin{aligned}
			\sum_{\sigma_{X,k}<\lambda}
			\langle\mathbf f,\phi_{X,k}\rangle_{L^2(P_X)}^2
			&=\sum_{\sigma_{X,k}<\lambda}
			\sigma_{X,k}^{2r}
			\langle u,\phi_{X,k}\rangle_{L^2(P_X)}^2\leq\lambda^{2r}\|u\|_{L^2(P_X)}^2
			\leq\lambda^{2r}.
		\end{aligned}
		\]
		Combining these two bounds, Parseval's identity gives
		\begin{equation}
			\left\|T_\lambda^{1/2}\mu^\star\right\|_K^2
			\geq\frac{1}{2\overline c}
			\left\{\|\mathbf f\|_{L^2(P_X)}^2-\lambda^{2r}\right\},
			\label{equ: population signal lower shift}
		\end{equation}
		Combining \eqref{equ: empirical population signal comparison shift}
		and~\eqref{equ: population signal lower shift}
		proves \eqref{equ: approximation error}. The proof is complete.
	\end{proof}
	The bounds in Proposition~\ref{prop: error decomposition}
	depend on $\mathscr P_{D,\lambda}$ and
	$\mathscr Q_{D,\lambda}$, defined in \eqref{equ: PDL} and \eqref{equ: QDL},
	which compare the operator $L_{K,0}$ and its empirical version $L_{K,D}$
	after regularization. In our main proofs, we need to  further  express these bounds in terms of the sample sizes and the regularization parameters. Hence,
	we conclude this section with the following lemma to control these similarity quantities. The proof is based on Hilbert-valued concentration  inequalities, similar
	to those used in \citet{ hagrass-two-sample-test,lin2020distributed, rudirosasco2017}.
	For completeness, we defer its   proof to Appendix~\ref{sec: addtional lemmas and proofs}.
	
	\begin{lemma}[Similarity moments]
		\label{lemma: similarity expectation shift}
		Suppose that
		Assumptions~\ref{assumption: auxiliary covariate distribution}
		and~\ref{assumption: eigenvalues and eigenfunctions} hold.
		If $\lambda\to0$ and
		$N\lambda^s/\log(\lambda^{-1})\to\infty$, then,
		for every fixed $u\in\mathbb Z^{+}$,
		\[
		\mathbb E(\mathscr P_{D,\lambda}^{u})=O(1),
		\qquad
		\mathbb E(\mathscr Q_{D,\lambda}^{u})
		=
		O\!\left[
		\left\{
		\frac{\log(1/\lambda)}{N\lambda^s}
		\right\}^{u/2}
		\right].
		\]
		The implicit constants may depend on $u$ but are
		independent of $N$ and $\lambda$.
	\end{lemma}

	\section{Proof of the Main Results}
	\label{sec: proofs}
	Recall that $T_\lambda =(L_{K,0}+\lambda I)^{-1}$ and $
	T_{D,\lambda} =(L_{K,D}+\lambda I)^{-1}.$ 
	Accordingly, we use the measure-specific kernel notation
	\begin{equation}
		\label{equ: some kernel notation 1}
		A_{\lambda,0}(X_1,X_2)
		=\left\langle T_\lambda^{1/2}K_{X_1},T_\lambda^{1/2}K_{X_2}\right\rangle_K,
		\qquad
		A_{\lambda,D}(X_1,X_2)
		=\left\langle T_{D,\lambda}^{1/2}K_{X_1},
		T_{D,\lambda}^{1/2}K_{X_2}\right\rangle_K.
	\end{equation}
	Define
	\begin{equation}
		\label{equ: some kernel notation 2}
		H_{\lambda,0}(Z_1,Z_2)
		=A_{\lambda,0}(X_1,X_2)\epsilon_1\epsilon_2,\qquad 
		R_{\lambda,0}(Z_1,Z_2)
		=\mathbb E\{H_{\lambda,0}(Z,Z_1)H_{\lambda,0}(Z,Z_2)\mid Z_1,Z_2\},
	\end{equation}
	where \(Z, Z_1, Z_2\) are independent labeled observations.
	\subsection{Proof of Theorem~1}
	The proof of Theorem~1    can be divided into two primary parts.\\
	
	\underline{\textit{Part 1. Asymptotic normality.}} 	Consider the oracle statistic
	\begin{equation}
		\label{equ: eta_lambda theta_0}
		\widehat{\eta}^{\mathrm{or}}_\lambda(\theta_0)
		:=
		\frac1{n(n-1)}
		\sum_{i\ne j}
		A_{\lambda,0}(X_i,X_j)\epsilon_i\epsilon_j
		=
		\frac1{n(n-1)}
		\sum_{i\ne j}
		H_{\lambda,0}(Z_i,Z_j).
	\end{equation}
	We decompose the empirical statistic as follows:
	\begin{equation}
		\label{equ: decomposition under H0}
		\begin{aligned}
			\widehat\eta_\lambda(\widehat\theta)
			={}&
			\widehat{\eta}^{\mathrm{or}}_\lambda(\theta_0)
			+
			\left\{
			\widehat\eta_\lambda(\widehat\theta)
			-\widehat\eta_\lambda(\theta_0)
			\right\} +
			\left\{
			\widehat\eta_\lambda(\theta_0)
			-\widehat{\eta}^{\mathrm{or}}_\lambda(\theta_0)
			\right\}.
		\end{aligned}
	\end{equation}
	
	We first establish the asymptotic normality of
	$\widehat{\eta}^{\mathrm{or}}_\lambda(\theta_0)$.
	The kernel $H_{\lambda,0}$ is symmetric, and under $H_0$,
	\[
	\begin{aligned}
		\mathbb E\{H_{\lambda,0}(Z_1,Z_2)\mid Z_1\} =
		\left\langle
		T_\lambda^{1/2}K_{X_1}\epsilon_1,
		T_\lambda^{1/2}\mathbb E(K_{X_2}\epsilon_2)
		\right\rangle_K =0.
	\end{aligned}
	\]
	Thus $\widehat{\eta}^{\mathrm{or}}_\lambda(\theta_0)$ is a degenerate
	order-two $U$-statistic.
	Following the martingale central limit argument of
	\citet{HALL19841},  if $\mathbb E\{H_{\lambda,0}^2(Z_1,Z_2)\}<\infty$ for each fixed $\lambda$, and as $n \to \infty$,
	\begin{equation}
		\label{equ: key condition}
		\frac{  \mathbb E\left\{R_{\lambda,0}^2(Z_1,Z_2)\right\} + n^{-1}\mathbb E \left\{H_{\lambda,0}^2(Z_1,Z_2)H_{\lambda,0}^2(Z_1,Z_3)\right\} + n^{-2}\mathbb E\left\{H_{\lambda,0}^4(Z_1,Z_2)\right\} }{\left[\mathbb E\left\{H_{\lambda,0}^2(Z_1,Z_2)\right\}\right]^2} \to 0,
	\end{equation}
	then 
	\begin{equation}
		\label{equ: asymptotic normality of the oracle statistic}
		\left[
		\frac{n(n-1)}{2\mathbb E\{H_{\lambda,0}^2(Z_1,Z_2)\}}
		\right]^{1/2}\widehat{\eta}^{\mathrm{or}}_\lambda(\theta_0)
		\leadsto N(0,1).
	\end{equation}
	It suffices to verify condition (\ref{equ: key condition}). By \eqref{equ: EH squared shift},
	\eqref{equ: ER squared shift},
	\eqref{equ: shared fourth shift},
	and~\eqref{equ: full fourth shift}, 
	$$
	\frac{\mathbb E\left\{R_{\lambda,0}^2(Z_1,Z_2)\right\}}{\left[\mathbb E\left\{H_{\lambda,0}^2(Z_1,Z_2)\right\}\right]^2} =O (\lambda^s),
	\quad
	\frac{\mathbb E \left\{H_{\lambda,0}^2(Z_1,Z_2)H_{\lambda,0}^2(Z_1,Z_3)\right\}}{n\left[\mathbb E\left\{H_{\lambda,0}^2(Z_1,Z_2)\right\}\right]^2} = O(n^{-1}),
	$$
	and 
	$$
	\frac{\mathbb E\left\{H_{\lambda,0}^4(Z_1,Z_2)\right\} }{n^2\left[\mathbb E\left\{H_{\lambda,0}^2(Z_1,Z_2)\right\}\right]^2} = O(n^{-2}\lambda^{-s}).
	$$
	Hence, if \(\lambda\to0\), \(n\to\infty\), and
	\(n\lambda^{s/2}\to\infty\), the condition~\eqref{equ: key condition} holds. Hence \eqref{equ: asymptotic normality of the oracle statistic} follows.

	We next show that the remaining terms in
	\eqref{equ: decomposition under H0}
	are negligible on the standard-deviation scale
	of the oracle statistic.
	By Lemma~\ref{lemma: similarity expectation shift},
	the conditions $\lambda\to0$ and
	$N\lambda^s/\log(\lambda^{-1})\to\infty$ imply
	$\mathbb E(\mathscr P_{D,\lambda}^2)=O(1)$.
	Since $\mathbf f=0$ under $H_0$,
	Proposition~\ref{prop: error decomposition} gives
	\begin{equation}
		\label{equ: parametric error under H0}
		\widehat\eta_\lambda(\widehat\theta)
		-\widehat\eta_\lambda(\theta_0)
		=
		O_p\!\left(n^{-1}+n^{-2}\lambda^{-s}\right)
		=
		o_p\!\left(n^{-1}\lambda^{-s/2}\right).
	\end{equation}
	The last equality follows because
	$\lambda\to0$ and
	$ n\lambda^{s/2} \to \infty$. The last term in \eqref{equ: decomposition under H0} satisfies
	\[
	\begin{aligned}
		\widehat\eta_\lambda(\theta_0)-\widehat{\eta}^{\mathrm{or}}_\lambda(\theta_0) =
		\frac1{n(n-1)}
		\sum_{i\ne j}
		\left\{
		A_{\lambda,D}(X_i,X_j)
		-A_{\lambda,0}(X_i,X_j)
		\right\}
		\epsilon_i\epsilon_j.
	\end{aligned}
	\]
	Conditional on $D$, this is a degenerate order-two
	$U$-statistic because
	$\mathbb E(\epsilon_i\mid X_i)=0$. By Lemma~\ref{lemma: conditional Hoeffding decomposition}-\eqref{equ: conditional variance projection},
	its conditional variance therefore satisfies
	\[
	\begin{aligned}
		&\quad\,\,  {n(n-1)}/2\cdot
		\mathbb E\left[
		\left\{
		\widehat\eta_\lambda(\theta_0)-\widehat{\eta}^{\mathrm{or}}_\lambda(\theta_0)
		\right\}^2
		\mid D\right]
		\\
		& =
		\mathbb E\left[
		\left\{
		A_{\lambda,D}(X_1,X_2)
		-A_{\lambda,0}(X_1,X_2)
		\right\}^2
		s_2(X_1,\theta_0)s_2(X_2,\theta_0)
		\mid D\right]
		\\[1mm]
		& \leq
		C_2^2\,
		\mathbb E\left[
		\left\{
		A_{\lambda,D}(X_1,X_2)
		-A_{\lambda,0}(X_1,X_2)
		\right\}^2
		\mid D\right]
		\\[1mm]
		& \lesssim
		\lambda^{-s}
		\mathscr P_{D,\lambda}^4
		\mathscr Q_{D,\lambda}^2,
	\end{aligned}
	\] 
	where the first step follows using iterated expectation formula, the second step follows from Lemma~\ref{lemma: residual moments at pseudo true}, and the last step follows from 
	\eqref{equ: merged kernel difference bound}. Integrating over $D$ and applying Cauchy--Schwarz gives
	\[
	\begin{aligned}
		\mathbb E\left[
		\left\{
		\widehat\eta_\lambda(\theta_0)-\widehat{\eta}^{\mathrm{or}}_\lambda(\theta_0)
		\right\}^2
		\right] &\lesssim
		n^{-2}\lambda^{-s}
		\left\{\mathbb E(\mathscr P_{D,\lambda}^8)\right\}^{1/2}
		\left\{\mathbb E(\mathscr Q_{D,\lambda}^4)\right\}^{1/2} \\
		& =
		O\!\left[
		n^{-2}\lambda^{-s}\cdot
		\frac{\log(1/\lambda)}{N\lambda^s}
		\right]
		\\
		& =
		o\!\left(n^{-2}\lambda^{-s}\right),
	\end{aligned}
	\]
	where the final two steps follow from
	Lemma~\ref{lemma: similarity expectation shift}
	and $N\lambda^s/\log(\lambda^{-1})\to\infty$.
	Chebyshev's inequality consequently yields
	\begin{equation}
		\label{equ: operator difference under H0}
		\widehat\eta_\lambda(\theta_0)-\widehat{\eta}^{\mathrm{or}}_\lambda(\theta_0)
		=
		o_p\!\left(n^{-1}\lambda^{-s/2}\right).
	\end{equation}
	
	Finally, the exact null standard deviation of
	$\widehat{\eta}^{\mathrm{or}}_\lambda(\theta_0)$ is
	\[
	\left[
	\frac{2\mathbb E\{H_{\lambda,0}^2(Z_1,Z_2)\}}
	{n(n-1)}
	\right]^{1/2}
	\asymp n^{-1}\lambda^{-s/2},
	\]
	which follows from
	\eqref{equ: EH squared shift}.
	Combining
	\eqref{equ: decomposition under H0},
	\eqref{equ: asymptotic normality of the oracle statistic},
	\eqref{equ: parametric error under H0},
	and~\eqref{equ: operator difference under H0},
	Slutsky's lemma gives
	\[
	\left[
	\frac{n(n-1)}
	{2\mathbb E\{H_{\lambda,0}^2(Z_1,Z_2)\}}
	\right]^{1/2}
	\widehat\eta_\lambda(\widehat\theta)
	\leadsto N(0,1).
	\]\\

	\underline{\textit{Part 2. Variance estimation.}} Recall that the  empirical estimator of $\mathbb{E}\{H_{\lambda,0}^2(Z_1,Z_2)\}$ is given as
	\begin{equation}
		\label{equ: hatvn2-rewrite}
		\widehat{v}_{n,N}^2
		=
		\frac{1}{n(n-1)}\sum_{i\neq j} A_{\lambda,D}^2(X_i,X_j)\widehat{\epsilon}_i^2\widehat{\epsilon}_j^2.
	\end{equation}
	Define
	\begin{equation}
		\label{equ: hatvn2}
		\widehat{v}_{1}^2
		:=	\frac{1}{n(n-1)}\sum_{i\neq j} 
		A_{\lambda,D}^2(X_i,X_j) \epsilon_i^2 \epsilon_j^2, 	
	\end{equation}
	and
	\begin{equation}
		\label{equ: vn2}
		\widehat{v}_{2}^2
		:=	\frac{1}{n(n-1)}\sum_{i\neq j} 
		A_{\lambda,0}^2(X_i,X_j) \epsilon_i^2\epsilon_j^2  	=	\frac{1}{n(n-1)}\sum_{i\neq j}H_{\lambda,0}^2(Z_i,Z_j).
	\end{equation}
	
	Then we can write
	$$
	\begin{aligned}
		\frac{\widehat{v}_{n,N}^2}{\mathbb E\{H_{\lambda,0}^2(Z_1,Z_2)\}} - 1&= \frac{\widehat{v}_{n,N}^2 -\widehat{v}_{1}^2}{\mathbb E\{H_{\lambda,0}^2(Z_1,Z_2)\}} + \frac{\widehat{v}_{1}^2-\widehat{v}_{2}^2   }{\mathbb E\{H_{\lambda,0}^2(Z_1,Z_2)\}} + \frac{\widehat{v}_{2}^2 - \mathbb E\{H_{\lambda,0}^2(Z_1,Z_2)\}}{\mathbb E\{H_{\lambda,0}^2(Z_1,Z_2)\}}\\[2mm]
		&=\mathfrak{V}_{n1} + \mathfrak{V}_{n2}+\mathfrak{V}_{n3}.
	\end{aligned}
	$$
	The consistency of $\widehat{v}_{n,N}^2$ can be obtained by showing $\mathfrak{V}_{ni} = o_p(1)$ for $i = 1,2,3$.\\
	
	\underline{\textit{Part 2-(i). Verification of $\mathfrak{V}_{n1}  = o_p(1)$.}} 
	Since
	$$
	\widehat{\epsilon}_i^2\widehat{\epsilon}_j^2 -   {\epsilon}_i^2 {\epsilon}_j^2 = 2\epsilon_i\epsilon_j\left(\widehat{\epsilon}_i\widehat{\epsilon}_j -   {\epsilon}_i  {\epsilon}_j  \right) +\left(\widehat{\epsilon}_i\widehat{\epsilon}_j -   {\epsilon}_i  {\epsilon}_j  \right)^2,
	$$
	the triangle inequality and Cauchy--Schwarz  inequality yield
	\begin{equation}
		\label{equ: vN-v1}
		\begin{aligned}
			\left|\widehat v_{n,N}^2-\widehat v_1^2\right|
			&\leq 
			2\left|
			\frac1{n(n-1)}
			\sum_{i\ne j}
			A_{\lambda,D}^2(X_i,X_j)\epsilon_i\epsilon_j
			(\widehat\epsilon_i\widehat\epsilon_j-\epsilon_i\epsilon_j)
			\right| +
			\frac1{n(n-1)}
			\sum_{i\ne j}
			A_{\lambda,D}^2(X_i,X_j)
			(\widehat\epsilon_i\widehat\epsilon_j-\epsilon_i\epsilon_j)^2
			\\ 
			& \leq   2\left(	\widehat{v}_{1}^2 \right)^{1/2} \left\{
			\frac{1}{n(n-1)}\sum_{i\ne j}A_{\lambda,D}^2\left(X_i,X_j\right)
			\left(\widehat\epsilon_i\widehat\epsilon_j-\epsilon_i\epsilon_j\right)^2
			\right\}^{1/2}\\
			& +\frac{1}{n(n-1)}\sum_{i\ne j}A_{\lambda,D}^2\left(X_i,X_j\right)
			\left(\widehat\epsilon_i\widehat\epsilon_j-\epsilon_i\epsilon_j\right)^2.
		\end{aligned}
	\end{equation} 
	The first-order integral formula in \eqref{equ: residual Taylor shift}, and the bounds in \eqref{equ: JH bound} imply
	\[
	\begin{aligned}
		\left|\widehat\epsilon_i-\epsilon_i\right|
		&=\left|\int_0^1
		J_g\{X_i,\theta_0+t (\widehat\theta-\theta_0 )\}^\top
		(\widehat\theta-\theta_0 )\,dt\right|\leq \|\widehat\theta-\theta_0 \|_2
		M_J(X_i).
	\end{aligned}
	\]
	
	Expanding the residual-product difference gives
	\[
	\begin{aligned}
		\widehat\epsilon_i\widehat\epsilon_j-\epsilon_i\epsilon_j=\left(\widehat\epsilon_i-\epsilon_i\right)\epsilon_j
		+\left(\widehat\epsilon_j-\epsilon_j\right)\epsilon_i
		+\left(\widehat\epsilon_i-\epsilon_i\right)
		\left(\widehat\epsilon_j-\epsilon_j\right).
	\end{aligned}
	\] 
	Using the inequality $(u+v+w)^2\leq 3(u^2 +v^2 +w^2)$, we obtain
	\begin{equation}
		\label{equ: step2i 0}
		\begin{aligned}
			&\quad \,\frac{1}{n(n-1)}\sum_{i\ne j}A_{\lambda,D}^2\left(X_i,X_j\right)
			\left(\widehat\epsilon_i\widehat\epsilon_j-\epsilon_i\epsilon_j\right)^2\\
			& \leq 6 \|\widehat\theta-\theta_0 \|_2^2
			\frac{1}{n(n-1)}\sum_{i\ne j}A_{\lambda,D}^2\left(X_i,X_j\right)
			M_J^2(X_i)
			\epsilon_j^2\\
			& +3 \|\widehat\theta-\theta_0\|_2^4
			\frac{1}{n(n-1)}\sum_{i\ne j}A_{\lambda,D}^2\left(X_i,X_j\right)
			M_J^2(X_i)
			M_J^2(X_j).
		\end{aligned}
	\end{equation}
	For the first pair average,  
	\[
	\begin{aligned}
		&\quad\,\, \mathbb E\left[
		\frac{1}{n(n-1)}\sum_{i\ne j}A_{\lambda,D}^2\left(X_i,X_j\right)
		M_J^2(X_i)
		\epsilon_j^2\,\middle|\,D\right]\\
		& =
		\mathbb E\left[
		A_{\lambda,D}^2(X_1,X_2)
		M_J^2(X_1)s_2(X_2,\theta_0)
		\mid D\right]\\[1mm]
		& \leq C_2\int M_J^2(x)
		\left\{\int A_{\lambda,D}^2\left(x,y\right)\,dP_X(y)\right\}dP_X(x)\\
		& \lesssim\mathscr P_{D,\lambda}^4\lambda^{-s}\|M_J\|_{L^2(P_X)}^2,
	\end{aligned}
	\]
	where the first step uses iterated expectation formula, the second step follows from Lemma~\ref{lemma: residual moments at pseudo true}, and the last inequality follows from
	\eqref{equ: empirical kernel row shift}.  For the second pair average,
	\eqref{equ: empirical kernel supremum squared shift}
	and independence of $X_1,X_2$ give
	\[
	\begin{aligned}
		\mathbb E\left[
		\frac1{n(n-1)}
		\sum_{i\ne j}
		A_{\lambda,D}^2(X_i,X_j)
		M_J^2(X_i)M_J^2(X_j)
		\mid D\right]   \lesssim 
		\mathscr P_{D,\lambda}^4\lambda^{-2s}
		\|M_J\|_{L^2(P_X)}^4.
	\end{aligned}
	\] 
	Integrating these two inequalities over $D$, using
	Lemma~\ref{lemma: similarity expectation shift}
	with $u=4$, and applying Markov's inequality yields
	\begin{align}
		\frac1{n(n-1)}
		\sum_{i\ne j}
		A_{\lambda,D}^2(X_i,X_j)
		M_J^2(X_i)\epsilon_j^2
		&=O_p(\lambda^{-s}),
		\label{equ: step2i 1}
		\\
		\frac1{n(n-1)}
		\sum_{i\ne j}
		A_{\lambda,D}^2(X_i,X_j)
		M_J^2(X_i)M_J^2(X_j)
		&=O_p(\lambda^{-2s}).
		\label{equ: step2i 2}
	\end{align}   
	Substituting   \eqref{equ: step2i 1} and \eqref{equ: step2i 2} into \eqref{equ: step2i 0},	and using
	Assumption~\ref{assumption: identibility and parametric estimation under H0}
	gives
	\begin{equation}
		\label{equ: empirical squared residual product rate}
		\frac1{n(n-1)}
		\sum_{i\ne j}
		A_{\lambda,D}^2(X_i,X_j)
		(\widehat\epsilon_i\widehat\epsilon_j-\epsilon_i\epsilon_j)^2
		=
		O_p\!\left(
		n^{-1}\lambda^{-s}+n^{-2}\lambda^{-2s}
		\right).
	\end{equation}
	
	Similarly, by iterated conditional expectation,
	the conditional variance bound, and
	\eqref{equ: merged empirical kernel squared norm},
	\[
	\begin{aligned}
		\mathbb E(\widehat v_1^2\mid D)
		&=
		\mathbb E\left[
		A_{\lambda,D}^2(X_1,X_2)
		s_2(X_1,\theta_0)s_2(X_2,\theta_0)
		\mid D\right]
		\\
		&\leq
		C_2^2
		\mathbb E\{A_{\lambda,D}^2(X_1,X_2)\mid D\}
		\\
		&\lesssim
		\mathscr P_{D,\lambda}^4\lambda^{-s}.
	\end{aligned}
	\]
	Integration over $D$,
	Lemma~\ref{lemma: similarity expectation shift},
	and Markov's inequality imply
	\begin{equation}
		\label{equ: v1Op}
		\widehat v_1^2=O_p(\lambda^{-s}).
	\end{equation}
	
	Finally, substituting \eqref{equ: empirical squared residual product rate} and \eqref{equ: v1Op} into \eqref{equ: vN-v1} yields that,  if $n \to \infty$ and $n\lambda^{s/2}\to \infty$,
	$$
	\widehat v_{n,N}^{\,2}
	- \widehat{v}_1^2 = O_p\left[\lambda^{-s}
	\left\{\frac{1}{\sqrt n}+\frac{1}{n\lambda^{s/2}}\right\}\right].
	$$
	Together with \eqref{equ: EH squared shift}, we obtain
	$$
	\mathfrak{V}_{n1} = \frac{\widehat{v}_{n,N}^2 -\widehat{v}_{1}^2}{\mathbb E\{H_{\lambda,0}^2(Z_1,Z_2)\}}  =O_p\left[ 
	\left\{\frac{1}{\sqrt n}+\frac{1}{n\lambda^{s/2}}\right\}\right] = o_p(1)
	$$
	provided that $\lambda \to 0$, $n\lambda^{s/2} \to \infty$ and $N\lambda^s/\log(\lambda^{-1}) \to \infty$.
	\\

	\underline{\textit{Part 2-(ii). Verification of $\mathfrak{V}_{n2}  = o_p(1)$.}} By definition,
	$$
	\widehat{v}_1^2 -\widehat{v}_2^2 =\frac{1}{n(n-1)}\sum_{i\neq j} 
	\left\{ A_{\lambda,D}^2(X_i,X_j)-A_{\lambda,0}^2(X_i,X_j)\right\} \epsilon_i^2 \epsilon_j^2.
	$$
	The triangle inequality gives
	\[
	\begin{aligned}
		\mathbb E\{|\widehat v_1^2-\widehat v_2^2|\mid D\}
		&\leq
		\mathbb E\left[
		\left|
		A_{\lambda,D}^2(X_1,X_2)
		-A_{\lambda,0}^2(X_1,X_2)
		\right|
		\epsilon_1^2\epsilon_2^2
		\mid D\right]
		\\
		&\leq
		C_2^2\mathbb E\left[
		\left|
		A_{\lambda,D}(X_1,X_2)+A_{\lambda,0}(X_1,X_2)
		\right| 
		\left|
		A_{\lambda,D}(X_1,X_2)-A_{\lambda,0}(X_1,X_2)
		\right|
		\mid D\right],
	\end{aligned}
	\]
	where the second step uses the iterated expectation formula and Lemma~\ref{lemma: residual moments at pseudo true}.  By the elementary inequality $(a+b)^2\leq2(a^2+b^2)$, and~\eqref{equ: merged empirical kernel squared norm},
	\[
	\mathbb E\left[
	\{A_{\lambda,D}(X_1,X_2)+A_{\lambda,0}(X_1,X_2)\}^2
	\mid D\right]
	\lesssim
	\lambda^{-s}(1+\mathscr P_{D,\lambda}^4).
	\]
	Moreover, \eqref{equ: merged kernel difference bound} gives
	\[
	\mathbb E\left[
	\{A_{\lambda,D}(X_1,X_2)-A_{\lambda,0}(X_1,X_2)\}^2
	\mid D\right]
	\lesssim
	\lambda^{-s}\mathscr P_{D,\lambda}^4
	\mathscr Q_{D,\lambda}^2.
	\]
	Hence, applying conditional Cauchy--Schwarz  inequality yields
	\[
	\begin{aligned}
		\mathbb E\{|\widehat v_1^2-\widehat v_2^2|\mid D\}
		& \leq C_2^2
		\left[\mathbb E\left[
		\left\{A_{\lambda,D}\left(X_1,X_2\right)
		+A_{\lambda,0}\left(X_1,X_2\right)\right\}^2\,\middle|\,D\right]\right]^{1/2} \\
		& \quad\quad    \cdot
		\left[\mathbb E\left[
		\left\{A_{\lambda,D}\left(X_1,X_2\right)
		-A_{\lambda,0}\left(X_1,X_2\right)\right\}^2\,\middle|\,D\right]\right]^{1/2}\\
		&\lesssim
		\lambda^{-s}
		(1+\mathscr P_{D,\lambda}^4)^{1/2}
		\mathscr P_{D,\lambda}^2\mathscr Q_{D,\lambda}.
	\end{aligned}
	\] 
	Then, integrating over \(D\) and applying  Cauchy--Schwarz inequality twice gives
	\[
	\begin{aligned}
		\mathbb E\left(| \widehat{v}_1^2 -\widehat{v}_2^2|  \right) &\lesssim\lambda^{-s}
		\left\{1+\mathbb E\left(\mathscr P_{D,\lambda}^4\right)\right\}^{1/2}
		\left\{\mathbb E\left(
		\mathscr P_{D,\lambda}^8  \right)\right\}^{1/4}\left\{\mathbb E\left(
		\mathscr Q_{D,\lambda}^4  \right)\right\}^{1/4}\\
		& =O\left[
		\lambda^{-s}\left\{\frac{\log(1/\lambda)}{N\lambda^s}\right\}^{1/2}
		\right],
	\end{aligned}
	\]
	where the last step follows from Lemma~\ref{lemma: similarity expectation shift} if  $\lambda \to 0$ and $N \lambda^{s} /\log (\lambda^{-1})\to \infty$.
	Markov's inequality consequently yields
	\begin{equation}
		\label{equ: variance operator replacement rate}
		\widehat{v}_1^2 -\widehat{v}_2^2
		=O_p\left[
		\lambda^{-s}\left\{\frac{\log(1/\lambda)}{N\lambda^s}\right\}^{1/2}
		\right].
	\end{equation}
	Together with \eqref{equ: EH squared shift}, we obtain
	$$
	\mathfrak{V}_{n2} =\frac{\widehat{v}_{1}^2 - \widehat{v}_{2}^2}{\mathbb E\{H_{\lambda,0}^2(Z_1,Z_2)\}} =  O_p\left[
	\left\{\frac{\log(1/\lambda)}{N\lambda^s}\right\}^{1/2}\right] =o_p(1)
	$$
	provided that $\lambda \to 0$ and $N \lambda^{s} /\log (\lambda^{-1})\to \infty$.
	\\

	\underline{\textit{Part 2-(iii). Verification of $\mathfrak{V}_{n3}  = o_p(1)$.}}
	Since
	$\mathbb E(\widehat v_2^2)
	=\mathbb E\{H_{\lambda,0}^2(Z_1,Z_2)\}$,
	the variance identity  in Lemma~\ref{lemma: conditional Hoeffding decomposition}-\eqref{equ: conditional variance kernel}, applied to the kernel $H_{\lambda,0}^2$, gives
	\[
	\begin{aligned}
		\operatorname{Var}(\widehat v_2^2)
		=&
		\frac{4(n-2)}{n(n-1)}
		\operatorname{Cov}\left(
		H_{\lambda,0}^2(Z_1,Z_2),
		H_{\lambda,0}^2(Z_1,Z_3)
		\right) +
		\frac{2}{n(n-1)}
		\operatorname{Var}\left(
		H_{\lambda,0}^2(Z_1,Z_2)
		\right)
		\\[1mm]
		\lesssim &
		n^{-1}
		\mathbb E\left\{
		H_{\lambda,0}^2(Z_1,Z_2)
		H_{\lambda,0}^2(Z_1,Z_3)
		\right\} +
		n^{-2}\mathbb E\{H_{\lambda,0}^4(Z_1,Z_2)\}
		\\[1mm]
		\lesssim &
		n^{-1}\lambda^{-2s}
		+n^{-2}\lambda^{-3s}.
	\end{aligned}
	\] 
	where the last step follows from \eqref{equ: shared fourth shift} and \eqref{equ: full fourth shift}.   
	Chebyshev's inequality then yields
	\begin{equation}
		\label{equ: oracle variance sample rate}
		\widehat{v}_2^2
		-\mathbb E\left\{H_{\lambda,0}^2\left(Z_1,Z_2\right)\right\}
		=O_p\left(n^{-1/2}\lambda^{-s} + n^{-1} \lambda^{-3s/2}\right).
	\end{equation}
	Together with \eqref{equ: EH squared shift}, we obtain
	$$
	\mathfrak{V}_{n3} =\frac{\widehat{v}_{2}^2 - \mathbb E\{H_{\lambda,0}^2(Z_1,Z_2)\}}{\mathbb E\{H_{\lambda,0}^2(Z_1,Z_2)\}} =O_p\left(n^{-1/2}  + n^{-1} \lambda^{-s/2}\right) =o_p(1).
	$$
	provided that $\lambda \to 0$,  $n^2\lambda^{s} \to \infty$, and $N \lambda^{s} /\log (\lambda^{-1})\to \infty$. This establishes the   consistency of the variance
	estimator.

	\subsection{Proof of Theorem 2}
	
	Theorem~\ref{theorem: uniform power} specifies conditions
	under which oracle MMR lacks nontrivial uniform power
	and the SRMR-based test is uniformly consistent.
	We prove these two results separately.
	For the oracle MMR lower bound, we construct a sequence
	of alternatives in the prescribed source ball along which
	its power converges to the nominal level $\alpha$.  The proof of the upper bound for the SRMR-based test relies
	primarily on Proposition~\ref{prop: error decomposition}.
	
	Before proceeding, we introduce two lemmas.
	Lemma~\ref{lemma: exact population projection} ensures that
	the constructed alternatives still have $g(\cdot,\theta_0)$
	as their population least-squares projection, so their
	conditional residuals equal the chosen perturbations.
	Lemma~\ref{lemma: general one sided framework power}
	provides sufficient separation conditions for uniform
	consistency based on one-sided signal bounds and
	control of the critical value.


	\begin{lemma}[Exact population projection]
		\label{lemma: exact population projection}
		Suppose that
		Assumption~\ref{assumption: regularities on parametric families}
		holds, and define
		\begin{equation}
			\label{equ: T0}
			\mathcal T_0
			:=
			\operatorname{span}\{
			J_{g,\ell}(\cdot,\theta_0):1\leq\ell\leq d
			\}.
		\end{equation}
		Let $f\in L^\infty(P_X)$ satisfy
		$f\perp\mathcal T_0$ and
		\begin{equation}
			\label{equ: lower perturbation amplitude}
			\|f\|_{L^\infty(P_X)}\,\mathbb E M_H(X)
			\leq
			\frac14\lambda_{\min}(Q_0).
		\end{equation}
		Suppose that there exists a radius $\varepsilon_0>0$
		such that
		\begin{equation}
			\label{equ: lower local quadratic bound}
			\|g(\cdot,\theta)-g(\cdot,\theta_0)\|_{L^2(P_X)}^2
			\geq
			\frac12\lambda_{\min}(Q_0)\|\theta-\theta_0\|_2^2
		\end{equation}
		for every $\theta\in\Theta$ with
		$0<\|\theta-\theta_0\|_2<\varepsilon_0$, and
		\begin{equation}
			\label{equ: lower perturbation global separation}
			\|f\|_{L^2(P_X)}
			<
			\frac12
			\inf_{\substack{\theta\in\Theta\\
					\|\theta-\theta_0\|_2\geq\varepsilon_0}}
			\|g(\cdot,\theta)-g(\cdot,\theta_0)\|_{L^2(P_X)}.
		\end{equation}
		If
		\[
		Y=g(X,\theta_0)+f(X)+e,
		\qquad
		\mathbb E(e\mid X)=0,
		\qquad
		\mathbb E e^2<\infty,
		\]
		then the population least-squares risk has the unique
		minimizer $\theta^\star=\theta_0$.
		Moreover,
		\[
		\mathbf f=m-g(\cdot,\theta^\star)=f
		\quad\text{in }L^2(P_X),
		\qquad
		\inf_{\theta\in\Theta}
		\|m-g(\cdot,\theta)\|_{L^2(P_X)}^2
		=
		\|f\|_{L^2(P_X)}^2.
		\]
	\end{lemma}
	
	\begin{proof}
		Under Assumption~\ref{assumption: regularities on parametric families}, Taylor's formula gives 
		\begin{equation}
			\label{equ: taylor expansion lower bounds}
			\begin{aligned}
				g(\cdot,\theta)-g(\cdot,\theta_0)
				&=
				(\theta-\theta_0)^\top
				\int_0^1
				J_g\{\cdot,\theta_0+t(\theta-\theta_0)\}\,dt
				\\
				&=
				(\theta-\theta_0)^\top J_g(\cdot,\theta_0) +
				(\theta-\theta_0)^\top
				\left[
				\int_0^1(1-t)
				H_g\{\cdot,\theta_0+t(\theta-\theta_0)\}\,dt
				\right]
				(\theta-\theta_0).
			\end{aligned}
		\end{equation}
		The gradient envelope condition  $\sup_{\theta \in \Theta}\|J_{g}(\cdot, \theta)\|_2 \leq M_J(\cdot)$ and the first equality in
		\eqref{equ: taylor expansion lower bounds} imply
		\[
		\|g(\cdot,\theta)-g(\cdot,\theta_0)\|_{L^2(P_X)}
		\leq
		\|\theta-\theta_0\|_2\|M_J\|_{L^2(P_X)}.
		\]
		Using $\mathbb E(e\mid X)=0$, we obtain
		\begin{equation}
			\label{equ: proof of theorem 3 decomposition at theta0}
			\begin{aligned}
				&\quad \,\,\mathbb E\{Y-g(X,\theta)\}^2
				-\mathbb E\{Y-g(X,\theta_0)\}^2\\
				& =
				\|g(\cdot,\theta)-g(\cdot,\theta_0)\|_{L^2(P_X)}^2
				-
				2\langle g(\cdot,\theta)-g(\cdot,\theta_0),f
				\rangle_{L^2(P_X)}.
			\end{aligned}
		\end{equation} 
		Since $f\perp\mathcal T_0$,   the second equality in
		\eqref{equ: taylor expansion lower bounds} yields
		\begin{equation}
			\label{equ: proof of theorem 3 decomposition at theta0 second bound}
			\begin{aligned}
				2\left|
				\langle g(\cdot,\theta)-g(\cdot,\theta_0),f
				\rangle_{L^2(P_X)}
				\right| 
				& =
				2\left|
				\mathbb E\left[
				f(X)(\theta-\theta_0)^\top
				\left\{
				\int_0^1(1-t)
				H_g\{X,\theta_0+t(\theta-\theta_0)\}\,dt
				\right\}
				(\theta-\theta_0)
				\right]
				\right|
				\\
				& \leq
				2\|f\|_{L^\infty(P_X)}
				\|\theta-\theta_0\|_2^2
				\mathbb E\int_0^1(1-t)M_H(X)\,dt
				\\
				& =
				\|f\|_{L^\infty(P_X)}\,\mathbb E M_H(X)
				\|\theta-\theta_0\|_2^2
				\\
				&\leq
				\frac14\lambda_{\min}(Q_0)
				\|\theta-\theta_0\|_2^2,
			\end{aligned}
		\end{equation}
		where the last inequality follows from
		\eqref{equ: lower perturbation amplitude}.
		
		Combining
		\eqref{equ: lower local quadratic bound}, 
		\eqref{equ: proof of theorem 3 decomposition at theta0},
		and~\eqref{equ: proof of theorem 3 decomposition at theta0 second bound},
		we obtain, for
		$0<\|\theta-\theta_0\|_2<\varepsilon_0$,
		\[
		\mathbb E\{Y-g(X,\theta)\}^2
		-\mathbb E\{Y-g(X,\theta_0)\}^2
		\geq
		\frac14\lambda_{\min}(Q_0)
		\|\theta-\theta_0\|_2^2
		>0.
		\] 
		For $\|\theta-\theta_0\|_2\geq\varepsilon_0$,
		Cauchy--Schwarz applied to
		\eqref{equ: proof of theorem 3 decomposition at theta0}
		instead gives
		\[
		\begin{aligned}
			&\quad \,\,\mathbb E\{Y-g(X,\theta)\}^2
			-\mathbb E\{Y-g(X,\theta_0)\}^2
			\\
			& \geq
			\|g(\cdot,\theta)-g(\cdot,\theta_0)\|_{L^2(P_X)}  
			\left\{
			\|g(\cdot,\theta)-g(\cdot,\theta_0)\|_{L^2(P_X)}
			-2\|f\|_{L^2(P_X)}
			\right\}
			>0,
		\end{aligned}
		\]
		where the last step follows from
		\eqref{equ: lower perturbation global separation}.
		
		These two cases cover every $\theta\in\Theta$
		with $\theta\ne\theta_0$.
		Thus the population least-squares risk is uniquely
		minimized at $\theta^\star=\theta_0$. Finally, $\mathbb E(e\mid X)=0$ implies
		$m=g(\cdot,\theta_0)+f$ in $L^2(P_X)$.
		Consequently,
		$\mathbf f=m-g(\cdot,\theta^\star)=f$.
		Moreover, $\mathbb E\{Y-g(X,\theta)\}^2
		=
		\mathbb E e^2
		+\|m-g(\cdot,\theta)\|_{L^2(P_X)}^2$,
		so the same parameter $\theta_0$ minimizes the
		squared distance to the parametric family. Hence
		\[
		\inf_{\theta\in\Theta}
		\|m-g(\cdot,\theta)\|_{L^2(P_X)}^2
		=
		\|m-g(\cdot,\theta_0)\|_{L^2(P_X)}^2
		=
		\|f\|_{L^2(P_X)}^2.
		\]
		This completes the proof. 
	\end{proof}
	\begin{lemma}
		\label{lemma: general one sided framework power}
		Let
		\[
		\phi_{S_\lambda}
		:=\mathbf 1\left\{\widehat S_{n,\lambda}
		>\widehat c_{\alpha,\lambda,n}\right\}
		\]
		be an   asymptotically $\alpha$-level test. Let \(c_0>0\)
		be fixed, and let \(\mathcal U_1,\mathcal U_2\) be positive deterministic
		functions of \(n,\lambda,\|\mathbf f\|_{L^2(P_X)}\).
		Assume the following two uniform probability bounds:
		\begin{align}
			&\lim_{M\to\infty}\limsup_{n\to\infty}
			\sup_{H_1(\mathcal F,\Delta_n)}
			\mathbb P\!\left[
			\left\{c_0\|\mathbf f\|_{L^2(P_X)}^2-\widehat S_{n,\lambda}\right\}_+
			>M\mathcal U_1\right]=0,
			\label{equ: general downward bound power}\\
			&\lim_{M\to\infty}\limsup_{n\to\infty}
			\sup_{H_1(\mathcal F,\Delta_n)}
			\mathbb P\!\left\{
			\left(\widehat c_{\alpha,\lambda,n}\right)_+
			>M\mathcal U_2 \right\}=0.
			\label{equ: general critical bound power}
		\end{align}
		where $
		H_1(\mathcal F,\Delta_n)
		= \{
		\mathbf f\in\mathcal F\subset L^2(P_X):
		\|\mathbf f\|_{L^2(P_X)}^2\geq\Delta_n
		\}$.  If
		\begin{equation}
			\label{equ: general one sided separation power}
			\inf_{H_1(\mathcal F,\Delta_n)}
			\frac{\|\mathbf f\|_{L^2(P_X)}^2}{\mathcal U_1+\mathcal U_2}
			\to\infty,
		\end{equation}
		then the uniform Type II error of $\phi_{S_\lambda}$    satisfies
		\[
		\limsup_{n\to\infty}
		\sup_{H_1(\mathcal F,\Delta_n)}
		\mathbb P(\phi_{S_\lambda} =0)
		= 0 .
		\]
	\end{lemma}
	The proof of Lemma~\ref{lemma: general one sided framework power} is similar to that of  \citet[Lemma~5.3]{chatalic2025scalable}; we defer its detailed proof to Appendix~\ref{sec: addtional lemmas and proofs}.    	\begin{proof}[Proof of Theorem 2] 
		The proof of Theorem~\ref{theorem: uniform power} is divided into two primary parts.\\
		
		\medskip
		\noindent
		\underline{\textit{Part 1. Lower bound for $T_{\mathrm{MMR}}^{\mathrm{or}}$.}} To establish the lower bound, we construct an admissible alternative sequence along which the rejection
		probability of  $T_{\mathrm{MMR}}^{\mathrm{or}}$  converges to $\alpha$.
		
		\medskip
		
		\emph{\underline{Part 1-(i). Construct the perturbation functions.}}
		Let $a_n:=n^{-2r/(2r+1)}$ and $\delta_n:=(\Delta_n a_n)^{1/2}$.
		Since $\Delta_n/a_n\to0$, we have
		$\delta_n\geq\Delta_n$ for all sufficiently large $n$,
		$\delta_n/a_n\to0$, and $\delta_n\to0$.  	Let $	M_n:=\left\lfloor c\delta_n^{-s/(2r)}\right\rfloor,$
		where $c>0$ is a sufficiently small fixed constant.
		We construct $f_n$ from the $d+1$ eigenfunctions
		$\phi_{X,M_n+1},\ldots,\phi_{X,M_n+d+1}$.  Specifically, define the $d\times(d+1)$ matrix
		\[
		A_n
		:=
		\left[
		\left\langle
		J_{g,j}(\cdot,\theta_0),\phi_{X,M_n+\ell}
		\right\rangle_{L^2(P_X)}
		\right]_{ 1\leq j\leq d, 1\leq\ell\leq d+1}.
		\]
		Since $A_n$ has more columns than rows, its null space is non-trivial, and thus there exists a   vector $v_n\in\mathbb R^{d+1}$ such that $A_nv_n = 0$ and $\|v_n\|_2=1$.
		Set
		\begin{equation}
			\label{equ: fn}
			f_n
			:=
			\sqrt{\delta_n}
			\sum_{\ell=1}^{d+1}
			v_{n,\ell}\phi_{X,M_n+\ell}.
		\end{equation}
		Since $\{\phi_{X,k}\}_{k\geq 1}$ forms an orthonormal basis of $L^2(P_X)$, direct calculations yield
		\begin{equation}
			\label{equ: L2norm of fn}
			\left\|f_n \right\|_{L^2(P_X)}^2 = \delta_n\cdot \|v_n\|_2^2 = \delta_n.
		\end{equation}
		For each $1\leq j \leq d$, it follows from $A_nv_n = 0$ that
		\begin{equation}
			\label{equ: fn orthogonal T0}
			\langle f_n, J_{g,j} (\cdot, \theta_0)\rangle_{L^2(P_X)} = 	\sqrt{\delta_n}
			\sum_{\ell=1}^{d+1}
			v_{n,\ell} \left\langle\phi_{X,M_n+\ell},J_{g,j} (\cdot, \theta_0)\right\rangle_{L^2(P_X)} =0.
		\end{equation}
		This implies $f_n\perp\mathcal T_0$,
		where $\mathcal{T}_0 $ is defined in \eqref{equ: T0}.
		Moreover, the uniform eigenfunction bound in Assumption~\ref{assumption: eigenvalues and eigenfunctions} implies
		\begin{equation}
			\label{equ: Linftynorm of fn}
			\|f_n\|_{L^\infty(P_X)}
			\leq
			\left(\sup_{k\geq1} \|\phi_{X,k}\|_{L^{\infty}(P_X)}\right)\sqrt{\delta_n}
			\sum_{\ell=1}^{d+1}|v_{n,\ell}|
			\leq
			\left(\sup_{k\geq1} \|\phi_{X,k}\|_{L^{\infty}(P_X)}\right)\sqrt{d+1}\sqrt{\delta_n}
			\to 0.
		\end{equation} 
		To verify the source condition, take
		\[
		u_n
		:=
		\sqrt{\delta_n}
		\sum_{\ell=1}^{d+1}
		v_{n,\ell}\sigma_{X,M_n+\ell}^{-r}
		\phi_{X,M_n+\ell}.
		\]
		Then $f_n=\mathcal L_{K,X}^r u_n$, and eigenvalue decay in Assumption~\ref{assumption: eigenvalues and eigenfunctions} gives
		\begin{equation}
			\label{equ: L2norm of un}
			\|u_n\|_{L^2(P_X)}^2
			=
			\delta_n
			\sum_{\ell=1}^{d+1}
			v_{n,\ell}^2\sigma_{X,M_n+\ell}^{-2r} \lesssim
			\delta_n(M_n+d+1)^{2r/s}
			\lesssim c^{2r/s} \leq 1,
		\end{equation}
		where the last step follows for sufficiently large $n$  by choosing $c$ sufficiently small.   
		Based on \eqref{equ: L2norm of fn}, \eqref{equ: fn orthogonal T0}, \eqref{equ: Linftynorm of fn}, and \eqref{equ: L2norm of un}, we obtain that, for all sufficiently large $n$,    the function $f_n$ in \eqref{equ: fn} satisfies
		\begin{equation}
			\label{equ: conclusion of step 1i}
			\|f_n\|_{L^2(P_X)}^2 \geq \Delta_n, \qquad f_n \perp\mathcal T_0,\quad \mathrm{and} \quad f_n \in \mathcal{F}_{r,B}.
		\end{equation}
		
		\medskip
		
		\emph{\underline{Part 1-(ii). Verify the population projection.}}
		The discrepancy measure in our alternative class is the $L^2(P_X)$ distance between the true regression function $m(\cdot)$ and its projection on the parametric family.   We therefore need to verify
		that adding $f_n$ to $g(\cdot,\theta_0)$ does not change
		that projection. Specifically, define	
		\begin{equation}
			\label{equ: constructed alternative model}
			m_n:=g(\cdot,\theta_0)+f_n,\qquad 
			Y=m_n(X)+e.
		\end{equation}
		We shall show that
		\begin{equation}
			\label{equ: conclusion of step1ii}	\theta_n^\star=\theta_0,
			\qquad
			\mathbf f_n=m_n-g(\cdot,\theta_n^\star)=f_n, \qquad \inf_{\theta\in\Theta}
			\|m_n-g(\cdot,\theta)\|_{L^2(P_X)}^2
			=
			\delta_n
			\geq\Delta_n.
		\end{equation}
		where $\theta_n^{\star}$ and $\mathbf{f}_n$ are defined through \eqref{equ: conditional residual function}.
		Combining \eqref{equ: conclusion of step 1i} and~\eqref{equ: conclusion of step1ii} shows that the constructed alternatives belong to
		$H_1(\mathcal F_{r,B},\Delta_n)$. 
		
		Using the first-order
		integral formula in~\eqref{equ: taylor expansion lower bounds}, Jensen's inequality gives
		\[
		\begin{aligned}
			&\quad \,\left\|g(\cdot,\theta)-g(\cdot,\theta_0)
			-J_g(\cdot,\theta_0)^\top(\theta-\theta_0)\right\|_{L^2(P_X)}\\
			& \leq
			\|\theta-\theta_0\|_2
			\left[
			\mathbb E\int_0^1
			\left\|J_g\{X,\theta_0+t(\theta-\theta_0)\}
			-J_g(X,\theta_0)\right\|_2^2\,dt
			\right]^{1/2}\\
			&=o(\|\theta-\theta_0\|_2),
		\end{aligned}
		\]
		where the last step follows from the dominated  convergence theorem, because the integrand converges pointwise to zero as
		\(\theta\to\theta_0\) and is bounded by \(4M_J^2(X)\),
		which is integrable.  Therefore, we obtain
		\begin{equation}
			\label{equ: proof of theorem 3 decomposition at theta0 first bound}
			\begin{aligned}
				\|g(\cdot,\theta)-g(\cdot,\theta_0)\|_{L^2(P_X)}^2 =
				(\theta-\theta_0)^\top
				Q_0
				(\theta-\theta_0)
				+o(\|\theta-\theta_0\|_2^2).
			\end{aligned}
		\end{equation} 
		Since $Q_0$ is positive definite, fix a sufficiently small
		radius $\varepsilon_0>0$ for which
		\eqref{equ: lower local quadratic bound} holds.
		The global identification condition in
		Assumption~\ref{assumption: regularities on parametric families}
		then gives
		\begin{equation}
			\label{equ: fixed radius global separation}
			\inf_{\substack{\theta\in\Theta\\
					\|\theta-\theta_0\|_2\geq\varepsilon_0}}
			\|g(\cdot,\theta)-g(\cdot,\theta_0)\|_{L^2(P_X)}
			>0.
		\end{equation}
		By \eqref{equ: Linftynorm of fn} and \eqref{equ: L2norm of fn},  $\|f_n\|_{L^{\infty}(P_X)}\to0$ and $\|f_n\|_{L^2(P_X)}\to0$. Hence,
		both \eqref{equ: lower perturbation amplitude}
		and~\eqref{equ: lower perturbation global separation}
		hold for all sufficiently large $n$. Then by Lemma~\ref{lemma: exact population projection},  \eqref{equ: conclusion of step1ii} follows.

		\medskip
		\emph{\underline{Part 1-(iii). Establish the indistinguishability.}}
		We now show that the oracle test $T_{\mathrm{MMR}}^{\mathrm{or}}$ has asymptotic rejection probability $\alpha$
		along the constructed alternatives in \eqref{equ: constructed alternative model}. The oracle MMR statistic has the form of 
		\[
		U_n(a,b)
		:=
		\frac1{n(n-1)}
		\sum_{i\ne j}a_iK(X_i,X_j)b_j.
		\]
		Under the null hypothesis,   $a_i = b_i =e_i$;
		under the constructed alternative in \eqref{equ: constructed alternative model},   $a_i = b_i  = e_i+f_n(X_i)$. We shall prove that the limiting distributions of $U_n(a,b)$ are the same under both hypotheses by bounding their difference: 
		\begin{equation}
			\label{equ: difference of 1and0}
			U_n(e+f_n,e+f_n)-U_n(e,e)
			=
			2U_n(e,f_n)+U_n(f_n,f_n).
		\end{equation} 
		
		The cross term $U_n(e,f_n)$ in
		\eqref{equ: difference of 1and0} has mean zero and can
		be written as a $U$-statistic:
		\[
		U_n(e,f_n)
		=
		\frac{2}{n(n-1)}
		\sum_{i<j}H_{e,f}(Z_i,Z_j),
		\qquad
		H_{e,f}(Z_1,Z_2)
		:=
		\frac12K(X_1,X_2)
		\{e_1f_n(X_2)+e_2f_n(X_1)\}.
		\]
		Since $\mathbb E(e\mid X)=0$, the first projection is
		\[
		\psi_{1,e,f}(Z_1)
		:=
		\mathbb E\{H_{e,f}(Z_1,Z_2)\mid Z_1\}
		=
		\frac12e_1(\mathcal L_{K,X}f_n)(X_1).
		\]
		By Assumption~\ref{assumption: conditional variance},
		\[
		\begin{aligned}
			\mathbb E\{\psi_{1,e,f}^2(Z_1)\}
			&=
			\frac14
			\mathbb E\left[
			e_1^2\{(\mathcal L_{K,X}f_n)(X_1)\}^2
			\right] =
			\frac14
			\mathbb E\left[
			\mathbb E(e_1^2\mid X_1)
			\{(\mathcal L_{K,X}f_n)(X_1)\}^2
			\right]
			\\
			&\leq
			\frac{C_e}{4}
			\|\mathcal L_{K,X}f_n\|_{L^2(P_X)}^2.
		\end{aligned}
		\]
		The second-order degenerate projection is
		\[
		\psi_{2,e,f}(Z_1,Z_2)
		:=
		H_{e,f}(Z_1,Z_2)-\psi_{1,e,f}(Z_1)-\psi_{1,e,f}(Z_2).
		\]
		Orthogonality of the Hoeffding components gives
		\[
		\begin{aligned}
			\mathbb E\{\psi_{2,e,f}^2(Z_1,Z_2)\}
			&=
			\mathbb E\{H_{e,f}^2(Z_1,Z_2)\}
			-2\mathbb E\{\psi_{1,e,f}^2(Z_1)\}
			\leq
			\mathbb E\{H_{e,f}^2(Z_1,Z_2)\}\\
			&=
			\frac14
			\mathbb E\left[
			K^2(X_1,X_2)
			\{e_1^2f_n^2(X_2)+e_2^2f_n^2(X_1)\}
			\right]
			\\
			&=
			\frac12
			\mathbb E\left[
			K^2(X_1,X_2)
			\mathbb E(e_1^2\mid X_1)f_n^2(X_2)
			\right]
			\\
			&\leq
			\frac{C_e\kappa^4}{2}\delta_n,
		\end{aligned}
		\]
		where the last inequality uses
		Assumption~\ref{assumption: conditional variance},
		boundedness of $K$, and~\eqref{equ: L2norm of fn}.
		Hoeffding's variance identity in Lemma~\ref{lemma: conditional Hoeffding decomposition}
		therefore yields
		\begin{equation}
			\label{equ: variance bound for cross term}
			\begin{aligned}
				\operatorname{Var}\{nU_n(e,f_n)\}
				&=
				4n\mathbb E\{\psi_{1,e,f}^2(Z_1)\}
				+
				\frac{2n}{n-1}
				\mathbb E\{\psi_{2,e,f}^2(Z_1,Z_2)\}
				\\
				&\leq
				C_e n\|\mathcal L_{K,X}f_n\|_{L^2(P_X)}^2
				+
				\frac{n}{n-1}C_e\kappa^4\delta_n
				\\
				&\lesssim
				n\|\mathcal L_{K,X}f_n\|_{L^2(P_X)}^2+\delta_n.
			\end{aligned}
		\end{equation}
		
		For the pure-signal term $U_n(f_n,f_n)$ in
		\eqref{equ: difference of 1and0}, write
		\[
		U_n(f_n,f_n)
		=
		\frac{2}{n(n-1)}
		\sum_{i<j}H_f(Z_i,Z_j),
		\qquad
		H_f(Z_1,Z_2)
		:=
		f_n(X_1)K(X_1,X_2)f_n(X_2).
		\]
		Its mean is 
		\begin{equation}
			\label{equ: mean of signal term}
			\mu_n
			:=
			\mathbb E\{H_f(Z_1,Z_2)\}
			=
			\langle f_n,\mathcal L_{K,X}f_n\rangle_{L^2(P_X)}.
		\end{equation} 
		The centered first projection of $H_f$ is
		\[
		\begin{aligned}
			\psi_{1,f}(Z_1)
			&:=
			\mathbb E\{H_f(Z_1,Z_2)\mid Z_1\}-\mu_n =
			f_n(X_1)(\mathcal L_{K,X}f_n)(X_1)-\mu_n.
		\end{aligned}
		\]
		Hence 
		\[
		\begin{aligned}
			\mathbb E\{\psi_{1,f}^2(Z_1)\}
			&=
			\operatorname{Var}\left\{
			f_n(X_1)(\mathcal L_{K,X}f_n)(X_1)
			\right\}\leq 
			\mathbb E\left[
			f_n^2(X_1)
			\{(\mathcal L_{K,X}f_n)(X_1)\}^2
			\right] \\
			&\leq 
			\|f_n\|_{L^\infty(P_X)}^2
			\|\mathcal L_{K,X}f_n\|_{L^2(P_X)}^2.
		\end{aligned}
		\]
		The second-order degenerate projection is
		\[
		\psi_{2,f}(Z_1,Z_2)
		:=
		H_f(Z_1,Z_2)-\mu_n
		-\psi_{1,f}(Z_1)-\psi_{1,f}(Z_2).
		\]
		Again, orthogonality of the Hoeffding components gives
		\[
		\begin{aligned}
			\mathbb E\{\psi_{2,f}^2(Z_1,Z_2)\}
			&=
			\mathbb E\{H_f^2(Z_1,Z_2)\}
			-\mu_n^2
			-2\mathbb E\{\psi_{1,f}^2(Z_1)\}
			\\
			&\leq
			\mathbb E\left[
			K^2(X_1,X_2)f_n^2(X_1)f_n^2(X_2)
			\right]
			\\
			&\leq
			\kappa^4
			\mathbb E\{f_n^2(X_1)f_n^2(X_2)\}
			\\
			&=
			\kappa^4\delta_n^2,
		\end{aligned}
		\]
		where the last equality uses independence of $X_1,X_2$
		and~\eqref{equ: L2norm of fn}.
		Applying the same variance identity yields
		\begin{equation}
			\label{equ: variance bound for pure signal term}
			\begin{aligned}
				\operatorname{Var}\{nU_n(f_n,f_n)\}
				&=
				4n\mathbb E\{\psi_{1,f}^2(Z_1)\}
				+
				\frac{2n}{n-1}
				\mathbb E\{\psi_{2,f}^2(Z_1,Z_2)\}
				\\
				&\lesssim
				n	\|f_n\|_{L^\infty(P_X)}^2
				\|\mathcal L_{K,X}f_n\|_{L^2(P_X)}^2+\delta_n^2.
			\end{aligned}
		\end{equation}
		
	 Based on the facts  $\mathbb{E}\{U_n(e,f_n)\mid X_1\ldots,X_n\}=0$ and $\mathbb{E}\{U_n(e,f_n) U_n(f_n,f_n)\} = 0$, we get from \eqref{equ: difference of 1and0},  \eqref{equ: variance bound for cross term}, \eqref{equ: mean of signal term}, and  \eqref{equ: variance bound for pure signal term} that
		
		$$
		\begin{aligned}
			&\quad \,\,n^2\, \mathbb{E}\left| U_n(e+f_n,e+f_n)-U_n(e,e)\right|^2\\
			&=4	\operatorname{Var}\{nU_n(e,f_n)\} + n^2\mu_n^2 + 	\operatorname{Var}\{nU_n(f_n,f_n)\} \\
			&\lesssim 	n\|\mathcal L_{K,X}f_n\|_{L^2(P_X)}^2+\delta_n + n^2\langle f_n,\mathcal L_{K,X}f_n\rangle_{L^2(P_X)}^2 + 	n	\|f_n\|_{L^\infty(P_X)}^2
			\|\mathcal L_{K,X}f_n\|_{L^2(P_X)}^2+\delta_n^2 .
		\end{aligned}
		$$
		By the eigenvalue decay in
		Assumption~\ref{assumption: eigenvalues and eigenfunctions},
		\eqref{equ: fn}, \eqref{equ: L2norm of fn}, \eqref{equ: Linftynorm of fn}
		and the definition of $M_n$, we have
		\[
		\max_{1\leq\ell\leq d+1}\sigma_{X,M_n+\ell}
		\lesssim M_n^{-1/s}
		\lesssim \delta_n^{1/(2r)},
		\qquad
		\|f_n\|_{L^\infty(P_X)}^2
		\lesssim \delta_n.
		\]
		Recall that $a_n=n^{-2r/(2r+1)}$ and
		$\delta_n/a_n\to0$. Since $f_n$ lies in $\mathrm{span}\{\phi_{X,M_n+1},\ldots,\phi_{X,M_n+d+1}\}$,
		\[
		\begin{aligned}
			n\|\mathcal L_{K,X}f_n\|_{L^2(P_X)}^2
			&\leq
			n\left(
			\max_{1\leq\ell\leq d+1}\sigma_{X,M_n+\ell}
			\right)^2
			\|f_n\|_{L^2(P_X)}^2 \lesssim n\delta_n^{1+1/r} =
			n^{-1/(2r+1)}
			\left(\frac{\delta_n}{a_n}\right)^{1+1/r}
			\to 0.
		\end{aligned}
		\]
		Similarly,
		\[
		\begin{aligned} 
			n^2\langle f_n,\mathcal L_{K,X}f_n
			\rangle_{L^2(P_X)}^2 \leq
			n^2\left(
			\max_{1\leq\ell\leq d+1}\sigma_{X,M_n+\ell}
			\right)^2
			\|f_n\|_{L^2(P_X)}^4 \lesssim n^2\delta_n^{2+1/r} =
			\left(\frac{\delta_n}{a_n}\right)^{2+1/r}
			\to 0.
		\end{aligned}
		\]
		Finally,
		\[
		\begin{aligned}
			n\|f_n\|_{L^\infty(P_X)}^2
			\|\mathcal L_{K,X}f_n\|_{L^2(P_X)}^2
			+\delta_n^2 \lesssim
			n\delta_n^{2+1/r}+\delta_n^2 
			=
			\frac1n
			\left(\frac{\delta_n}{a_n}\right)^{2+1/r}
			+\delta_n^2
			\to 0.
		\end{aligned}
		\]
		Since $\delta_n\to0$, the preceding  bounds therefore give $	n^2\mathbb E\left|
		U_n(e+f_n,e+f_n)-U_n(e,e)
		\right|^2
		\to 0$. Markov's inequality further gives
		\[
		n\{U_n(e+f_n,e+f_n)-U_n(e,e)\}
		\xrightarrow{p}0.
		\]
		It has been established by  \citet{muandet2020kernel} that
		\[
		n U_n(e,e)
		\leadsto
		\sum_{k\geq1}\tau_k(W_k^2-1)
		=:W,
		\qquad
		W_k\stackrel{\mathrm{i.i.d.}}{\sim}N(0,1).
		\]
		Here the  weights  $\{\tau_k\}_{k\geq1}$  depend on the kernel, the parametric model and the data distribution. Hence, Slutsky's lemma shows that
		$nU_n(e+f_n,e+f_n)$ has the same limit $W$.
		Therefore,
		\[
		\lim_{n\to\infty}\mathbb P_{f_n}(T_{\mathrm{MMR}}^{\mathrm{or}}=0)
		=
		\mathbb P(W\leq q_{1-\alpha,W})
		=
		1-\alpha.
		\]
		where $q_{1-\alpha,W}$ denotes the $(1-\alpha)$-th quantile of $W$.
		Since the constructed alternatives belong to
		$H_1(\mathcal F_{r,B},\Delta_n)$ for all sufficiently large $n$,
		\[
		\liminf_{n\to\infty}
		\sup_{H_1(\mathcal F_{r,B},\Delta_n)}
		\mathbb P(T_{\mathrm{MMR}}^{\mathrm{or}}=0)
		\geq1-\alpha.
		\]
		This establishes the lower bound for oracle MMR.
		
		\medskip
		\noindent
		\underline{\textit{Part 2. Detection boundary of
				$T_{\mathrm{SRMR}}$.}}
		We now study the detection boundary of the proposed test.
		We first show that uniformly over 	$H_1(\mathcal F_{r,B},\Delta_n)$,
		\begin{equation}
			\label{equ: combined one sided rate power}
			\begin{aligned}
				\left\{
				\frac{\|\mathbf f\|_{L^2(P_X)}^2}{3\overline c}
				-\widehat\eta_\lambda(\widehat\theta)
				\right\}_+
				&=O_p\!\left\{
				\lambda^{2r}
				+\frac{\lambda^{-s/2}}n
				+\frac1n
				+\frac{\lambda^{-s}}{n^2}
				+\frac{\|\mathbf f\|_{L^2(P_X)}}{\sqrt n}
				\right\}.
			\end{aligned}
		\end{equation} 
		Under $\lambda\to0$ and
		$N\lambda^s/\log(\lambda^{-1})\to\infty$,
		Lemma~\ref{lemma: similarity expectation shift} gives
		$\mathbb E(\mathscr P_{D,\lambda}^2)=O(1)$. It follows from \eqref{equ: total parametric rate shift} and \eqref{equ: sample error rate alternatives shift} that 
		\begin{equation}
			\label{equ: proof theorem 2 pe+se}
			\begin{aligned}
				\left|
				\widehat\eta_\lambda(\widehat\theta)
				-\eta_\lambda^s(\theta^\star)
				\right|
				&\leq
				\left|
				\widehat\eta_\lambda(\widehat\theta)
				-\widehat\eta_\lambda(\theta^\star)
				\right|
				+
				\left|
				\widehat\eta_\lambda(\theta^\star)
				-\eta_\lambda^s(\theta^\star)
				\right|\\
				&=O_p\!\left\{
				\frac1n
				+\frac{\lambda^{-s}}{n^2}
				+\frac{\lambda^{-s/2}}n
				+\frac{\|\mathbf f\|_{L^2(P_X)}}{\sqrt n}
				\right\}.
			\end{aligned}
		\end{equation}
		On the event $\{\mathscr Q_{D,\lambda}\leq1/2\}$, it follows from
		\eqref{equ: approximation error} that,  
		\[
		\frac{\|\mathbf f\|_{L^2(P_X)}^2}{3\overline c}
		-\eta_\lambda^s(\theta^\star)
		\leq
		\frac{\lambda^{2r}}{3\overline c}.
		\]
		Consequently, on this event,
		\[
		\left\{
		\frac{\|\mathbf f\|_{L^2(P_X)}^2}{3\overline c}
		-\widehat\eta_\lambda(\widehat\theta)
		\right\}_+
		\leq
		\frac{\lambda^{2r}}{3\overline c}
		+
		\left|
		\widehat\eta_\lambda(\widehat\theta)
		-\eta_\lambda^s(\theta^\star)
		\right|.
		\]
		Hence, for any $t>0$,
		\begin{equation}
			\label{equ: proof theorem 2 three error relation}
			\mathbb P\left[
			\left\{
			\frac{\|\mathbf f\|_{L^2(P_X)}^2}{3\overline c}
			-\widehat\eta_\lambda(\widehat\theta)
			\right\}_+
			>
			\frac{\lambda^{2r}}{3\overline c}+t
			\right]
			\leq
			\mathbb P(\mathscr Q_{D,\lambda}>1/2)
			+
			\mathbb P\left\{
			\left|
			\widehat\eta_\lambda(\widehat\theta)
			-\eta_\lambda^s(\theta^\star)
			\right|>t
			\right\}.
		\end{equation}
		By Lemma~\ref{lemma: similarity expectation shift},
		if $\lambda\to0$ and
		$N\lambda^s/\log(\lambda^{-1})\to\infty$, then
		$\mathscr Q_{D,\lambda}=o_p(1)$. 
		Thus,
		\eqref{equ: combined one sided rate power}
		follows  from
		\eqref{equ: proof theorem 2 pe+se}
		and~\eqref{equ: proof theorem 2 three error relation}. The uniformity follows since the constants in the proof of Proposition~\ref{prop: error decomposition}
		depend only on the common derivative-envelope, noise-moment,
		and operator bounds,  and the uniform parametric-estimation error rate in Assumption~\ref{assumption: uniform parametric estimation under H1}.
		
		Under the alternatives, define $\widehat v_1^2$ as
			in~\eqref{equ: hatvn2}, with
			$\epsilon_i=\epsilon(Z_i,\theta^\star)$.
			Using the common residual-moment bounds in
			Lemma~\ref{lemma: residual moments at pseudo true}
			and the uniform parameter-estimation rate in
			Assumption~\ref{assumption: uniform parametric estimation under H1},
			the arguments for
			\eqref{equ: empirical squared residual product rate}
			and~\eqref{equ: v1Op} apply uniformly over
			$H_1(\mathcal F_{r,B},\Delta_n)$.
			Consequently, \eqref{equ: vN-v1} gives, under the stated conditions,
			\[
			\widehat v_{n,N}^2
			=
			\widehat v_1^2+o_p(\lambda^{-s})
			=
			O_p(\lambda^{-s})
			\]
			uniformly over this class. Thus,
			\begin{equation}
				\label{equ: proof of theorem 2 critical value bound}
				\sqrt{2}\,z_{1-\alpha}\widehat v_{n,N}/n
				=
				O_p\left(n^{-1}\lambda^{-s/2}\right)
			\end{equation}
			also holds uniformly over the alternatives.
			
%
		Take
		$\widehat S_{n,\lambda}
		=\widehat\eta_\lambda(\widehat\theta)$
		and $c_0=(3\overline c)^{-1}$.
		In view of
		\eqref{equ: combined one sided rate power}
		and~\eqref{equ: proof of theorem 2 critical value bound},
		take
		\[
		\begin{aligned}
			\mathcal U_1\{n,\lambda,\|\mathbf f\|_{L^2(P_X)}\}
			&=
			\lambda^{2r}
			+\frac{\lambda^{-s/2}}n
			+\frac1n
			+\frac{\lambda^{-s}}{n^2}
			+\frac{\|\mathbf f\|_{L^2(P_X)}}{\sqrt n},\\
			\mathcal U_2\{n,\lambda,\|\mathbf f\|_{L^2(P_X)}\}
			&=\frac{\lambda^{-s/2}}n.
		\end{aligned}
		\]
		{ Since
			$\|\mathbf f\|_{L^2(P_X)}^2\geq\Delta_n$ on the alternative class,
			\[
			\sup_{H_1(\mathcal F_{r,B},\Delta_n)}
			\frac{\mathcal U_1+\mathcal U_2}{\|\mathbf f\|_{L^2(P_X)}^2}
			\leq
			\frac{\lambda^{2r}+2n^{-1}\lambda^{-s/2}+n^{-1}+n^{-2}\lambda^{-s}}
			{\Delta_n}
			+\frac{1}{\sqrt{n\Delta_n}}.
			\]
		}
		It follows from
		Lemma~\ref{lemma: general one sided framework power}
		that if $\lambda\to0$,
		$n\lambda^{s/2}\to\infty$,
		$N\lambda^s/\log(\lambda^{-1})\to\infty$, and
		\begin{equation}
			\label{equ: sufficient separation power}
			\frac{\lambda^{2r}+n^{-1}\lambda^{-s/2}}{\Delta_n}
			\to0,
		\end{equation}
		then
		\begin{equation}
			\label{equ: uniform type two power}
			\limsup_{n\to\infty}
			\sup_{H_1(\mathcal F_{r,B},\Delta_n)}
			\mathbb P(T_{\mathrm{SRMR}}=0)
			=0.
		\end{equation}
		Choosing $\lambda\asymp n^{-2/(4r+s)}$,
		condition~\eqref{equ: sufficient separation power}
		becomes $\Delta_n n^{4r/(4r+s)}\to\infty$.
		The corresponding auxiliary-sample condition is
		$Nn^{-2s/(4r+s)}/\log n\to\infty$.
		The proof is complete.
	\end{proof}
	\subsection{Proof of Theorem 3}
	We prove the lower bound by constructing a mixture of
	alternatives that is asymptotically indistinguishable from
	a fixed member of the null family.
	The proof proceeds in four steps.
	
In \emph{Step 1}, we construct bounded orthonormal functions
that are orthogonal to the gradient components.
In \emph{Step 2}, we combine these functions to obtain
perturbations in $\mathcal F_{r,B}$ with squared norm
$\Delta_n$. In \emph{Step 3}, we assign a prior to these alternatives, and in \emph{Step 4}, we show that the resulting mixture
cannot be distinguished from the null asymptotically
by proving that its likelihood ratio has second moment
converging to one. 

	{\medskip
		\noindent
		\underline{\textit{Step 1. Construct bounded orthonormal directions.}}
		We construct an orthonormal family of functions in \(L^2(P_X)\)
		that are orthogonal to the tangent space \(\mathcal T_0\) defined
		in \eqref{equ: T0}.  
		
		Let \(M\geq d+1\) be an integer and set $R:=\left\lfloor {M}/({d+1})\right\rfloor.$
		For \(1\leq j\leq R\) and \(1\leq q\leq d+1\), write
		\[
		k_{j,q}:=(j-1)(d+1)+q,
		\qquad
		B_j:=\{k_{j,1},\ldots,k_{j,d+1}\}.
		\]
		The sets \(B_1,\ldots,B_R\) are pairwise disjoint subsets of
		\(\{1,\ldots,M\}\). Since \(d\) is fixed, \(R\asymp M\) as
		\(M\to\infty\). 	For each \(1\leq j\leq R\), define the \(d\times(d+1)\) matrix
		\[
		A_j
		:=
		\left[ 
		\left\langle
		J_{g,\ell}(\cdot,\theta_0),\phi_{X,k_{j,q}}
		\right\rangle_{L^2(P_X)}
		\right]_{ {1\leq\ell\leq d, 1\leq q\leq d+1}}.
		\]
		Since $A_j$ has more columns than rows, its null space is non-trivial, and thus there exists a   vector $v_j = (v_{j,1},\ldots,v_{j,d+1})^{\top}\in\mathbb R^{d+1}$ such that $A_jv_j = 0$ and $\|v_j\|_2=1$. Set
		\[
		\psi_{M,j}
		:=
		\sum_{q=1}^{d+1}v_{j,q}\phi_{X,k_{j,q}},\qquad 1\leq j \leq R.
		\]
		For every \(1\leq\ell\leq d\), it follows from $A_jv_j =0$ that
		\[
		\left\langle
		J_{g,\ell}(\cdot,\theta_0),\psi_{M,j}
		\right\rangle_{L^2(P_X)}
		=
		\sum_{q=1}^{d+1}v_{j,q}
		\left\langle
		J_{g,\ell}(\cdot,\theta_0),\phi_{X,k_{j,q}}
		\right\rangle_{L^2(P_X)} =0.
		\]
		This implies  \(\psi_{M,j}\perp\mathcal T_0\) for each $1\leq j \leq R$.  Moreover,  by orthonormality
		of the eigenfunctions,   the disjointness of \(\{B_j\}_{j=1}^R\), and $\|v_j\|_2 =1$, we have
		\[\|\psi_{M,j}\|_{L^2(P_X)}^2=\sum_{q=1}^{d+1}v_{j,q}^2=1,\qquad \langle\psi_{M,j},\psi_{M,j'}\rangle_{L^2(P_X)}=0,\quad j\neq j'.
		\] 
		Thus \(\{\psi_{M,j}\}_{j=1}^R\) is an orthonormal family. 	Finally,
		by Cauchy--Schwarz inequality and \(\|v_j\|_2=1\),
		\begin{equation}
			\label{equ: cpsi}
			\begin{aligned}
				\max_{1\leq j\leq R}\|\psi_{M,j}\|_{L^\infty(P_X)}
				&\leq \left(\sup_{k\geq1}\|\phi_{X,k}\|_{L^\infty(P_X)}\right)
				\max_{1\leq j\leq R}\sum_{q=1}^{d+1}|v_{j,q}|\\
				&
				\leq \left(\sup_{k\geq1}\|\phi_{X,k}\|_{L^\infty(P_X)}\right)\sqrt{d+1}
				=:C_\psi.
			\end{aligned}
		\end{equation}
		The first step follows from the uniform boundedness condition in Assumption~\ref{assumption: eigenvalues and eigenfunctions}. The constant \(C_\psi\) is independent of \(M\) and \(j\), and
		every \(\psi_{M,j}\) belongs to
		\(\operatorname{span}\{\phi_{X,1},\ldots,\phi_{X,M}\}\).}

	\medskip
	\noindent
	\underline{\textit{Step 2: Construct alternatives with prescribed separation.}} {Recall that  $R:=\left\lfloor {M}/({d+1})\right\rfloor.$}
	 For fixed constants $c_s,c_b>0$, to be chosen sufficiently
	small, set
	\[
	M=\left\lfloor c_s\Delta_n^{-s/(2r)}\right\rfloor,
	\qquad
	k=\min\left\{
	R,\left\lfloor c_b/\Delta_n\right\rfloor
	\right\},
	\qquad
	a=\sqrt{\Delta_n/k}.
	\] 
	
	Let $S$ be a uniformly distributed $k$-element subset of
	$\{1,\ldots,R\}$, and $\zeta_1,\ldots,\zeta_R$ be
	independent Rademacher random variables.
	Define
	\begin{equation}
		\label{equ: constructed perturbation}
		f_{S,\zeta}
		:=
		a\sum_{j\in S}\zeta_j\psi_{M,j}.
	\end{equation}
Since  $f_{S,\zeta}$ is a linear combination of  $\{\psi_{M,\ell}\}_{\ell = 1}^R$, we have $f_{S,\zeta}\perp\mathcal T_0$ for  every $S,\zeta$.	On the other hand, by orthonormality of $\{\psi_{M,\ell}\}_{\ell = 1}^R$,
	\begin{equation}
		\label{equ: constructed perturbation L2 norm}
		\|f_{S,\zeta}\|_{L^2(P_X)}^2
		=
		a^2\sum_{j\in S}\zeta_j^2
		=
		a^2k
		=
		\Delta_n.
	\end{equation}
	Since \(f_{S,\zeta} \in \mathrm{span}\{\psi_{M,1},\ldots,\psi_{M,R}\}\) and   $\psi_{M,\ell} \in \mathrm{span}\{\phi_{X,1},\ldots,\phi_{X,M}\}$ for $1\leq \ell \leq R$,
\begin{equation}
	\label{equ: lower source control directions}
	\begin{aligned}
		\left\|\mathcal L_{K,X}^{-r}f_{S,\zeta}\right\|_{L^2(P_X)}^2
		&\leq\left(\min_{1\leq j\leq M}\sigma_{X,j}\right)^{-2r} 
		\|f_{S,\zeta}\|_{L^2(P_X)}^2\\[2mm]
		&
		\lesssim  M^{2r/s}\|f_{S,\zeta}\|_{L^2(P_X)}^2 \leq c_s^{2r/s} \leq 1.
	\end{aligned}
\end{equation} 
where the second step follows from the eigenvalue decay condition in Assumption~\ref{assumption: eigenvalues and eigenfunctions}, and the last step is satisfied for every $S,\zeta$ by choosing \(c_s\) sufficiently small. 
	The bound in \eqref{equ: cpsi} and $k\Delta_n\leq c_b$ give
	\begin{equation}
		\label{equ: constructed perturbation Linfty norm}
		\begin{aligned}
			\|f_{S,\zeta}\|_{L^\infty(P_X)}
			&\leq
			a\sum_{j\in S}\|\psi_{M,j}\|_{L^\infty(P_X)}
		\leq
			akC_\psi
			=
			C_\psi\sqrt{k\Delta_n}
			\leq
			C_\psi\sqrt{c_b}.
		\end{aligned}
	\end{equation}
	Choose $c_b$ sufficiently small that
	\begin{equation}
		\label{equ: constant conditions}
			C_\psi\sqrt{c_b}\leq\min\{B,1/2\},
		\qquad
		C_\psi\sqrt{c_b}\,\mathbb E M_H(X)
		\leq\frac14\lambda_{\min}(Q_0).
	\end{equation}
	
	Based on \eqref{equ: constructed perturbation L2 norm}, \eqref{equ: lower source control directions}, \eqref{equ: constructed perturbation Linfty norm}, and \eqref{equ: constant conditions}, it follows that, uniformly over $S,\zeta$,
	\[
	f_{S,\zeta}\in\mathcal F_{r,B},
	\qquad
	f_{S,\zeta}\perp\mathcal T_0,
	\qquad
	\|f_{S,\zeta}\|_{L^2(P_X)}^2=\Delta_n.
	\] 
	Furthermore,	for each $S,\zeta$, consider
	\begin{equation}
		\label{equ: constructed alternatives}
			Y=g(X,\theta_0)+f_{S,\zeta}(X)+e,
		\qquad
		\mathbb E(e\mid X)=0,
		\qquad
		\mathbb E e^2<\infty.
	\end{equation}
	The  condition
	\eqref{equ: lower perturbation amplitude} follows from \eqref{equ: constructed perturbation Linfty norm} and  \eqref{equ: constant conditions}. 	Use the fixed radius $\varepsilon_0$ in
	\eqref{equ: fixed radius global separation}, for which
	\eqref{equ: lower local quadratic bound} holds.
	Since this radius is independent of $n,S,\zeta$ and $\|f_{S,\zeta}\|_{L^2(P_X)}
	=
	\sqrt{\Delta_n}
	\to 0$,
	we have, for all sufficiently large $n$,
	\[
	\|f_{S,\zeta}\|_{L^2(P_X)}
	<
	\frac12
	\inf_{\substack{\theta\in\Theta\\
			\|\theta-\theta_0\|_2\geq\varepsilon_0}}
	\|g(\cdot,\theta)-g(\cdot,\theta_0)\|_{L^2(P_X)}
	\]
	simultaneously for every $S,\zeta$.
	Thus \eqref{equ: lower perturbation global separation}
	also holds uniformly over the constructed perturbations.
	Lemma~\ref{lemma: exact population projection} now gives
	$\theta^\star=\theta_0$ uniquely and
	\[
	\mathbf f(\cdot)
	=
	m-g(\cdot,\theta^\star)
	=
	f_{S,\zeta}(\cdot)
	\quad\text{in }L^2(P_X).
	\]
	Consequently,
	\[
	\inf_{\theta\in\Theta}
	\|m-g(\cdot,\theta)\|_{L^2(P_X)}^2
	=
	\|f_{S,\zeta}\|_{L^2(P_X)}^2
	=
	\Delta_n.
	\]

 \medskip
 \noindent\underline{\textit{Step 3: Choice of a Bayesian prior.}}
 We now work in the Gaussian regression model
 \[
 Y=m(X)+e,\qquad e\mid X\sim N(0,1).
 \]
 Let \(P_m\) denote the law of   \((X,Y)\) given a fixed regression function $m$.
 Define the null prior by
 \[
 \Pi_0\{m=g(\cdot,\theta_0)\}=1.
 \]
 For the alternative prior, start with the uniform draws of
 \(S,\zeta\) used in \emph{Step 2}.  Since $	\mathbb E\zeta_j=0, 	\mathbb E\zeta_j^2
 =
 \mathbb E\zeta_j^4=1,$  direct calculation yields
 \[
 \begin{aligned}
 	\mathbb E_\zeta\left[
 	\left\{\sum_{j\in S}\zeta_j\psi_{M,j}(x)\right\}^4
 	\right] &=
 	\sum_{j\in S}\psi^4_{M,j}(x)
 	+
 	6\sum_{\substack{j,\ell\in S\\j<\ell}}
 	\psi^2_{M,j}(x)\psi^2_{M,\ell}(x)\\
 	& =
 	3\left\{\sum_{j\in S}\psi_{M,j}^2(x)\right\}^2
 	-
 	2\sum_{j\in S}\psi_{M,j}^4(x)\\
 	& \leq
 	3\left\{\sum_{j\in S}\psi_{M,j}^2(x)\right\}^2.
 \end{aligned}
 \] 
 Since \(|S|=k\) and
 \(\|\psi_{M,j}\|_{L^\infty(P_X)}\leq C_\psi\) by \eqref{equ: cpsi}, we also have
 \[
 \left\{\sum_{j\in S}\psi_{M,j}^2(x)\right\}^2
 \leq
 \left(kC_\psi^2\right)^2
 =
 k^2C_\psi^4.
 \]
 Therefore, a direct calculation based on   \eqref{equ: constructed perturbation} gives,
 \[
 \begin{aligned}
 	\mathbb E_{S,\zeta}
 	\|f_{S,\zeta}\|_{L^4(P_X)}^4
 	&=
 	a^4\mathbb E_S\int
 	\mathbb E_\zeta\left[
 	\left\{\sum_{j\in S}\zeta_j\psi_{M,j}(x)\right\}^4
 	\right]\,dP_X(x)\\
 	&\leq
 	3a^4\mathbb E_S\int
 	\left\{\sum_{j\in S}\psi_{M,j}^2(x)\right\}^2
 	\,dP_X(x)\\
 	&\leq
 	3a^4\mathbb E_S\int k^2C_\psi^4\,dP_X(x)\\
 	&=
 	3C_\psi^4a^4k^2
 	=
 	3C_\psi^4\Delta_n^2,
 \end{aligned}
 \]
 where the last equality follows from \(a^2k=\Delta_n\). Both separation regimes in Theorem~\ref{theorem: minimax optimality} imply
 $\sqrt n\,\Delta_n\to0$.
 Consequently, Markov's inequality yields
 \begin{equation}
 	\label{equ: restricted probability to one}
 	\mathbb P_{S,\zeta}\left(
 	\|f_{S,\zeta}\|_{L^4(P_X)}^4
 	>\frac{\Delta_n}{\sqrt n}
 	\right)
 	\leq3C_\psi^4\sqrt n\,\Delta_n=o(1).  
 \end{equation}
 {
 	Define $\Pi_{1n}$ by conditioning these draws on the event $\{(S,\zeta):
 	\|f_{S,\zeta}\|_{L^4(P_X)}^4\leq\Delta_n/\sqrt n\}$ :
 \[
 \Pi_{1n}\{m=g(\cdot,\theta_0)+f_{S,\zeta}\}
 =p_{S,\zeta},
 \]
 where
 \[
 p_{S,\zeta}
 =
 \frac{
 	\mathds 1\left\{
 	\|f_{S,\zeta}\|_{L^4(P_X)}^4
 	\leq\Delta_n/\sqrt n
 	\right\}
 }{
 	\binom Rk\,2^k\,
 	\mathbb P_{S,\zeta}\left(
 	\|f_{S,\zeta}\|_{L^4(P_X)}^4
 	\leq\Delta_n/\sqrt n
 	\right).
 }
 \]
 This probability mass is well defined for all sufficiently large $n$ by \eqref{equ: restricted probability to one}.}
 Let \(p_m(\mathbf y,\mathbf x)\) denote the density of
 \(P_m^{\otimes n}\) with respect to
 \(d\mathbf y\,dP_X^{\otimes n}(\mathbf x)\).  Under the Gaussian model,
 \begin{equation}
 	\label{equ: pm}
 	p_m(\mathbf y,\mathbf x)
 	=
 	(2\pi)^{-n/2}
 	\exp\left[
 	-\frac12\sum_{i=1}^n\{Y_i-m(X_i)\}^2
 	\right].
 \end{equation}
 Define
 \[
 p_0=\int p_m\,d\Pi_0(m),
 \qquad
 p_{1n}=\int p_m\,d\Pi_{1n}(m),
 \qquad
 Z_n=\frac{p_{1n}}{p_0}.
 \]
 Since the covariate distribution is common to all regression
 functions,
 \[
 Z_n
 =
 \frac{p_{1n}(\mathbf y,\mathbf x)}
 {p_0(\mathbf y,\mathbf x)}
 =
 \frac{p_{1n}(\mathbf y\mid\mathbf x)}
 {p_0(\mathbf y\mid\mathbf x)}.
 \] 
  Write $\mathbb E_0$ for expectation under $p_0$. Let
 	$T_n$ be  any  asymptotically  $\alpha$-level test. 
 	Since $\Pi_0$ is concentrated on $g(\cdot,\theta_0)$, we have 
 	\begin{equation}
 		\label{equ: lower bound}
 		\begin{aligned}
 			&\quad \,\, P_{g(\cdot,\theta_0)}^{\otimes n}(T_n=1)
 			+\sup_{S,\zeta}
 			P_{g(\cdot,\theta_0)+f_{S,\zeta}}^{\otimes n}(T_n=0)
 			\\
 			& \geq
 			\int P_m^{\otimes n}(T_n=1)\,d\Pi_0(m)
 			+\int P_m^{\otimes n}(T_n=0)\,d\Pi_{1n}(m)
 			\\
 			& =
 			\mathbb E_0\{T_n+Z_n(1-T_n)\}
 			\\
 			& \geq
 			\mathbb E_0\min(1,Z_n)
 			\\
 			&  	=
 			1-\frac12\mathbb E_0|Z_n-1|.
 		\end{aligned}
 	\end{equation}
 	Since $\mathbb E_0Z_n=1$, Cauchy--Schwarz gives
 	\begin{equation}
 		\label{equ: condition tranformation}
 		\mathbb E_0|Z_n-1|
 		\leq
 		\{\mathbb E_0(Z_n-1)^2\}^{1/2}
 		=
 		\{\mathbb E_0(Z_n^2)-1\}^{1/2}.
 	\end{equation} 
 	All constructed alternatives belong to
 	$H_1(\mathcal F_{r,B},\Delta_n)$ for sufficiently large $n$.
 	Consequently,
 	\eqref{equ: lower bound}
 	and~\eqref{equ: condition tranformation} imply
 	\[
 	\sup_{H_1(\mathcal F_{r,B},\Delta_n)}
 	\mathbb P(T_n=0)
 	\geq
 	1-P_{g(\cdot,\theta_0)}^{\otimes n}(T_n=1)
 	-\frac12
 	\{\mathbb E_0(Z_n^2)-1\}^{1/2}.
 	\]
 If we can establish
 \begin{equation}
 	\label{equ: squared lr}
 	\mathbb E_0(Z_n^2)=1+o(1),
 \end{equation}
 then
 \[
 \begin{aligned}
 	\liminf_{n\to\infty}
 	\sup_{H_1(\mathcal F_{r,B},\Delta_n)}
 	\mathbb P(T_n=0)
 	&\geq
 	1-\limsup_{n\to\infty}
 	P_{g(\cdot,\theta_0)}^{\otimes n}(T_n=1) \geq1-\alpha.
 \end{aligned}
 \]
 We verify~\eqref{equ: squared lr} in the next step.
 	
 	\medskip
 \noindent\underline{\textit{Step 4: Study the Gaussian likelihood ratio.}}
 For \(m(\cdot)=g(\cdot,\theta_0)+f_{S,\zeta}(\cdot)\), it follows from \eqref{equ: pm} that 
 \[
 \begin{aligned}
 	p_m(\mathbf{y}\mid \mathbf{x}) &= (2\pi)^{-n/2} \exp\left[-\frac{1}{2}\sum_{i=1}^n\{Y_i - m(X_i)\}^2 \right]\\
 	&= (2\pi)^{-n/2} \exp\left[-\frac{1}{2}\sum_{i=1}^n\{Y_i -g(X_i,\theta_0) -f_{S,\zeta}(X_i)\}^2 \right]\\
 	&= (2\pi)^{-n/2} \exp\left[-\frac{1}{2}\sum_{i=1}^n\{ \epsilon(Z_i,\theta_0)\}^2 \right]\exp\left[-\frac{1}{2} \sum_{i=1}^n f_{S,\zeta}^2(X_i) + \sum_{i=1}^n \epsilon(Z_i,\theta_0)f_{S,\zeta}(X_i)  \right]\\
 	&= p_0(\mathbf{y} \mid \mathbf{x})\cdot \exp\left[-\frac{1}{2} \sum_{i=1}^n f_{S,\zeta}^2(X_i) + \sum_{i=1}^n \epsilon(Z_i,\theta_0)f_{S,\zeta}(X_i)  \right].
 \end{aligned}
 \]
 Consequently,
 \[
 Z_n=\int
 \exp\left\{\sum_{i=1}^n\epsilon(Z_i,\theta_0)f_{S,\zeta}(X_i)
 -\frac12\sum_{i=1}^nf^2_{S,\zeta}(X_i)\right\}\,d\Pi_{1n}(m).
 \]
 
 Under the null, the \(\epsilon(Z_i,\theta_0)\) are conditionally
 independent standard Gaussian variables. 
 Using the  Gaussian moment-generating
 function $\mathbb{E}\exp{tX} = \exp(t^2/2)$ for $X\sim N(0,1)$, direct calculation yields 
 \begin{equation}
 	\label{equ: Gaussian likelihood pair identity}
 	\begin{aligned}
 		\mathbb E_0(Z_n^2\mid\mathbf x)
 		&=\iint\exp\left\{\sum_{i=1}^nf(X_i)f'(X_i)\right\}
 		\,d\Pi_{1n}(m)\,d\Pi_{1n}(m'),\\
 		\mathbb E_0(Z_n^2)
 		&=\iint\left[\mathbb E_X\exp\{f(X)f'(X)\}\right]^n
 		\,d\Pi_{1n}(m)\,d\Pi_{1n}(m').
 	\end{aligned}
 \end{equation}
 where \(m(\cdot)=g(\cdot,\theta_0)+f\) and
 \(m'(\cdot)=g(\cdot,\theta_0)+f'\) denote two independent prior
 draws. Here, we write $f = f_{S,\zeta}$ and $f^{\prime} = f_{S^{\prime},\zeta^{\prime}}$ for short. 
 By \eqref{equ: constructed perturbation Linfty norm} and \eqref{equ: constant conditions}, we have \(\|f\|_{L^{\infty}(P_X)}, \|f^{\prime}\|_{L^{\infty}(P_X)} \leq 1/2\).
 Using \(e^t\leq1+t+Ct^2\) for \(|t|\leq1/4\) gives
 \[
 \mathbb E_X\exp\{f(X)f'(X)\}
 \leq1+\langle f,f'\rangle_{L^2(P_X)}
 +C\int f(x)^2f'(x)^2\,dP_X(x),
 \]
 where $C>0$ is a fixed constant.
 Cauchy--Schwarz and the prior restriction imply
 \[
 \int f(x)^2f'(x)^2\,dP_X(x)
 \leq\|f\|_{L^4(P_X)}^2\|f'\|_{L^4(P_X)}^2
 \leq\frac{\Delta_n}{\sqrt n}.
 \]
 Thus
 \begin{equation}
 	\label{equ: Gaussian likelihood pair identity inner part}
 	\left[\mathbb E_X\exp\{f(X)f'(X)\}\right]^n
 	\leq\exp\left\{
 	n\langle f,f'\rangle_{L^2(P_X)}+C\sqrt n\,\Delta_n\right\}.
 \end{equation} 
 By \eqref{equ: restricted probability to one}, the two conditioned priors introduce a normalization factor at most
 \((1-3C_\psi^4\sqrt n\,\Delta_n)^{-2}\).
 After dropping their restriction indicators from the nonnegative
 integral, we get from  \eqref{equ: Gaussian likelihood pair identity} and \eqref{equ: Gaussian likelihood pair identity inner part} that
 \[
 \begin{aligned}
 	\mathbb E_0(Z_n^2)
 	&\leq
 	\frac{\exp(C\sqrt n\,\Delta_n)}
 	{(1-3C_\psi^4\sqrt n\,\Delta_n)^2} \cdot
 	\mathbb E_{S,\zeta,S',\zeta'}
 	\exp\{n\langle f_{S,\zeta},f_{S',\zeta'}\rangle_{L^2(P_X)}\}\\
 	& = \frac{\exp(C\sqrt n\,\Delta_n)}
 	{(1-3C_\psi^4\sqrt n\,\Delta_n)^2} \cdot
 	\mathbb E_{S,\zeta,S',\zeta'}
 	\exp\left\{n a^2\sum_{j\in S\cap S'}\zeta_j\zeta_j'\right\}\\
 	& = \frac{\exp(C\sqrt n\,\Delta_n)}
 	{(1-3C_\psi^4\sqrt n\,\Delta_n)^2} \cdot
 	\mathbb E_{S,S'}\left[ \left\{\cosh(na^2)\right\}^{|S\cap S'|}\right].
 \end{aligned}
 \]
 where the second step follows from the orthonormality of $\{\psi_{M,\ell}, 1\leq \ell \leq R\}$, and the last step follows from the independence of the Rademacher random variables. 
 Here, $\cosh(x) = (e^{x} + e^{-x})/2$.
 For \(t\geq1\), the hypergeometric intersection satisfies   
      	\[
      	\begin{aligned}
 	     		\mathbb E\{t^{|S\cap S'|}\}
 	     		&=\sum_{j=0}^k(t-1)^j
 	     		\binom{k}{j}\frac{k(k-1)\ldots(k-j+1)}{R(R-1)\ldots(R-j+1)}\\
 	     		&\leq\{1+(k/R)(t-1)\}^k
 	     		\leq\exp\{(k^2/R)(t-1)\},
 	     	\end{aligned}
      	\]
 Taking \(t=\cosh(na^2)\) yields
 \begin{equation}
 	\label{equ: lower sparse likelihood bound}
 	\mathbb E_0(Z_n^2)
 	\leq\frac{\exp(C\sqrt n\,\Delta_n)}
 	{(1-3C_\psi^4\sqrt n\,\Delta_n)^2}
 	\exp\left[
 	\frac{k^2}{R}\left\{\cosh\left(\frac{n\Delta_n}{k}\right)-1\right\}
 	\right].
 \end{equation}
 
 If \(r\geq s/4\), then \(s/(2r)\leq2\),
 \(R\asymp\Delta_n^{-s/(2r)}\), and
 \(k^{-1}\lesssim R^{-1}+\Delta_n\). Thus
 \[
 \frac{n\Delta_n}{k}
 \lesssim n\Delta_n^{1+s/(2r)}+n\Delta_n^2=o(1)
 \]
 under the  case of $\Delta_n = o(n^{-4r/(4r+s)})$.
 Using \(\cosh(t)-1\lesssim t^2\) for \(|t|\leq1\),
 \eqref{equ: lower sparse likelihood bound} gives
 \[
 \mathbb E_0(Z_n^2)
 \leq\frac{\exp\left\{C\sqrt n\,\Delta_n
 	+C_1n^2\Delta_n^{2+s/(2r)}\right\}}
 {\left(1-3C_\psi^4\sqrt n\,\Delta_n\right)^2}=1+o(1),
 \]
 where $C_1>0$ is a constant independent of $n$.
 
 If \(0<r<s/4\), then \(s/(2r)>2\) and
 \(k\asymp\Delta_n^{-1}\). If \(\Delta_n=o(n^{-1/2})\), then
 \(n\Delta_n/k\lesssim n\Delta_n^2=o(1)\). The hypergeometric exponent
 in \eqref{equ: lower sparse likelihood bound} is therefore at most
 \[
 C_1n^2\Delta_n^{2+s/(2r)}=o(1),
 \]
 since \(n^2(n^{-1/2})^{2+s/(2r)}=n^{1-s/(4r)}\to0\).
 Also \(\sqrt n\,\Delta_n=o(1)\), so \(\mathbb E_0Z_n^2=1+o(1)\). The proof is complete.
 %
 
	\section{Additional Lemmas, Proofs and Examples}
	\label{sec: addtional lemmas and proofs}
	\begin{proof}[Proof of Proposition~\ref{prop: computation}]
		For Tikhonov regularization,
		$T_{D,\lambda}=(L_{K,D}+\lambda I)^{-1}$.
		Define the scaled sampling operator
		$S_D:\mathcal H_K\to\mathbb R^N$ by
		\[
		S_Df
		:=
		N^{-1/2}
		\bigl(f(X_{01}),\ldots,f(X_{0N})\bigr)^\top,
		\qquad f\in\mathcal H_K.
		\]
		Its adjoint is
		\[
		S_D^\ast\mathbf a
		=
		N^{-1/2}\sum_{\ell=1}^N a_\ell K_{X_{0\ell}},
		\qquad
		\mathbf a=(a_1,\ldots,a_N)^\top\in\mathbb R^N.
		\]
		Consequently, $	S_D^\ast S_D=L_{K,D}$ and $
		S_DS_D^\ast=N^{-1}K_{NN}.$ Applying the functional-calculus identity in \eqref{equ: functional calculus identity} 
		with $A=S_D^\ast$ and $h(t)=(t+\lambda)^{-1}$ gives
		\[
		(S_D^\ast S_D+\lambda I)^{-1}S_D^\ast
		=
		S_D^\ast(S_DS_D^\ast+\lambda I)^{-1}.
		\]
		Furthermore, taking $A = L_{K,D} +\lambda I$ and $B = \lambda I$ in the equation
		$	A^{-1}   = B^{-1}+ A^{-1}(B-A)B^{-1}$,  we obtain
		\[
		\begin{aligned}
			T_{D,\lambda}
			&=(L_{K,D}+\lambda I)^{-1}\\
			&=\lambda^{-1}
			\left\{
			I-(L_{K,D}+\lambda I)^{-1}L_{K,D}
			\right\}\\
			&=\lambda^{-1}
			\left\{
			I-S_D^\ast(S_DS_D^\ast+\lambda I)^{-1}S_D
			\right\}\\
			&=\lambda^{-1}
			\left\{
			I-S_D^\ast
			(N^{-1}K_{NN}+\lambda I)^{-1}S_D
			\right\}.
		\end{aligned}
		\]	Thus, for any $f,h\in\mathcal H_K$,
		\begin{equation}
			\label{equ: inner product transformation}
			\begin{aligned}
				\langle T_{D,\lambda}f,h\rangle_K
				= 
				\lambda^{-1}\langle f,h\rangle_K -
				\lambda^{-1}(S_Df)^\top
				(N^{-1}K_{NN}+\lambda I)^{-1}S_Dh.
			\end{aligned}
		\end{equation}
		Since $	S_DK_{X_i}
		=
		N^{-1/2} (K(X_{01},X_i),\ldots,K(X_{0N},X_i) )^\top$, 
		substituting $f=K_{X_i}$ and $h=K_{X_j}$ into
		\eqref{equ: inner product transformation}, and using
		$(N^{-1}K_{NN}+\lambda I)^{-1}
		=N(K_{NN}+N\lambda I)^{-1}$, gives
		\[
		\begin{aligned}
			\left\langle
			T_{D,\lambda}^{1/2}K_{X_i},
			T_{D,\lambda}^{1/2}K_{X_j}
			\right\rangle_K =
			\langle T_{D,\lambda}K_{X_i},K_{X_j}\rangle_K=
			\lambda^{-1}
			\left[
			K_{nn}
			-K_{nN}(K_{NN}+N\lambda I)^{-1}K_{nN}^\top
			\right]_{ij}.
		\end{aligned}
		\]
		Therefore, with $	R_n
		=
		\operatorname{diag}\{
		\epsilon(Z_1,\widehat\theta),\ldots,
		\epsilon(Z_n,\widehat\theta)
		\},$
		the matrix   $M_n$ satisfies
		\[
		(M_n)_{ij}
		=
		\left\langle
		T_{D,\lambda}^{1/2}K_{X_i},
		T_{D,\lambda}^{1/2}K_{X_j}
		\right\rangle_K
		\epsilon(Z_i,\widehat\theta)
		\epsilon(Z_j,\widehat\theta).
		\]
		It then follows from
		\eqref{equ: totally empirical statistic} that
		\[
		\begin{aligned}
			n(n-1)\widehat\eta_\lambda(\widehat\theta)
			=
			\sum_{i\ne j}(M_n)_{ij} =
			\mathbf1_n^\top M_n\mathbf1_n
			-\operatorname{Tr}(M_n).
		\end{aligned}
		\]
		Dividing both sides by $n(n-1)$ proves the result.
	\end{proof}

	\begin{proof}[Proof of Lemma~\ref{lemma: similarity expectation shift}] Recall that $\mathscr{P}_{D,\lambda}$ and $\mathscr{Q}_{D,\lambda}$ are defined in \eqref{equ: PDL} and \eqref{equ: QDL}.  We first consider $\mathscr{Q}_{D,\lambda}$.
		By \eqref{equ: Tb12Kx} with $b = 0$,
		and the equivalence of $P_0$ and $P_X$ in Assumption~\ref{assumption: auxiliary covariate distribution},
		\[
		\operatorname*{ess\,sup}_{X\sim P_0}
		\left\|T_\lambda^{1/2}K_X\right\|_K^2
		\lesssim\lambda^{-s}.
		\]
		Note that
		$L_{K,0}$ is positive trace class, with
		$\operatorname{Tr}(L_{K,0}) \leq\kappa^2$.
		Applying the Bernstein inequality in
		\citet[Proposition~6]{rudirosasco2017}
		to the independent RKHS elements
		$\{K_{X_{0\ell}}\}_{\ell=1}^N$,
		and taking $\delta=e^{-t}/2$,
		we obtain that,
		for all sufficiently small $\lambda>0$ and every $t\geq0$,
		\begin{equation}
			\label{equ: similarity Bernstein tail shift}
			\mathbb P\left[
			\mathscr Q_{D,\lambda}>
			C\left\{
			\sqrt{\frac{\log(C/\lambda)+t}{N\lambda^s}}
			+\frac{\log(C/\lambda)+t}{N\lambda^s}
			\right\}
			\right]
			\leq e^{-t}.
		\end{equation}
		where $C>0$ is a  fixed constant. 
		Integrating this tail bound gives
		\[
		\begin{aligned}
			\mathbb E(\mathscr Q_{D,\lambda}^u)
			&\leq
			C_u\int_0^\infty
			\left[
			\left\{
			\frac{\log(C/\lambda)+t}{N\lambda^s}
			\right\}^{u/2}
			+
			\left\{
			\frac{\log(C/\lambda)+t}{N\lambda^s}
			\right\}^{u}
			\right]e^{-t}\,dt\\
			&\leq
			C_u\left[
			\left\{
			\frac{\log(C/\lambda)}{N\lambda^s}
			\right\}^{u/2}
			+
			\left\{
			\frac{\log(C/\lambda)}{N\lambda^s}
			\right\}^{u}
			\right]\\
			&=
			O\!\left[
			\left\{
			\frac{\log(1/\lambda)}{N\lambda^s}
			\right\}^{u/2}
			\right],
		\end{aligned}
		\]
		where the last step uses
		$\log(C/\lambda)\asymp\log(1/\lambda)$ and
		$\log(1/\lambda)/(N\lambda^s)\to0$. 
		We next control $\mathscr P_{D,\lambda}$.
		By the definition in \eqref{equ: TLTDL},
		\[
		T_\lambda^{1/2}(L_{K,D}+\lambda I)T_\lambda^{1/2}
		=
		I-T_\lambda^{1/2}
		(L_{K,0}-L_{K,D})T_\lambda^{1/2}.
		\]
		Consequently, it follows from~\eqref{equ: PDL} that
		\[
		\begin{aligned}
			\mathscr P_{D,\lambda}^2
			=
			\|T_\lambda^{-1/2}T_{D,\lambda}T_\lambda^{-1/2}\| =
			\left\|
			\left\{
			I-T_\lambda^{1/2}
			(L_{K,0}-L_{K,D})T_\lambda^{1/2}
			\right\}^{-1}
			\right\|.
		\end{aligned}
		\]
		Thus $\mathscr P_{D,\lambda}^2\leq2$ on
		$\{\mathscr Q_{D,\lambda}\leq1/2\}$.
		On the other hand, we also have
		\[
		\mathscr P_{D,\lambda}^2
		\leq 
		\frac{\|L_{K,0}\|+\lambda}{\lambda}
		\leq1+\frac{\kappa^2}{\lambda}.
		\] 
		Choose a fixed $c>0$ sufficiently small that
		$C(\sqrt{2c}+2c)\leq1/2$.
		Since $N\lambda^s/\log(1/\lambda)\to\infty$,
		we have $\log(C/\lambda)\leq cN\lambda^s$ eventually.
		Taking $t=cN\lambda^s$ in
		\eqref{equ: similarity Bernstein tail shift} therefore yields
		\[
		\mathbb P(\mathscr Q_{D,\lambda}>1/2)
		\leq e^{-cN\lambda^s}.
		\]
		By the total expectation formula,
		\[
		\begin{aligned}
			\mathbb E(\mathscr P_{D,\lambda}^u)
			& = \mathbb E(\mathscr P_{D,\lambda}^u \mid \mathscr Q_{D,\lambda}>1/2)	\mathbb P(\mathscr Q_{D,\lambda}>1/2) + \mathbb E(\mathscr P_{D,\lambda}^u \mid \mathscr Q_{D,\lambda}\leq 1/2)	\mathbb P(\mathscr Q_{D,\lambda}\leq 1/2)  \\
			&\leq
			2^{u/2}
			+
			\left(1+\frac{\kappa^2}{\lambda}\right)^{u/2}
			\mathbb P(\mathscr Q_{D,\lambda}>1/2) \leq
			2^{u/2}
			+
			\left(1+\frac{\kappa^2}{\lambda}\right)^{u/2}
			e^{-cN\lambda^s}
			=O(1),
		\end{aligned}
		\]
		because $N\lambda^s/\log(1/\lambda)\to\infty$.
		This completes the proof.
	\end{proof}
	\begin{proof}[Proof of Lemma~\ref{lemma: general one sided framework power}]
		Fix $M>0$. By
		\eqref{equ: general one sided separation power}, for all
		sufficiently large $n$,
		\[
		c_0\|\mathbf f\|_{L^2(P_X)}^2
		>
		M(\mathcal U_1+\mathcal U_2)
		\qquad
		\text{for every }\mathbf f\in H_1(\mathcal F,\Delta_n).
		\]
		On the nonrejection event
		$\{\phi_{S_\lambda}=0\}
		=\{\widehat S_{n,\lambda}
		\leq\widehat c_{\alpha,\lambda,n}\}$,
		\[
		\begin{aligned}
			c_0\|\mathbf f\|_{L^2(P_X)}^2
			&=
			c_0\|\mathbf f\|_{L^2(P_X)}^2
			-\widehat S_{n,\lambda}
			+\widehat S_{n,\lambda}\leq
			\left\{
			c_0\|\mathbf f\|_{L^2(P_X)}^2
			-\widehat S_{n,\lambda}
			\right\}_+
			+
			(\widehat c_{\alpha,\lambda,n})_+.
		\end{aligned}
		\]
		Consequently, for all sufficiently large $n$,
		\[
		\begin{aligned}
			\{\phi_{S_\lambda}=0\}
			\subseteq&
			\left\{
			\left\{
			c_0\|\mathbf f\|_{L^2(P_X)}^2
			-\widehat S_{n,\lambda}
			\right\}_+
			>M\mathcal U_1
			\right\}\cup
			\left\{
			(\widehat c_{\alpha,\lambda,n})_+
			>M\mathcal U_2
			\right\}.
		\end{aligned}
		\]
		The union bound therefore gives
		\[
		\begin{aligned}
			\limsup_{n\to\infty}
			\sup_{H_1(\mathcal F,\Delta_n)}
			\mathbb P(\phi_{S_\lambda}=0)
			&\leq
			\limsup_{n\to\infty}
			\sup_{H_1(\mathcal F,\Delta_n)}
			\mathbb P\!\left[
			\left\{
			c_0\|\mathbf f\|_{L^2(P_X)}^2
			-\widehat S_{n,\lambda}
			\right\}_+
			>M\mathcal U_1
			\right]\\
			&+
			\limsup_{n\to\infty}
			\sup_{H_1(\mathcal F,\Delta_n)}
			\mathbb P\!\left[
			(\widehat c_{\alpha,\lambda,n})_+
			>M\mathcal U_2
			\right].
		\end{aligned}
		\]
		Letting $M\to\infty$ and applying
		\eqref{equ: general downward bound power}
		and~\eqref{equ: general critical bound power}, we obtain
		\[
		\limsup_{n\to\infty}
		\sup_{H_1(\mathcal F,\Delta_n)}
		\mathbb P(\phi_{S_\lambda}=0)=0.
		\]
		The proof is complete.
	\end{proof}
	
	The following two lemmas collect the moment and operator
	bounds used in our proofs.
	In particular, Lemma~\ref{lemma: empirical kernel row shift} relates $L_{K,0}$ to
	$L_{K,X}$ and controls the regularized embeddings and kernels,
	with $\mathscr P_{D,\lambda}$ and $\mathscr Q_{D,\lambda}$
	tracking the effect of operator estimation.
	These estimates are used in the error analysis of
	Proposition~\ref{prop: error decomposition}
	and in establishing asymptotic normality and
	variance-estimator consistency in
	Theorem~\ref{theorem: asymptotic null distribution}.

	\begin{lemma}
		\label{lemma: residual moments at pseudo true}  
		Under Assumption~\ref{assumption: conditional variance}, if
		\(\|\mathbf f\|_{L^\infty(P_X)}\leq B\) for some $B >0$, then
		\begin{equation}
			\label{equ: residual moments}
			c_2\leq s_2(x,\theta^\star)\leq C_2,
			\qquad
			s_4(x,\theta^\star)\leq C_4,
			\qquad P_X\text{-a.e. }x.
		\end{equation}
		for   constants $c_2,C_2,C_4>0$ depending only on $c_e,C_e, B$. Here,  $s_{k}(x,\theta) :=
		\mathbb E\{|\epsilon(Z,\theta)|^k\mid X=x\}$.  
	\end{lemma}
	
	\begin{proof}
		Recall that $\epsilon(Z,\theta) = Y -g(X,\theta)$ and $e = Y - m(X)$. Then the identity \(\epsilon(Z,\theta^\star)=e+\mathbf f(X)\) and
		\(\mathbb E(e\mid X)=0\) give
		\[
		s_2(x,\theta^\star)=\mathbb E(e^2\mid X=x)+\mathbf f^2(x).
		\]
		The  second-moment bounds in \eqref{equ: residual moments} follow from \eqref{equ: conditional noise moments}  by setting $c_2 := c_e$ and $C_2 : = C_e + B^2$. Moreover, the inequality
		\(|a+b|^4\leq8(|a|^4+|b|^4)\) gives
		\[
		s_4(x,\theta^\star)
		\leq8\{\mathbb E(e^4\mid X=x)+|\mathbf f(x)|^4\}
		\leq8(C_{e}^2+B^4)=:C_4.
		\]
		This completes the proof.
	\end{proof}

	\begin{lemma}[Spectral, operator, and kernel moment bounds]
		\label{lemma: spectral quantities shift}
		\label{lemma: null kernel moments shift}
		\label{lemma: tikhonov comparison shift}
		\label{lemma: empirical kernel row shift}
		Suppose that
		Assumptions~\ref{assumption: auxiliary covariate distribution},
		\ref{assumption: conditional variance},
		and~\ref{assumption: eigenvalues and eigenfunctions} hold,
		and $0<\lambda<1$.
		Recall $T_\lambda$ and $T_{D,\lambda}$ in~\eqref{equ: TLTDL},
		and $\mathscr P_{D,\lambda}$ and $\mathscr Q_{D,\lambda}$
		in~\eqref{equ: PDL} and \eqref{equ: QDL}.
		Then
		\begin{equation}
			\underline c L_{K,X}
			\preceq L_{K,0}
			\preceq \overline c L_{K,X}.
			\label{equ: two operators are comparable}
		\end{equation}
		Moreover, for every $h\in L^2(P_X)$,
		\begin{equation}
			\begin{aligned}
				\left\|T_{D,\lambda}^{1/2}I_{K,X}^{\ast}h\right\|_K
				\lesssim \mathscr P_{D,\lambda}
				\|h\|_{L^2(P_X)},\qquad 
				\operatorname*{ess\,sup}_{X\sim P_X}
				\left\|T_{D,\lambda}^{1/2}K_X\right\|_K^2
				\lesssim
				\mathscr P_{D,\lambda}^2\lambda^{-s}.
			\end{aligned}
			\label{equ: empirical feature bound shift}
		\end{equation}
		For $b\in\{0,D\}$, the kernels
		in~\eqref{equ: some kernel notation 1} satisfy
		\begin{align}
			\mathbb E\left\{
			A_{\lambda,b}^2(X_1,X_2)\mid D
			\right\}
			&\lesssim
			\mathscr P_{b,\lambda}^4\lambda^{-s},
			\label{equ: merged empirical kernel squared norm}\\
			\operatorname*{ess\,sup}_{X\sim P_X}
			\int A_{\lambda,b}^2(X,x)\,dP_X(x)
			&\lesssim
			\mathscr P_{b,\lambda}^4\lambda^{-s},
			\label{equ: empirical kernel row shift}\\
			\operatorname*{ess\,sup}_{X,X'\sim P_X}
			A_{\lambda,b}^2(X,X')
			&\lesssim
			\mathscr P_{b,\lambda}^4\lambda^{-2s},
			\label{equ: empirical kernel supremum squared shift}
		\end{align}
		where $\mathscr P_{0,\lambda}:=1$.
		In addition,
		\begin{equation}
			\mathbb E\left[
			\left\{
			A_{\lambda,D}(X_1,X_2)
			-A_{\lambda,0}(X_1,X_2)
			\right\}^2\mid D
			\right]
			\lesssim
			\mathscr P_{D,\lambda}^4
			\mathscr Q_{D,\lambda}^2\lambda^{-s}.
			\label{equ: merged kernel difference bound}
		\end{equation}
		
		If $\|\mathbf f\|_{L^\infty(P_X)}\leq B$
		for some fixed $B\geq0$, then
		\begin{equation}
			\begin{aligned}
				\left\|
				T_{D,\lambda}^{1/2}
				\mathbb E(\omega\otimes\omega)
				T_{D,\lambda}^{1/2}
				\right\|
				&\lesssim \mathscr P_{D,\lambda}^2,\\
				\operatorname{Tr}\left\{
				T_{D,\lambda}^{1/2}
				\mathbb E(\omega\otimes\omega)
				T_{D,\lambda}^{1/2}
				\right\}
				&\lesssim
				\mathscr P_{D,\lambda}^2\lambda^{-s},
			\end{aligned}
			\label{equ: covariance empirical shift}
		\end{equation}
		and the kernels
		in~\eqref{equ: some kernel notation 2} satisfy
		\begin{align}
			\mathbb E\left\{
			H_{\lambda,0}^2(Z_1,Z_2)
			\right\}
			&\asymp\lambda^{-s},
			\label{equ: EH squared shift}\\
			\mathbb E\left\{
			R_{\lambda,0}^2(Z_1,Z_2)
			\right\}
			&\lesssim\lambda^{-s},
			\label{equ: ER squared shift}\\
			\mathbb E\left\{
			H_{\lambda,0}^2(Z_1,Z_2)
			H_{\lambda,0}^2(Z_1,Z_3)
			\right\}
			&\lesssim\lambda^{-2s},
			\label{equ: shared fourth shift}\\
			\mathbb E\left\{
			H_{\lambda,0}^4(Z_1,Z_2)
			\right\}
			&\lesssim\lambda^{-3s}.
			\label{equ: full fourth shift}
		\end{align}
	\end{lemma}
	
	\begin{proof}
		For every $u\in\mathcal H_K$,
		\[
		\underline c\langle u,L_{K,X}u\rangle_K
		\leq
		\langle u,L_{K,0}u\rangle_K
		=
		\int u^2(X)w(X)\,dP_X(X)
		\leq
		\overline c\langle u,L_{K,X}u\rangle_K,
		\]
		which proves~\eqref{equ: two operators are comparable}. 
		The min--max characterization of eigenvalues of positive operators ~\citep{bhatia2013matrix} therefore gives 
		$\underline c\,\sigma_{X,k}
		\leq\sigma_{0,k}\leq\overline c\,\sigma_{X,k}$.
		With $m_\lambda:=\lceil\lambda^{-s}\rceil$,
		the eigenvalue decay in Assumption~\ref{assumption: eigenvalues and eigenfunctions} yields
		\begin{equation}
			\label{equ: effective dimension comparison}
			\begin{aligned}
				\lambda^{-s}
				&\lesssim
				\sum_{k\leq m_\lambda}
				\left(\frac{\sigma_{0,k}}{\sigma_{0,k}+\lambda}\right)^2
				\leq
				\sum_{k\geq1}
				\left(\frac{\sigma_{0,k}}{\sigma_{0,k}+\lambda}\right)^2 
				\leq  
				\sum_{k\geq1}
				\frac{\sigma_{0,k}}{\sigma_{0,k}+\lambda} \\
				&\leq
				m_\lambda+\lambda^{-1}\sum_{k>m_\lambda}\sigma_{0,k}
				\lesssim
				m_\lambda+\lambda^{-1}
				\int_{m_\lambda}^{\infty}t^{-1/s}\,dt
				\lesssim\lambda^{-s}.
			\end{aligned}
		\end{equation}
		The first inequality follows because
		$\sigma_{0,k}\gtrsim m_\lambda^{-1/s}\asymp\lambda$
		for $k\leq m_\lambda$, so each summand is bounded below
		by a positive constant.
		The fourth inequality follows  because
		$\sigma_{0,k}/(\sigma_{0,k}+\lambda)
		\leq\min\{1,\sigma_{0,k}/\lambda\}$.
		The last step follows since
		$\int_{m_\lambda}^{\infty}t^{-1/s}\,dt
		\asymp m_\lambda^{1-1/s}\asymp  \lambda^{1-s}$ for $0<s<1$.
		
		Based on \eqref{equ: two operators are comparable} and \eqref{equ: effective dimension comparison}, we have
		\begin{equation}
			\begin{aligned}
				\left\|T_\lambda^{1/2}L_{K,X}T_\lambda^{1/2}\right\|
				\leq\underline c^{-1} \left\|L_{K,0}T_{\lambda} \right\|	\leq\underline c^{-1}, 
				\qquad  
			\end{aligned}
			\label{equ: merged regularized operator bound 1}
		\end{equation}
		and
		\begin{equation}
			\label{equ: merged regularized operator bound 2} 	\operatorname{Tr}\left(T_\lambda^{1/2}L_{K,X}T_\lambda^{1/2}\right)	\leq\underline c^{-1} \mathrm{Tr}\left(L_{K,0}T_{\lambda}\right) =\underline c^{-1}	\sum_{k\geq1}
			\frac{\sigma_{0,k}}{\sigma_{0,k}+\lambda} \lesssim \lambda^{-s}.
		\end{equation}
		Write $T_{0,\lambda}:=T_\lambda$.
		For $b\in\{0,D\}$, 
		\[
		\begin{aligned}
			\left\|T_{b,\lambda}^{1/2}I_{K,X}^{\ast}h\right\|_K^2 & = 	\left\|T_{b,\lambda}^{1/2}T_{\lambda}^{-1/2}T_\lambda^{1/2}I_{K,X}^{\ast}h\right\|^2\\
			&\leq
			\mathscr P_{b,\lambda}^2
			\left\|T_\lambda^{1/2}I_{K,X}^{\ast}\right\|^2
			\|h\|_{L^2(P_X)}^2\\
			&=
			\mathscr P_{b,\lambda}^2
			\left\|T_\lambda^{1/2}L_{K,X}T_\lambda^{1/2}\right\|
			\|h\|_{L^2(P_X)}^2\\
			&
			\leq
			\underline c^{-1}\mathscr P_{b,\lambda}^2
			\|h\|_{L^2(P_X)}^2,
		\end{aligned}
		\]
		where the second step uses the definition of $\mathscr{P}_{b,\lambda}$, and the last step follows using \eqref{equ: merged regularized operator bound 1}. This proves the first bound in \eqref{equ: empirical feature bound shift}.

		Since $\{(\sigma_{X,k},\phi_{X,k})\}_{k\geq 1}$ are normalized eigenpairs of $\mathcal{L}_{K,X}$,  the functions  
		\(\{\sqrt{\sigma_{X,k}}\phi_{X,k}\}_{k\geq 1}\) form an orthonormal eigenbasis of
		\(\mathcal H_K\), and  
		\begin{equation}
			\label{equ: LKonHK}
			\begin{aligned}
				L_{K,X} = \sum_{k\geq 1} \sigma_{X,k} \left(  \sqrt{\sigma_{X,k}}\phi_{X,k} \right)\otimes \left(  \sqrt{\sigma_{X,k}}\phi_{X,k} \right) \qquad \mathrm{on} \ \ \mathcal{H}_K.
			\end{aligned}
		\end{equation}
		Hence, 
		\begin{equation}
			\label{equ: Tb12Kx}
			\begin{aligned}
				\left\|T_{b,\lambda}^{1/2}K_X\right\|_K^2  &= 		\left\|T_{b,\lambda}^{1/2}T_{\lambda}^{-1/2}T_{\lambda}^{1/2}K_X\right\|_K^2  \leq
				\mathscr P_{b,\lambda}^2
				\langle K_X,T_\lambda K_X\rangle_K\\
				& \leq 	\underline c^{-1} 	\mathscr P_{b,\lambda}^2
				\left\langle K_X,(L_{K,X} + \lambda I)^{-1} K_X\right\rangle_K\\
				&=
				\underline c^{-1}\mathscr P_{b,\lambda}^2
				\sum_{k\geq1}
				\frac{\sigma_{X,k}}{\sigma_{X,k}+\lambda}
				\phi_{X,k}^2(X)\\
				&\lesssim
				\mathscr P_{b,\lambda}^2
				\sum_{k\geq1}
				\frac{\sigma_{0,k}}{\sigma_{0,k}+\lambda}
				\lesssim
				\mathscr P_{b,\lambda}^2\lambda^{-s}.
			\end{aligned}
		\end{equation}
		The second step follows from the definition of $\mathscr{P}_{b,\lambda}$. The third step follows from \eqref{equ: two operators are comparable}. The fourth step uses the spectral decomposition in~\eqref{equ: LKonHK}, and the reproducing property. 	The fifth step uses the uniform eigenfunction
		bound in Assumption~\ref{assumption: eigenvalues and eigenfunctions}, and $\sigma_{0,k}\asymp\sigma_{X,k}$, while the final
		inequality follows from
		\eqref{equ: effective dimension comparison}.
		Taking $b=D$ in \eqref{equ: Tb12Kx} proves the second bound in~\eqref{equ: empirical feature bound shift}.
		
		Using the bounds in \eqref{equ: empirical feature bound shift}, by the definitions of $A_{\lambda,0}$ and $A_{\lambda,D}$ in \eqref{equ: some kernel notation 1},
		we obtain
		\begin{equation}
			\label{equ: AbL bounds}
			\begin{aligned}
				\mathbb E\left\{
				A_{\lambda,b}^2(X_1,X_2)\mid D
				\right\}
				& \leq
				\operatorname*{ess\,sup}_{X\sim P_X}
				\int A_{\lambda,b}^2(X,x)\,dP_X(x)\\
				& =
				\operatorname*{ess\,sup}_{X\sim P_X}
				\left\langle
				T_{b,\lambda}^{1/2}K_X,
				T_{b,\lambda}^{1/2}L_{K,X}T_{b,\lambda}^{1/2}
				T_{b,\lambda}^{1/2}K_X
				\right\rangle_K\\
				&\leq 
				\left\|T_{b,\lambda}^{1/2}I_{K,X}^{\ast}\right\|^2
				\operatorname*{ess\,sup}_{X\sim P_X}
				\left\|T_{b,\lambda}^{1/2}K_X\right\|_K^2\\
				&
				\lesssim
				\mathscr P_{b,\lambda}^4\lambda^{-s},
			\end{aligned}
		\end{equation}
		and 
		\[
		\operatorname*{ess\,sup}_{X,X'\sim P_X }
		A_{\lambda,b}^2(X,X')
		\leq
		\left\{
		\operatorname*{ess\,sup}_{X\sim P_X}
		\left\|T_{b,\lambda}^{1/2}K_X\right\|_K^2
		\right\}^2
		\lesssim
		\mathscr P_{b,\lambda}^4\lambda^{-2s}.
		\]
		These prove
		\eqref{equ: merged empirical kernel squared norm}, \eqref{equ: empirical kernel row shift}, and 
		\eqref{equ: empirical kernel supremum squared shift}.
		
		The identity  $B^{-1} - A^{-1} = B^{-1}(A-B)A^{-1}$  with $A =L_{K,0} + \lambda I $ and $B = L_{K,D} + \lambda I $ gives
		\begin{equation}
			\label{equ: resolvent identity}
			\begin{aligned}
				&T_\lambda^{-1/2}
				(T_{D,\lambda}-T_\lambda)T_\lambda^{-1/2} =
				\left(T_\lambda^{-1/2}T_{D,\lambda}T_\lambda^{-1/2}\right)
				\left\{
				T_\lambda^{1/2}(L_{K,0}-L_{K,D})T_\lambda^{1/2}
				\right\}.
			\end{aligned}
		\end{equation}
		By the self-adjointness of $T_{\lambda}$ and $T_{D,\lambda}$,
		$$
		A_{\lambda,D}(X_1,X_2)
		-A_{\lambda,0}(X_1,X_2) = \left\langle  (T_{D,\lambda} - T_{\lambda})
		K_{X_1}
		,K_{X_2}\right\rangle_K.
		$$
		Then by Lemma~\ref{lemma: Balazs2005},
		\[
		\begin{aligned}
			\left\{A_{\lambda,D}(X_1,X_2)
			-A_{\lambda,0}(X_1,X_2) \right\}^2 = 	\left\langle (T_{D,\lambda} - T_{\lambda}) K_{X_1}\otimes K_{X_1}(T_{D,\lambda} - T_{\lambda}),  K_{X_2}\otimes K_{X_2}\right\rangle_{HS}.
		\end{aligned}
		\]
		This, together with the independence of $D$ and labeled observations, implies
		\[
		\begin{aligned}
			& \quad \,\,\mathbb E\left[
			\left\{
			A_{\lambda,D}(X_1,X_2)
			-A_{\lambda,0}(X_1,X_2)
			\right\}^2\mid D
			\right]\\
			&= \left\langle (T_{D,\lambda} - T_{\lambda})
			L_{K,X}(T_{D,\lambda} - T_{\lambda})
			,L_{K,X}\right\rangle_{HS}\\
			&=
			\left\|
			L_{K,X}^{1/2} T_{\lambda}^{1/2}T_{\lambda}^{-1/2}
			(T_{D,\lambda}-T_\lambda)T_{\lambda}^{-1/2}T_{\lambda}^{1/2}L_{K,X}^{1/2}
			\right\|_{HS}^2\\
			&\leq
			\left\|
			L_{K,X}^{1/2} T_{\lambda}^{1/2}\right\|_{HS}^2 \left\|T_{\lambda}^{-1/2}
			(T_{D,\lambda}-T_\lambda)T_{\lambda}^{-1/2}\right\|^2 \left\|T_{\lambda}^{1/2}L_{K,X}^{1/2}
			\right\|^2\\
			&\leq 
			\operatorname{Tr}\left(
			T_\lambda^{1/2}L_{K,X}T_\lambda^{1/2}
			\right)
			\left\|T_\lambda^{1/2}L_{K,X}T_\lambda^{1/2}\right\|
			\left\|T_\lambda^{-1/2}T_{D,\lambda}T_\lambda^{-1/2}\right\|^2
			\left\|
			T_\lambda^{1/2}(L_{K,0}-L_{K,D})T_\lambda^{1/2}
			\right\|^2\\
			&
			\lesssim
			\mathscr P_{D,\lambda}^4
			\mathscr Q_{D,\lambda}^2\lambda^{-s}.
		\end{aligned}
		\]
		The third step uses the inequality $\|ABC\|_{HS} \leq \|A\|_{HS}\|B\|\|C\|$ for a Hilbert-Schmidt operator $A$ and bounded operators $B$ and $C$. The fourth step follows from \eqref{equ: resolvent identity}. The last step uses \eqref{equ: merged regularized operator bound 1} and \eqref{equ: merged regularized operator bound 2},  and the definitions of $\mathscr{P}_{D,\lambda}$ and $\mathscr{Q}_{D,\lambda}$ in \eqref{equ: PDL} and \eqref{equ: QDL}.  This
		proves~\eqref{equ: merged kernel difference bound}.

		Now suppose $\|\mathbf f\|_{L^\infty(P_X)}\leq B$.
		By Lemma~\ref{lemma: residual moments at pseudo true},	 
		\[
		\mathbb E\left(\omega\otimes \omega\right)
		=\mathbb E\left\{
		s_2\left(X,\theta^\star\right)K_X\otimes K_X\right\}
		-\mu^\star\otimes\mu^\star
		\preceq C_2L_{K,X}.
		\]
		Sandwiching this inequality and using
		\eqref{equ: merged regularized operator bound 1} gives
		\[
		\left\|T_\lambda^{1/2}\mathbb E\left(\omega\otimes\omega\right)
		T_\lambda^{1/2}\right\|\leq C_2\underline c^{-1},
		\]
		and
		\[
		\begin{aligned}
			\left\|T_{D,\lambda}^{1/2}\mathbb E(\omega\otimes \omega)
			T_{D,\lambda}^{1/2}\right\| 
			&=
			\left\|T_{D,\lambda}^{1/2}T_\lambda^{-1/2}
			\left\{T_\lambda^{1/2}\mathbb E(\omega\otimes \omega)
			T_\lambda^{1/2}\right\}
			T_\lambda^{-1/2}T_{D,\lambda}^{1/2}\right \|\\
			& \leq
			\left\|T_{D,\lambda}^{1/2}T_\lambda^{-1/2}\right\|^2\left
			\|T_\lambda^{1/2}\mathbb E(\omega\otimes \omega)
			T_\lambda^{1/2}\right\|\\
			&
			\leq C_2\underline c^{-1}\mathscr P_{D,\lambda}^2,
		\end{aligned}
		\]
		where the last step follows from \eqref{equ: PDL}. Similarly,
		\[
		\begin{aligned}
			\operatorname{Tr}\left\{
			T_{D,\lambda}^{1/2}\mathbb E(\omega\otimes\omega)
			T_{D,\lambda}^{1/2}\right\}
			&\leq C_2\operatorname{Tr}\left\{
			T_{D,\lambda}^{1/2} L_{K,X}
			T_{D,\lambda}^{1/2}\right\}\\
			& =  C_2\operatorname{Tr}\left\{
			T_{D,\lambda}^{1/2}T_{\lambda}^{-1/2}(T_{\lambda}^{1/2} L_{K,X}T_{\lambda}^{1/2})T_{\lambda}^{-1/2} 
			T_{D,\lambda}^{1/2}\right\}\\
			& \leq  C_2\operatorname{Tr}\left\{
			T_{\lambda}^{1/2} L_{K,X}T_{\lambda}^{1/2} \right\} \left\|T_{D,\lambda}^{1/2}T_{\lambda}^{-1/2} \right\|^2\\
			& \lesssim \mathscr{P}_{D,\lambda}^2\mathcal  \lambda^{-s},  
		\end{aligned}
		\]
		where the third step follows from Lemma~\ref{lemma: trace inequality}, and the last step follows from \eqref{equ: merged regularized operator bound 2} and \eqref{equ: PDL}.
		These prove \eqref{equ: covariance empirical shift}. 
		
		For the remaining moment bounds, define
			\[
			C_\lambda
			:=
			T_\lambda^{1/2}
			\mathbb E\left\{
			s_2(X,\theta^\star)K_X\otimes K_X
			\right\}
			T_\lambda^{1/2}.
			\]
			By~\eqref{equ: residual moments},
			\[
			c_2T_\lambda^{1/2}L_{K,X}T_\lambda^{1/2}
			\preceq C_\lambda
			\preceq
			C_2T_\lambda^{1/2}L_{K,X}T_\lambda^{1/2}.
			\]
			Combining this with
			\eqref{equ: two operators are comparable} gives
			\[
			\frac{c_2}{\overline c}L_{K,0}T_\lambda
			\preceq C_\lambda
			\preceq
			\frac{C_2}{\underline c}L_{K,0}T_\lambda.
			\]
			In particular,
			\[
			\|C_\lambda\|
			\leq
			\frac{C_2}{\underline c}\|L_{K,0}T_\lambda\|
			\leq
			\frac{C_2}{\underline c}.
			\]
			Both $C_\lambda$ and $L_{K,0}T_\lambda$ are positive
			trace-class operators.
			The min--max characterization of eigenvalues ~\citep{bhatia2013matrix} therefore
			yields
			\[
			\left(\frac{c_2}{\overline c}\right)^2
			\sum_{k\geq1}
			\left(
			\frac{\sigma_{0,k}}{\sigma_{0,k}+\lambda}
			\right)^2
			\leq
			\operatorname{Tr}(C_\lambda^2)
			\leq
			\left(\frac{C_2}{\underline c}\right)^2
			\sum_{k\geq1}
			\left(
			\frac{\sigma_{0,k}}{\sigma_{0,k}+\lambda}
			\right)^2.
			\]
			Consequently, \eqref{equ: effective dimension comparison}
			implies
			\[
			\operatorname{Tr}(C_\lambda^2)
			\asymp
			\sum_{k\geq1}
			\left(
			\frac{\sigma_{0,k}}{\sigma_{0,k}+\lambda}
			\right)^2
			\asymp\lambda^{-s}.
			\]
		 
		By the definitions of kernels in \eqref{equ: some kernel notation 2}, we have
		$$
		\begin{aligned}
			& \quad \,\,	\mathbb{E}\left\{H_{\lambda,0}^2(Z_1,Z_2)\right\} \\&=\mathbb E  \left(\left\langle  T_\lambda^{1/2}K_{X_1},T_\lambda^{1/2}K_{X_2}\right\rangle_K^2\epsilon_1^2\epsilon_2^2\right)\\
			&  =\mathbb E  \left\{\left\langle  T_\lambda^{1/2}K_{X_1},T_\lambda^{1/2}K_{X_2}\right\rangle_K^2s_2(X_1,\theta^{\star})s_2(X_2,\theta^{\star})\right\}\\
			& = \left\langle T_\lambda^{1/2}
			\mathbb E\left\{
			s_2(X_1,\theta^\star)K_{X_1}\otimes K_{X_1}
			\right\}
			T_\lambda^{1/2},T_\lambda^{1/2}
			\mathbb E\left\{
			s_2(X_2,\theta^\star)K_{X_2}\otimes K_{X_2}
			\right\}
			T_\lambda^{1/2} \right\rangle_{HS}\\
			& = \mathrm{Tr}(C_{\lambda}^2)\asymp \lambda^{-s},
		\end{aligned}
		$$
		where the second step uses the iterated expectation formula, and the  third step uses Lemma~\ref{lemma: Balazs2005}. A similar calculation gives
		$$
		\begin{aligned}
			R_{\lambda,0}(Z_1,Z_2)
			& = \mathbb E  \left( \left\langle  T_\lambda^{1/2}K_{X},T_\lambda^{1/2}K_{X_1}\right\rangle_K\left\langle  T_\lambda^{1/2}K_{X},T_\lambda^{1/2}K_{X_2}\right\rangle_K\epsilon^2\epsilon_1\epsilon_2\mid Z_1,Z_2\right)\\
			& =  \left\langle T_\lambda^{1/2}
			\mathbb E\left\{
			s_2(X,\theta^\star)K_{X}\otimes K_{X}
			\right\}
			T_\lambda^{1/2},T_\lambda^{1/2} K_{X_1}\otimes K_{X_2}
			T_\lambda^{1/2} \right\rangle_{HS}\epsilon_1\epsilon_2\\
			&=	\left\langle
			T_\lambda^{1/2}K_{X_1}\epsilon_1,
			C_\lambda T_\lambda^{1/2}K_{X_2}\epsilon_2
			\right\rangle_K,
		\end{aligned}
		$$ 
		and thus
		\[
		\begin{aligned}
			\mathbb E\left\{
			R_{\lambda,0}^2(Z_1,Z_2)
			\right\} &=
			\left\langle  T_\lambda^{1/2}\mathbb{E} (K_{X_1} \otimes K_{X_1}\epsilon_1^2)T_\lambda^{1/2},C_{\lambda}T_\lambda^{1/2}\mathbb{E} (K_{X_2} \otimes K_{X_2}\epsilon_2^2)T_\lambda^{1/2}C_{\lambda}\right\rangle_{HS} \\
			&=
			\operatorname{Tr}(C_\lambda^4)
			\leq
			\|C_\lambda\|^2\operatorname{Tr}(C_\lambda^2)
			\lesssim\lambda^{-s}.
		\end{aligned}
		\]
		where the third step follows from Lemma~\ref{lemma: trace inequality}.
		These prove~\eqref{equ: EH squared shift}
		and~\eqref{equ: ER squared shift}.

		Finally, iterated expectation formula  and the bounds in 	\eqref{equ: merged empirical kernel squared norm}, \eqref{equ: empirical kernel row shift}, and 
		\eqref{equ: empirical kernel supremum squared shift} with $b=0$
		give
		\[
		\begin{aligned}
			\mathbb E\left\{
			H_{\lambda,0}^2(Z_1,Z_2)
			H_{\lambda,0}^2(Z_1,Z_3)
			\right\}
			&=
			\mathbb E\left[
			s_4(X_1,\theta^\star)
			\left\{
			\int A_{\lambda,0}^2(X_1,x)
			s_2(x,\theta^\star)\,dP_X(x)
			\right\}^2
			\right]\\
			&\leq
			C_4C_2^2
			\left\{
			\operatorname*{ess\,sup}_{X\sim P_X}
			\int A_{\lambda,0}^2(X,x)\,dP_X(x)
			\right\}^2
			\lesssim\lambda^{-2s},
		\end{aligned}
		\]
		and 
		\begin{align*}
			\mathbb E\left\{
			H_{\lambda,0}^4(Z_1,Z_2)
			\right\}
			&=
			\mathbb E\left\{
			A_{\lambda,0}^4(X_1,X_2)
			s_4(X_1,\theta^\star)s_4(X_2,\theta^\star)
			\right\}\\
			&\leq
			C_4^2
			\operatorname*{ess\,sup}_{X,X'\sim P_X }
			A_{\lambda,0}^2(X,X')
			\mathbb E\left\{A_{\lambda,0}^2(X_1,X_2)\right\}
			\lesssim\lambda^{-3s}.
		\end{align*}
		These establish~\eqref{equ: shared fourth shift}
		and~\eqref{equ: full fourth shift}. This completes the proof.
	\end{proof}
	
	The  following two lemmas provide  standard
	identities for rank-one operators on a real Hilbert space
	\citep{balazs2008hilbert} and a trace inequality for positive
	operators \citep{simon2005trace}.
 
	\begin{lemma}
		\label{lemma: Balazs2005}
		For any $f, g, u,v \in \mathcal{H}$,   $\langle f\otimes   g, u \otimes  v\rangle_{HS} = \langle f,u \rangle_{\mathcal{H}}\langle g,v \rangle_{\mathcal{H}}$. For $ B :=f\otimes f $, $\|B\| = \|B\|_{HS} = \mathrm{Tr}(B) = \|f\|_{\mathcal{H}}^2$. For a bounded operator $A$ on $\mathcal{H}$, $\langle Af,g \rangle_{\mathcal{H}}^2 = \langle A(f\otimes f)A^{\ast} g,g \rangle_{\mathcal{H}}  $.
	\end{lemma}
	\begin{lemma}
		\label{lemma: trace inequality}
		Let $A$ and $B$ be  bounded positive operators. If $A$ is trace-class, then $	\mathrm{Tr}(AB)  \leq     \|B\|\ \mathrm{Tr}(A)  .$
		
	\end{lemma}
	
	The following lemma follows from Hoeffding's decomposition
	and the variance identities in
	\citet[Chapter~5]{serfling2009approximation}.
	\begin{lemma}
		\label{lemma: conditional Hoeffding decomposition}
		Let $n\geq2$, and let $Z_1,\ldots,Z_n$ be i.i.d. observations, independent of $D$.
		Let $H$ be a measurable symmetric kernel, possibly depending
		on $D$, such that
		$\mathbb E\{H^2(Z_1,Z_2)\mid D\}<\infty$ almost surely.
		Define
		\[
		U_n:=\frac{2}{n(n-1)}\sum_{i<j}H(Z_i,Z_j),
		\qquad
		\mu_D:=\mathbb E\{H(Z_1,Z_2)\mid D\},
		\]
		and define the conditional projections  by
		\[
		\begin{aligned}
			\psi_1(Z_1)
			&:=
			\mathbb E\{H(Z_1,Z_2)\mid D,Z_1\}-\mu_D,\\
			\psi_2(Z_1,Z_2)
			&:=
			H(Z_1,Z_2)-\mu_D-\psi_1(Z_1)-\psi_1(Z_2).
		\end{aligned}
		\]
		Moreover, the following two variance identities hold:
		\begin{align}
			\operatorname{Var}(U_n\mid D)
			&=
			\frac4n\mathbb E\{\psi_1^2(Z_1)\mid D\}
			+
			\frac{2}{n(n-1)}
			\mathbb E\{\psi_2^2(Z_1,Z_2)\mid D\},
			\label{equ: conditional variance projection}
			\\
			&=
			\frac{4(n-2)}{n(n-1)}
			\mathbb E\{\psi_1^2(Z_1)\mid D\}
			+
			\frac{2}{n(n-1)}
			\left[
			\mathbb E\{H^2(Z_1,Z_2)\mid D\}-\mu_D^2
			\right].
			\label{equ: conditional variance kernel}
		\end{align}
		If $\mu_D=0$, then
		$\operatorname{Var}(U_n\mid D)=\mathbb E(U_n^2\mid D)$.
	\end{lemma}
	
%
	The following two examples verify the parameter-estimation
	conditions in
	Assumptions~\ref{assumption: identibility and parametric estimation under H0}
	and~\ref{assumption: uniform parametric estimation under H1}
	for linear and nonlinear least-squares estimators. In both examples, the observations are i.i.d., $P_X$ is fixed,
	and $e:=Y-m(X)$ satisfies
	$\mathbb E(e\mid X)=0$ and
	$\mathbb E(e^2\mid X)\leq C_e$,
	with a common constant $C_e$.
	
	\begin{example}[Linear least squares]
		\label{example: linear}
		Let $g(x,\theta)=v(x)^\top\theta$, where
		$v:\mathcal X\to\mathbb R^d$,
		and $\Theta=\mathbb R^d$.
		Suppose that
		$Q:=\mathbb E\{v(X)v(X)^\top\}$ satisfies
		$\lambda_{\min}(Q)\geq q_0>0$ and
		$\mathbb E\|v(X)\|_2^2<\infty$.
		Consider regression models satisfying
		\[
		m(x)=v(x)^\top\theta^\star+f(x),
		\qquad
		\mathbb E\{v(X)f(X)\}=0,
		\qquad
		\|f\|_{L^\infty(P_X)}\leq B.
		\]
		For $R(\theta):=\mathbb E\{Y-g(X,\theta)\}^2$,
		\[
		R(\theta)-R(\theta^\star)
		=
		(\theta-\theta^\star)^\top Q(\theta-\theta^\star)
		\geq q_0\|\theta-\theta^\star\|_2^2.
		\]
		Hence $\theta^\star$ is the unique population
		least-squares minimizer.
		
		Let $Q_n:=n^{-1}\sum_{i=1}^n v(X_i)v(X_i)^\top$.
		Use ordinary least squares when $Q_n$ is nonsingular
		and set $\widehat\theta=0$ otherwise.
		On the nonsingularity event,
		\[
		\begin{aligned}
			\widehat\theta-\theta^\star
			=
			Q_n^{-1}
			\left\{\frac1n\sum_{i=1}^n v(X_i)Y_i
			-Q_n\theta^\star\right\}=
			Q_n^{-1}\frac1n\sum_{i=1}^n
			v(X_i)\{f(X_i)+e_i\}.
		\end{aligned}
		\]
		Thus, on $\{\lambda_{\min}(Q_n)\geq q_0/2\}$,
		\begin{equation}
			\label{equ: increment linear}
			\sqrt n\|\widehat\theta-\theta^\star\|_2
			\leq
			\frac{2}{q_0}
			\left\|
			\frac1{\sqrt n}\sum_{i=1}^n
			v(X_i)\{f(X_i)+e_i\}
			\right\|_2.
		\end{equation}
		The summands are centered, and independence gives
		\[
		\begin{aligned}
			&\mathbb E
			\left\|
			\frac1{\sqrt n}\sum_{i=1}^n
			v(X_i)\{f(X_i)+e_i\}
			\right\|_2^2=
			\mathbb E\left[
			\|v(X)\|_2^2
			\{f(X)^2+\mathbb E(e^2\mid X)\}
			\right]
			\leq
			(B^2+C_e)\mathbb E\|v(X)\|_2^2.
		\end{aligned}
		\]
		For every $M>0$, we get from \eqref{equ: increment linear} that
		\[
		\begin{aligned}
			\{\sqrt n\|\widehat\theta-\theta^\star\|_2>M\}\subseteq
			\{\lambda_{\min}(Q_n)<q_0/2\}
			\cup
			\left\{
			\left\|
			\frac1{\sqrt n}\sum_{i=1}^n
			v(X_i)\{f(X_i)+e_i\}
			\right\|_2>\frac{q_0M}{2}
			\right\}.
		\end{aligned}
		\]
		Taking probabilities and applying Markov's inequality therefore
		yield, for every $M>0$,
		\[
		\mathbb P\{
		\sqrt n\|\widehat\theta-\theta^\star\|_2>M
		\}
		\leq
		\mathbb P\{\lambda_{\min}(Q_n)<q_0/2\}
		+
		\frac{4(B^2+C_e)\mathbb E\|v(X)\|_2^2}
		{q_0^2M^2}.
		\]
		This bound is uniform over the stated models.
		Since $P_X$ is fixed, the law of large numbers gives
		$\|Q_n-Q\|\xrightarrow{\mathbb P}0$, so the first
		probability tends to zero.
		Taking first $n\to\infty$ and then $M\to\infty$
		verifies
		Assumption~\ref{assumption: uniform parametric estimation under H1}.
		Setting $f=0$ and $\theta^\star=\theta_0$ verifies
		Assumption~\ref{assumption: identibility and parametric estimation under H0}.
		
		For a concrete example, take $d=2$, $v(x)=(1,x)^\top$,
		$P_X=\operatorname{Unif}(-1,1)$, and the periodic Sobolev kernel
		\begin{equation}
			\label{equ: periodic Sobolev kernel examples}
			K(x,x')
			:=
			1+2\sum_{j=1}^{\infty}
			j^{-2}\cos\{\pi j(x-x')\},
			\qquad x,x'\in[-1,1].
		\end{equation}
		Under $P_X$, its normalized eigenfunctions are
		$1$, $\sqrt{2}\cos(\pi jx)$, and $\sqrt{2}\sin(\pi jx)$,
		$j\geq1$, with corresponding eigenvalues
		$1$, $j^{-2}$, and $j^{-2}$.
		For $f(x)=\delta\cos(\pi x)$,
		\[
		Q=\operatorname{diag}(1,1/3),
		\qquad
		\mathbb E\{v(X)f(X)\}=0,
		\qquad
		\mathcal L_{K,X}f=f.
		\]
		Consequently, $f=\mathcal L_{K,X}^{\,r}u$ with $u=f$ for every $r>0$,
		and
		\[
		\|u\|_{L^2(P_X)}=\frac{|\delta|}{\sqrt{2}},
		\qquad
		\|f\|_{L^\infty(P_X)}=|\delta|.
		\]
		Hence $f\in\mathcal F_{r,B}$ for every $r>0$ whenever
		$0<|\delta|\leq\min\{B,\sqrt{2}\}$.
	\end{example}
	
	\begin{example}[Nonlinear least squares]
		\label{example: non-linear}
		Let $g(x,\theta)=\exp(\theta x)$,
		$\Theta=[-a,a]$ with $a>0$, and suppose that
		$|X|<1$ and $q:=\mathbb EX^2>0$.
		Fix $0<B<\exp(-3a)$ and consider regression models
		satisfying
		\[
		m(x)=\exp(\theta^\star x)+f(x),
		\qquad
		\theta^\star\in[-a,a],
		\qquad
		\|f\|_{L^\infty(P_X)}\leq B,
		\qquad
		\mathbb E\{f(X)X\exp(\theta^\star X)\}=0.
		\]
		Put $c_a:=\exp(-2a)-B\exp(a)>0$ and
		$d_\theta(x):=\exp(\theta x)-\exp(\theta^\star x)$.
		Taylor's formula gives
		\begin{equation}
			\label{equ: taylor formula}
			\begin{aligned}
				d_{\theta}(x) &= x(\theta - \theta^{\star})\int_0^1 \exp\left[\left\{ \theta^{\star} + t(\theta - \theta^{\star})\right\}x\right]dt \\
				& = x(\theta - \theta^{\star})  \exp (\theta^{\star}x) +x^2(\theta - \theta^{\star})^2\int_0^1(1-t) \exp\left[\left\{ \theta^{\star} + t(\theta - \theta^{\star})\right\}x\right]dt.
			\end{aligned}
		\end{equation}
		Since $|x| \leq 1$ and $|\theta|  \leq a$, we   have
		\begin{equation}
			\label{equ: taylor bound}
				\begin{aligned}
				|d_\theta(x)|
				\geq
				\exp(-a)|\theta-\theta^\star||x|,\quad 
				|d_\theta(x)
				-(\theta-\theta^\star)x\exp(\theta^\star x)|
				\leq
				\frac12\exp(a)(\theta-\theta^\star)^2x^2.
			\end{aligned}
		\end{equation}
		Hence, for $R(\theta):=\mathbb E\{Y-g(X,\theta)\}^2$,
		using $Y=g(X,\theta^\star)+f(X)+e$, we obtain
		\[
		\begin{aligned}
			R(\theta)-R(\theta^\star)
			&=
			\mathbb E\left[
			\{f(X)+e-d_\theta(X)\}^2-\{f(X)+e\}^2
			\right]\\
			&=
			\mathbb E\{d_\theta(X)^2\}
			-2\mathbb E\{f(X)d_\theta(X)\},
		\end{aligned}
		\]
		where we use $	\mathbb E\{e\,d_\theta(X)\}
		=
		\mathbb E\{d_\theta(X)\mathbb E(e\mid X)\}
		=0.$
		The   equation
		$\mathbb E\{f(X)X\exp(\theta^\star X)\}=0$
		and \eqref{equ: taylor bound} imply
		\[
		\begin{aligned}
			2\left|\mathbb E\{f(X)d_\theta(X)\}\right|
			&=
			2\left|
			\mathbb E\left[
			f(X)\{d_\theta(X)
			-(\theta-\theta^\star)X\exp(\theta^\star X)\}
			\right]
			\right|\\
			&\leq
			2B\,\mathbb E\left|
			d_\theta(X)
			-(\theta-\theta^\star)X\exp(\theta^\star X)
			\right|\\
			&\leq
			B\exp(a)q(\theta-\theta^\star)^2,
		\end{aligned}
		\]
	and
		\[
		\mathbb E\{d^2_\theta(X) \}
		\geq
		\exp(-2a)q(\theta-\theta^\star)^2.
		\]
		Combining these bounds gives, for every $\theta\in[-a,a]$,
		\[
		\begin{aligned}
			R(\theta)-R(\theta^\star)
			&\geq
			\mathbb E\{d_\theta(X)^2\}
			-2\left|\mathbb E\{f(X)d_\theta(X)\}\right| \geq
			\{\exp(-2a)-B\exp(a)\}
			q(\theta-\theta^\star)^2 =c_aq(\theta-\theta^\star)^2.
		\end{aligned}
		\]
		Since $c_aq>0$, every $\theta\neq\theta^\star$ has strictly
		larger population risk. Thus $\theta^\star$ is the unique
		population least-squares minimizer.

For a measurable function $H$, write
$\mathbb P_nH:=n^{-1}\sum_{i=1}^nH(X_i,Y_i)$.
Let $\widehat\theta$ be a measurable global minimizer of $R_n(\theta)
:=
\mathbb P_n\{Y-g(X,\theta)\}^2$, $ \theta\in[-a,a],$
and define
\[
U_n
:=
\sup_{|t|\leq a}
\left|\mathbb P_n\{eX^2\exp(tX)\}\right|,
\qquad
A_n:=c_a\mathbb P_nX^2-U_n.
\] 
Since $Y-g(X,\theta)=f(X)+e-d_\theta(X)$, expanding
the squared residuals gives
\[
\begin{aligned}
	R_n(\theta)-R_n(\theta^\star)
	&=
	\mathbb P_n\left[
	\{f(X)+e-d_\theta(X)\}^2-\{f(X)+e\}^2
	\right]\\
	&=
	\mathbb P_n\{d_\theta(X)^2\}
	-2\mathbb P_n\{f(X)d_\theta(X)\}
	-2\mathbb P_n\{e\,d_\theta(X)\}\\
	&=
	\mathbb P_n\{d_\theta(X)^2\}
	-2(\theta-\theta^\star)
	\mathbb P_n\{(f(X)+e)X\exp(\theta^\star X)\}\\
	&\quad-
	2\mathbb P_n\left[
	f(X)\{d_\theta(X)
	-(\theta-\theta^\star)X\exp(\theta^\star X)\}
	\right]\\
	&\quad-
	2\mathbb P_n\left[
	e\{d_\theta(X)
	-(\theta-\theta^\star)X\exp(\theta^\star X)\}
	\right].
\end{aligned}
\] 
The  lower bound for $|d_\theta(X)|$ in \eqref{equ: taylor bound} gives $\mathbb P_n\{d^2_\theta(X)\}
\geq
\exp(-2a)(\theta-\theta^\star)^2\mathbb P_nX^2.$ Using $\|f\|_{L^\infty(P_X)}\leq B$ and the upper bound in \eqref{equ: taylor bound}, we obtain
\[
\begin{aligned}
	2\left|
	\mathbb P_n\left[
	f(X)\{d_\theta(X)
	-(\theta-\theta^\star)X\exp(\theta^\star X)\}
	\right]\right|& \leq
	2B\,\mathbb P_n\left|
	d_\theta(X)
	-(\theta-\theta^\star)X\exp(\theta^\star X)
	\right|\\
	& \leq
	B\exp(a)(\theta-\theta^\star)^2\mathbb P_nX^2.
\end{aligned}
\]
Furthermore, since  $\theta,\theta^\star\in[-a,a]$, the integral Taylor formula
in~\eqref{equ: taylor formula} yields   
\[
\begin{aligned}
	&2\left|
	\mathbb P_n\left[
	e\{d_\theta(X)
	-(\theta-\theta^\star)X\exp(\theta^\star X)\}
	\right]\right|\\
	&\quad\leq
	2(\theta-\theta^\star)^2
	\int_0^1(1-t)
	\left|
	\mathbb P_n\left[
	eX^2
	\exp\{(\theta^\star+t(\theta-\theta^\star))X\}
	\right]\right|dt\\
	&\quad\leq
	2(\theta-\theta^\star)^2U_n
	\int_0^1(1-t)\,dt =
	(\theta-\theta^\star)^2U_n.
\end{aligned}
\]
Combining these bounds, we obtain, almost surely, for every
$\theta\in[-a,a]$,
\[
\begin{aligned}
	R_n(\theta)-R_n(\theta^\star)
	&\geq
	\exp(-2a)(\theta-\theta^\star)^2\mathbb P_nX^2\\
	&\quad-
	B\exp(a)(\theta-\theta^\star)^2\mathbb P_nX^2
	-(\theta-\theta^\star)^2U_n\\
	&\quad-
	2|\theta-\theta^\star|
	\left|
	\mathbb P_n\{(f(X)+e)X\exp(\theta^\star X)\}
	\right|.
\end{aligned}
\]
Recalling $c_a=\exp(-2a)-B\exp(a)$ and
$A_n=c_a\mathbb P_nX^2-U_n$, this becomes
\begin{equation}
	\label{equ: empirical risk nonlinear}
	\begin{aligned}
		R_n(\theta)-R_n(\theta^\star)
		 \geq
		A_n(\theta-\theta^\star)^2 -
		2|\theta-\theta^\star|
		\left|
		\mathbb P_n\{(f(X)+e)X\exp(\theta^\star X)\}
		\right|.
	\end{aligned}
\end{equation} 
 
	Since $R_n(\widehat\theta)\leq R_n(\theta^\star)$, 
	on $\{A_n>0\}$, we have
	\begin{equation}
		\label{equ: increment nonlinear}
		\sqrt n|\widehat\theta-\theta^\star|
		\leq
		\frac{2}{A_n}
		\left|
		\frac1{\sqrt n}\sum_{i=1}^n
		\{f(X_i)+e_i\}X_i\exp(\theta^\star X_i)
		\right|.
	\end{equation}
	 The summands are centered, and independence of the observations gives
	\[
	\begin{aligned}
		 \mathbb E\left\{
		\left(
		\frac1{\sqrt n}\sum_{i=1}^n
		\{f(X_i)+e_i\}X_i\exp(\theta^\star X_i)
		\right)^2
		\right\} &=
		\mathbb E\left[
		\{f(X)+e\}^2X^2\exp(2\theta^\star X)
		\right]\\
		& =
		\mathbb E\left[
		\{f(X)^2+\mathbb E(e^2\mid X)\}
		X^2\exp(2\theta^\star X)
		\right]\\
		& \leq
		\exp(2a)(B^2+C_e)q.
	\end{aligned}
	\] 
	To control $A_n=c_a\mathbb P_nX^2-U_n$, we   bound $U_n$.
	Let $h_n(t):=\mathbb P_n\{eX^2\exp(tX)\}$ and $
	h_n'(t)=\mathbb P_n\{eX^3\exp(tX)\}.$
		The fundamental theorem of calculus gives
 \begin{equation}
 		\label{equ: fundamental theorem}
 	U_n
 	=
 	\sup_{|t|\leq a}|h_n(t)|
 	\leq
 	|h_n(0)|+\int_{-a}^a|h_n'(t)|\,dt.
 \end{equation}
	Since $\mathbb E(e\mid X)=0$, both $h_n(0)$ and
	$h_n'(t)$ are centered. Independence of the observations
	therefore gives
	\[
	\begin{aligned}
		\mathbb E\{h_n^2(0) \}
		&=
		\frac1n\mathbb E(e^2X^4)
		\leq\frac{C_e}{n}\mathbb EX^4,\\
		\mathbb E\left[\{h_n'(t)\}^2\right]
		&=
		\frac1n\mathbb E\{e^2X^6\exp(2tX)\}
		\leq
		\frac{C_e\exp(2a)}n\mathbb EX^6,
		\qquad |t|\leq a.
	\end{aligned}
	\]
	Taking expectations in  \eqref{equ: fundamental theorem}
	and applying the Cauchy--Schwarz inequality, we obtain
	\[
	\begin{aligned}
		\mathbb EU_n
		&\leq
		\mathbb E|h_n(0)|
		+\int_{-a}^a\mathbb E|h_n'(t)|\,dt \leq
		\{\mathbb E h^2_n(0) \}^{1/2}
		+\int_{-a}^a
		[\mathbb E\{ h_n'(t)\}^2]^{1/2}\,dt\\
		&\leq
		\sqrt{\frac{C_e}{n}}
		\left\{
		(\mathbb EX^4)^{1/2}
		+2a\exp(a)(\mathbb EX^6)^{1/2}
		\right\} \leq\frac{C}{\sqrt n},
	\end{aligned}
	\]
	where the last inequality uses $|X|<1$, and $C$
	depends only on $a$ and $C_e$.

	On the event $	\{|\mathbb P_nX^2-q|\leq q/4,\;
	U_n\leq c_aq/4\},$
	we have $	A_n
	\geq  c_aq/2.$
	Together with~\eqref{equ: increment nonlinear}, this implies
	that, for every $M>0$,
	\[
	\begin{aligned}
		\{\sqrt n|\widehat\theta-\theta^\star|>M\}
		&\subseteq
		\{|\mathbb P_nX^2-q|>q/4\}
		\cup
		\{U_n>c_aq/4\}\\
		&\quad\cup
		\left\{
		\left|
		\frac1{\sqrt n}\sum_{i=1}^n
		\{f(X_i)+e_i\}X_i\exp(\theta^\star X_i)
		\right|
		>\frac{c_aqM}{4}
		\right\}.
	\end{aligned}
	\]
	Taking probabilities and applying Markov's inequality  yields
	\[
	\begin{aligned}
		\mathbb P\{
		\sqrt n|\widehat\theta-\theta^\star|>M
		\}
		&\leq
		\mathbb P\{|\mathbb P_nX^2-q|>q/4\}
		+\frac{C}{\sqrt n}
		+\frac{C}{M^2},
	\end{aligned}
	\]
	where the constant $C>0$ may depend on $a,B,C_e,q$,
	but is independent of the particular model in the stated
	collection.
		The first probability tends to zero by the law of large numbers.
		Taking first $n\to\infty$ and then $M\to\infty$
		verifies
		Assumption~\ref{assumption: uniform parametric estimation under H1}.
		Setting $f=0$ and $\theta^\star=\theta_0$ verifies
		Assumption~\ref{assumption: identibility and parametric estimation under H0}.
		
		For a concrete example, take $a=1/5$, $B=1/10$,
		$P_X=\operatorname{Unif}(-1,1)$, $\theta^\star=0$,
		and the kernel in~\eqref{equ: periodic Sobolev kernel examples}.
		Then $q=1/3$ and $B<\exp(-3a)$.
		For $f(x)=\delta\cos(\pi x)$ with
		$0<|\delta|\leq1/10$, symmetry gives
		$\mathbb E\{f(X)X\exp(\theta^\star X)\}
		=\mathbb E\{Xf(X)\}=0$.
		The calculation in Example~\ref{example: linear} also shows that
		$f\in\mathcal F_{r,B}$ for every $r>0$.
	\end{example}
	
\end{document}